\documentclass[aps,amsfonts,amssymb,10pt,letterpaper, 
final, nofootinbib, longbibliography, 
onecolumn,notitlepage,hyperref,twoside,allowtoday,accepted=2020-05-13]{quantumarticle}
\pdfoutput=1
\usepackage[table]{xcolor}
\usepackage{etoolbox}
\makeatletter
\patchcmd{\@hex@@Hex}{f\else}{F\else}{\typeout{Patching xcolor}}{}
\makeatother

\usepackage{adjustbox}
\usepackage{soul}
\usepackage{graphicx,amsmath,amsfonts,amssymb,amsthm,bm,bbm}
\usepackage{tikz,pgfplots}
\pgfplotsset{compat=newest}
\pgfplotsset{plot coordinates/math parser=false}
\usetikzlibrary{arrows,decorations.pathmorphing,backgrounds,positioning,fit,calc,patterns,shapes,decorations.markings,shadings,snakes}
\usetikzlibrary{external}
\usepackage{ifthen}
\usepackage{multirow}
\usepackage{diagbox}
\usepackage{stmaryrd}
\usepackage{tabularx}
\usepackage{booktabs}
\usepackage{rotating}
\newcommand{\greyhline}{\arrayrulecolor[gray]{.9}\hline\arrayrulecolor{black}}
\newcommand{\greycline}[1]{\arrayrulecolor[gray]{.8}\cline{#1}\arrayrulecolor{black}}

\renewcommand{\eqref}[1]{Eqn.~\ref{#1}}

\makeatletter
\newcommand\xleftrightarrow[2][]{%
	\ext@arrow 9999{\longleftrightarrowfill@}{#1}{#2}}
\newcommand\longleftrightarrowfill@{%
	\arrowfill@\leftarrow\relbar\rightarrow}
\makeatother

\usepackage{pdftexcmds}

\makeatletter
\ifcase\pdf@shellescape
\or
\tikzset{external/system call={pdflatex \tikzexternalcheckshellescape -halt-on-error -interaction=batchmode -jobname "\image" "\texsource";rm "\image".log && rm "\image".dpth}}
\or
\fi
\makeatother

\newtheoremstyle{exampleA}
{\topsep}
{10pt}
{}
{}
{\bfseries}
{\\\\}
{ 10mm }
{\thmname{#1}\thmnumber{ #2}\thmnote{ (#3)}}

\theoremstyle{exampleA}

\usepackage[pdfpagelabels,pdftex,bookmarks,breaklinks]{hyperref}
\definecolor{darkblue}{RGB}{0,0,127} 
\definecolor{darkgreen}{RGB}{0,180,0}
\definecolor{darkred}{RGB}{180,0,0}
\hypersetup{
	colorlinks,
	linkcolor=darkblue,
	citecolor=darkgreen,
	filecolor=red,
	urlcolor=blue,
	pdftitle={Computing data for Levin-Wen with defects},
	pdfauthor={Jacob C Bridgeman and Daniel Barter}
}

\definecolor{tctwistcolor}{RGB}{255,0,0}
\definecolor{tcmcolor}{RGB}{0,0,180}
\definecolor{tcecolor}{RGB}{0,180,0}
\definecolor{tcTTppcolor}{RGB}{184,134,11}
\definecolor{tcTTpmcolor}{RGB}{252,15,192}

\theoremstyle{plain}
\newtheorem{theorem}{Theorem}

\newtheorem{proposition}[theorem]{Proposition}

\theoremstyle{definition}
\newtheorem{definition}[theorem]{Definition}

\newtheorem{example}[theorem]{Example}

\newcommand{\ZZ}[1]{\mathbb{Z}/#1\mathbb{Z}}

\newcommand{\onlinecite}[1]{Ref.~\cite{#1}}
\newcommand{\onlinecites}[1]{Refs.~\cite{#1}}

\newcommand{\ket}[1]{|{#1}\rangle}

\newcommand{\D}[1]{\mathcal{D}{} }

\newcommand{\restrict}[1]{\raise-.2ex\hbox{\ensuremath|}_{#1}}

\definecolor{dandark}{HTML}{666666}
\definecolor{tensorblue}{rgb}{0.8,0.8,1}
\definecolor{tensorred}{rgb}{1,0.5,0.5}
\definecolor{tensorgreen}{rgb}{0.6,1,0.6}
\definecolor{tensorpurp}{rgb}{1,0.5,1}

\definecolor{pointdefectcolor}{HTML}{000000} 
\definecolor{boundarycolor}{HTML}{93a1a1} 
\definecolor{bulkcolor}{HTML}{e5e5e5} 

\tikzset{ten/.style={fill=tensorblue}}
\tikzset{tenred/.style={fill=tensorred}}
\tikzset{tengreen/.style={fill=tensorgreen}}
\tikzset{tenpurp/.style={fill=tensorpurp}}

\makeatletter
\newcommand{\vast}{\bBigg@{4}}
\newcommand{\Vast}{\bBigg@{9}}
\makeatother

\def\Put(#1,#2)#3{\leavevmode\makebox(0,0){\put(#1,#2){#3}}}

\makeatletter
\def\pgf@plot@curveto@handler@finish{%
  \ifpgf@plot@started%
    \pgfpathcurvebetweentimecontinue{0}{0.995}{\pgf@plot@curveto@first}{\pgf@plot@curveto@first@support}{\pgf@plot@curveto@second}{\pgf@plot@curveto@second}%
  \fi%
}
\makeatother

\newlength\figureheight
\newlength\figurewidth

\newcommand{\includeTikz}[2]
{
	\tikzifexternalizing
	{
		\includeTikzrm{#1}{#2}
	}
	{
		\IfFileExists{figures/#1.pdf}{
			\includegraphics{figures/#1}
		}
		{
			\includeTikzrm{#1}{#2}
		}
	}
}

\newcommand{\includeTikzrm}[2]{
	\tikzset{external/remake next}
	\tikzsetnextfilename{#1}
	#2
}

  \definecolor{nicegreena}{RGB}{1,115,16}
  \definecolor{nicegreenb}{RGB}{1,240,16}

  \definecolor{nicegreen}{RGB}{60,183,82}

    \colorlet{ccred}{red!20}
    \colorlet{ccgreen}{green!50}
    \colorlet{ccblue}{blue!20}

  \tikzset{hexr/.style= {shape=regular polygon,regular polygon sides=6,minimum size=1cm, draw,inner sep=0,anchor=center,fill=red!50}}
  \tikzset{hexg/.style= {shape=regular polygon,regular polygon sides=6,minimum size=1cm, draw,inner sep=0,anchor=center,fill=green!50}}
  \tikzset{hexb/.style= {shape=regular polygon,regular polygon sides=6,minimum size=1cm, draw,inner sep=0,anchor=center,fill=blue!50}}

  \definecolor{tensorblue}{rgb}{0.8,0.8,1}
  \definecolor{tensorred}{rgb}{1,0.5,0.5}
  \definecolor{tensorpurp}{rgb}{1,0.5,1}

  \tikzset{nonesty/.style={fill=none,draw=none}}
  \tikzset{ten/.style={fill=tensorblue}}
  \tikzset{tenred/.style={fill=tensorred}}
  \tikzset{tengreen/.style={fill=green!50!black!50}}
  \tikzset{tenpurp/.style={fill=tensorpurp}}
  \tikzset{tengrey/.style={fill=black!20}}
  \tikzset{tenorange/.style={fill=orange!30}}
  \tikzset{u/.style={fill=blue!20,draw=black}}
  \tikzset{w/.style={fill=green!50!black!50,draw=black}}

\usetikzlibrary{calc}
\tikzstyle{inline text}=[text height=1.1ex, text depth=0.1ex, yshift=-.1ex]

\newcommand{\inflationalhs}[3]{
	\ifthenelse{\equal{#1}{}}{}
	{
		\draw[blue](-.75,1)--(0,.5);
		\node[above,inline text,blue] at (-.75,1) {#1};
	}
	\ifthenelse{\equal{#3}{}}{}
	{
		\draw[blue](.75,1)--(0,-.5);
		\node[above,inline text,blue] at (.75,1) {#3};
	}
	\draw[red,ultra thick] (0,-1)--(0,1);
	\node[below,inline text,red] at (0,-1) {#2};
}

\newcommand{\inflationarhs}[5]{
	\ifthenelse{\equal{#1}{}}{}
	{
		\draw[blue](-1.25,1)--(-.5,.5);
		\node[above,inline text,blue] at (-1.25,1) {#1};
	}
	\ifthenelse{\equal{#3}{}}{}
	{
		\draw[blue](-.5,-.75)--(.5,-.25);
		\node[above,inline text,blue] at (0,-.5) {#3};
	}
	\ifthenelse{\equal{#5}{}}{}
	{
		\draw[blue](1.25,1)--(.5,0);
		\node[above,inline text,blue] at (1.25,1) {#5};
	}
	\draw[red,ultra thick] (-.5,-1)--(-.5,1);
	\draw[orange,ultra thick] (.5,-1)--(.5,1);
	\node[below,inline text,red] at (-.5,-1) {#2};
	\node[below,inline text,orange] at (.5,-1) {#4};
}

\newcommand{\inflationarhss}[5]{
	\ifthenelse{\equal{#1}{}}{}
	{
		\draw[blue](-1.25,1)--(-.5,.5);
		\node[above,inline text,blue] at (-1.25,1) {#1};
	}
	\ifthenelse{\equal{#3}{}}{}
	{
		\draw[blue](-.5,-.75)--(.5,-.25);
		\node[above,inline text,blue] at (0,-.5) {#3};
	}
	\ifthenelse{\equal{#5}{}}{}
	{
		\draw[blue](1.25,1)--(.5,0);
		\node[above,inline text,blue] at (1.25,1) {#5};
	}
	\draw[red,ultra thick] (-.5,-1)--(-.5,1);
	\draw[red,ultra thick] (.5,-1)--(.5,1);
	\node[below,inline text,red] at (-.5,-1) {#2};
	\node[below,inline text,red] at (.5,-1) {#4};
}

\newcommand{\inflationblhs}[3]{
	\ifthenelse{\equal{#1}{}}{}
	{
		\draw[blue](-.75,-1)--(0,-.5);
		\node[below,inline text,blue] at (-.75,-1) {#1};
	}
	\ifthenelse{\equal{#3}{}}{}
	{
		\draw[blue](.75,-1)--(0,.5);
		\node[below,inline text,blue] at (.75,-1) {#3};
	}
	\draw[red,ultra thick] (0,-1)--(0,1);
	\node[above,inline text,red] at (0,1) {#2};
}

\newcommand{\inflationbrhs}[5]{
	\ifthenelse{\equal{#1}{}}{}
	{
		\draw[blue](-1.25,-1)--(-.5,-.5);
		\node[below,inline text,blue] at (-1.25,-1) {#1};
	}
	\ifthenelse{\equal{#3}{}}{}
	{
		\draw[blue](-.5,-.25)--(.5,.25);
		\node[above,inline text,blue] at (0,0) {#3};
	}
	\ifthenelse{\equal{#5}{}}{}
	{
		\draw[blue](1.25,-1)--(.5,.5);
		\node[below,inline text,blue] at (1.25,-1) {#5};
	}
	\draw[red,ultra thick] (-.5,-1)--(-.5,1);
	\draw[orange,ultra thick] (.5,-1)--(.5,1);
	\node[above,inline text,red] at (-.5,1) {#2};
	\node[above,inline text,orange] at (.5,1) {#4};
}

\newcommand{\inflationbrhss}[5]{
	\ifthenelse{\equal{#1}{}}{}
	{
		\draw[blue](-1.25,-1)--(-.5,-.5);
		\node[below,inline text,blue] at (-1.25,-1) {#1};
	}
	\ifthenelse{\equal{#3}{}}{}
	{
		\draw[blue](-.5,-.25)--(.5,.25);
		\node[above,inline text,blue] at (0,0) {#3};
	}
	\ifthenelse{\equal{#5}{}}{}
	{
		\draw[blue](1.25,-1)--(.5,.5);
		\node[below,inline text,blue] at (1.25,-1) {#5};
	}
	\draw[red,ultra thick] (-.5,-1)--(-.5,1);
	\draw[red,ultra thick] (.5,-1)--(.5,1);
	\node[above,inline text,red] at (-.5,1) {#2};
	\node[above,inline text,red] at (.5,1) {#4};
}

\newcommand{\annss}[2]{
	\def\ra{.5};
	\def\rb{1.5};
	\ifthenelse{\equal{#1}{}}{}
		{
			\node[left,inline text,blue] at ($(0, 0) + (180:.75 and .75)$() {\footnotesize#1};
			\draw[blue,xscale=2.5] (0,-.75)to[out=90+45,in=270-45] (0,.75);
		}
	\ifthenelse{\equal{#2}{}}{}
	{
		\node[right,inline text,blue] at ($(0, 0) + (0:.75 and .75)$() {\footnotesize#2};
		\draw[blue,xscale=1.5] (0,-1.25)to[out=90-45,in=270+45] (0,1.25);
	}
	\draw[red,ultra thick] (0,\ra)--(0,\rb);
	\draw[red,ultra thick] (0,-\ra)--(0,-\rb);
	\node[below,inline text,red] at ($(0, 0) + (270:1.5 and 1.5)$() {\footnotesize\ta};
	\node[above,inline text,red] at ($(0, 0) + (90:1.5 and 1.5)$() {\footnotesize\tb};
	\node[above,inline text,red] at ($(0, 0) + (270:.5 and .5)$() {\footnotesize\tap};
	\node[below,inline text,red] at ($(0, 0) + (90:.5 and .5)$() {\footnotesize\tbp};
	\draw (0,0) circle (\ra);
	\draw (0,0) circle (\rb);
}

\newcommand{\annst}[2]{
	\def\ra{.5};
	\def\rb{1.5};
	\ifthenelse{\equal{#1}{}}{}
	{
		\node[left,inline text,blue] at ($(0, 0) + (180:.75 and .75)$() {\footnotesize#1};
		\draw[blue,xscale=2.5] (0,-.75)to[out=90+45,in=270-45] (0,.75);
	}
	\ifthenelse{\equal{#2}{}}{}
	{
		\node[right,inline text,blue] at ($(0, 0) + (0:.75 and .75)$() {\footnotesize#2};
		\draw[blue,xscale=1.5] (0,-1.25)to[out=90-45,in=270+45] (0,1.25);
	}
	\draw[orange,ultra thick] (0,\ra)--(0,\rb);
	\draw[red,ultra thick] (0,-\ra)--(0,-\rb);
	\node[below,inline text,red] at ($(0, 0) + (270:1.5 and 1.5)$() {\footnotesize\ta};
	\node[above,inline text,orange] at ($(0, 0) + (90:1.5 and 1.5)$() {\footnotesize\tb};
	\node[above,inline text,red] at ($(0, 0) + (270:.5 and .5)$() {\footnotesize\tap};
	\node[below,inline text,orange] at ($(0, 0) + (90:.5 and .5)$() {\footnotesize\tbp};
	\draw (0,0) circle (\ra);
	\draw (0,0) circle (\rb);
}

\newcommand{\annstpq}[4]{
	\ifthenelse{\equal{#1}{}}{}
	{
		\node[left,inline text,blue] at ($(180:.6 and .75)$() {\footnotesize#1};
		\coordinate (A) at ({.5*cos(-140)},{1*sin(140)});
		\draw[blue] ({.5*cos(-140)},{1*sin(-140)}) to[out=90+45,in=270-45] (A);
	}
	\ifthenelse{\equal{#2}{}}{}
	{
		\node[above,inline text,blue] at ($(90:.5 and .9)$() {\footnotesize#2};
		\coordinate (B) at ({.5*cos(40)},{1.7*sin(40)});
		\draw[blue] ({.5*cos(140)},{1.2*sin(140)}) to[out=45,in=270-45] (B);
	}
	\ifthenelse{\equal{#3}{}}{}
	{
		\node[right,inline text,blue] at ($(0,.3)+(0:.75 and .75)$() {\footnotesize#3};
		\coordinate (C) at ({.5*cos(40)},{2*sin(40)});
		\draw[blue] ({.5*cos(40)},{1*sin(-40)}) to[out=90-45,in=270+45] (C);
	}
	\ifthenelse{\equal{#4}{}}{}
	{
		\node[above,inline text,blue] at ($(270:.5 and 1.1)$() {\footnotesize#4};
		\coordinate (A) at ({.5*cos(-40)},{1.5*sin(-40)});
		\draw[blue] ({.5*cos(-140)},{2*sin(-140)}) to[out=45,in=270-45] (A);
	}
	\draw[red,ultra thick] ({.5*cos(-140)},{.5*sin(-140)+.1})--({.5*cos(-140)},{-1.5*sin(acos(1/3*cos(-140)))});
	\draw[orange,ultra thick] ({.5*cos(-40)},{.5*sin(-40)+.1})--({.5*cos(-40)},{-1.5*sin(acos(1/3*cos(-40)))});
	\draw[darkgreen,ultra thick] ({.5*cos(140)},{.5*sin(140)-.1})--({.5*cos(140)},{1.5*sin(acos(1/3*cos(140)))});
	\draw[darkred,ultra thick] ({.5*cos(40)},{.5*sin(40)-.1})--({.5*cos(40)},{1.5*sin(acos(1/3*cos(40)))});
	\draw[fill=white] (0,0) ellipse (.5 and .5);
	\draw (0,0) circle (1.5 and 1.5);
	\node[below,inline text,red] at ({.5*cos(-140)},{-1.5*sin(acos(1/3*cos(-140)))}) {\footnotesize\ta};
	\node[below,inline text,orange] at ({.5*cos(-40)},{-1.5*sin(acos(1/3*cos(-40)))}) {\footnotesize\tb};
	\node[above,inline text,darkgreen] at ({.5*cos(140)},{1.5*sin(acos(1/3*cos(140)))}) {\footnotesize\tc};
	\node[above,inline text,darkred] at ({.5*cos(40)},{1.5*sin(acos(1/3*cos(40)))}) {\footnotesize\td};
}
\newcommand{\annsspq}[4]{
	\ifthenelse{\equal{#1}{}}{}
	{
		\node[left,inline text,blue] at ($(180:.6 and .75)$() {\footnotesize#1};
		\coordinate (A) at ({.5*cos(-140)},{1*sin(140)});
		\draw[blue] ({.5*cos(-140)},{1*sin(-140)}) to[out=90+45,in=270-45] (A);
	}
	\ifthenelse{\equal{#2}{}}{}
	{
		\node[above,inline text,blue] at ($(90:.5 and .9)$() {\footnotesize#2};
		\coordinate (B) at ({.5*cos(40)},{1.7*sin(40)});
		\draw[blue] ({.5*cos(140)},{1.2*sin(140)}) to[out=45,in=270-45] (B);
	}
	\ifthenelse{\equal{#3}{}}{}
	{
		\node[right,inline text,blue] at ($(0,.3)+(0:.75 and .75)$() {\footnotesize#3};
		\coordinate (C) at ({.5*cos(40)},{2*sin(40)});
		\draw[blue] ({.5*cos(40)},{1*sin(-40)}) to[out=90-45,in=270+45] (C);
	}
	\ifthenelse{\equal{#4}{}}{}
	{
		\node[above,inline text,blue] at ($(270:.5 and 1.1)$() {\footnotesize#4};
		\coordinate (A) at ({.5*cos(-40)},{1.5*sin(-40)});
		\draw[blue] ({.5*cos(-140)},{2*sin(-140)}) to[out=45,in=270-45] (A);
	}
	\draw[red,ultra thick] ({.5*cos(-140)},{.5*sin(-140)+.1})--({.5*cos(-140)},{-1.5*sin(acos(1/3*cos(-140)))});
	\draw[red,ultra thick] ({.5*cos(-40)},{.5*sin(-40)+.1})--({.5*cos(-40)},{-1.5*sin(acos(1/3*cos(-40)))});
	\draw[darkgreen,ultra thick] ({.5*cos(140)},{.5*sin(140)-.1})--({.5*cos(140)},{1.5*sin(acos(1/3*cos(140)))});
	\draw[darkred,ultra thick] ({.5*cos(40)},{.5*sin(40)-.1})--({.5*cos(40)},{1.5*sin(acos(1/3*cos(40)))});
	\draw[fill=white] (0,0) ellipse (.5 and .5);
	\draw (0,0) circle (1.5 and 1.5);
	\node[below,inline text,red] at ({.5*cos(-140)},{-1.5*sin(acos(1/3*cos(-140)))}) {\footnotesize\ta};
	\node[below,inline text,red] at ({.5*cos(-40)},{-1.5*sin(acos(1/3*cos(-40)))}) {\footnotesize\tb};
	\node[above,inline text,darkgreen] at ({.5*cos(140)},{1.5*sin(acos(1/3*cos(140)))}) {\footnotesize\tc};
	\node[above,inline text,darkred] at ({.5*cos(40)},{1.5*sin(acos(1/3*cos(40)))}) {\footnotesize\td};
}
\newcommand{\annstpp}[4]{
	\ifthenelse{\equal{#1}{}}{}
	{
		\node[left,inline text,blue] at ($(180:.6 and .75)$() {\footnotesize#1};
		\coordinate (A) at ({.5*cos(-140)},{1*sin(140)});
		\draw[blue] ({.5*cos(-140)},{1*sin(-140)}) to[out=90+45,in=270-45] (A);
	}
	\ifthenelse{\equal{#2}{}}{}
	{
		\node[above,inline text,blue] at ($(90:.5 and .9)$() {\footnotesize#2};
		\coordinate (B) at ({.5*cos(40)},{1.7*sin(40)});
		\draw[blue] ({.5*cos(140)},{1.2*sin(140)}) to[out=45,in=270-45] (B);
	}
	\ifthenelse{\equal{#3}{}}{}
	{
		\node[right,inline text,blue] at ($(0,.3)+(0:.75 and .75)$() {\footnotesize#3};
		\coordinate (C) at ({.5*cos(40)},{2*sin(40)});
		\draw[blue] ({.5*cos(40)},{1*sin(-40)}) to[out=90-45,in=270+45] (C);
	}
	\ifthenelse{\equal{#4}{}}{}
	{
		\node[above,inline text,blue] at ($(270:.5 and 1.1)$() {\footnotesize#4};
		\coordinate (A) at ({.5*cos(-40)},{1.5*sin(-40)});
		\draw[blue] ({.5*cos(-140)},{2*sin(-140)}) to[out=45,in=270-45] (A);
	}
	\draw[red,ultra thick] ({.5*cos(-140)},{.5*sin(-140)+.1})--({.5*cos(-140)},{-1.5*sin(acos(1/3*cos(-140)))});
	\draw[orange,ultra thick] ({.5*cos(-40)},{.5*sin(-40)+.1})--({.5*cos(-40)},{-1.5*sin(acos(1/3*cos(-40)))});
	\draw[darkgreen,ultra thick] ({.5*cos(140)},{.5*sin(140)-.1})--({.5*cos(140)},{1.5*sin(acos(1/3*cos(140)))});
	\draw[darkgreen,ultra thick] ({.5*cos(40)},{.5*sin(40)-.1})--({.5*cos(40)},{1.5*sin(acos(1/3*cos(40)))});
	\draw[fill=white] (0,0) ellipse (.5 and .5);
	\draw (0,0) circle (1.5 and 1.5);
	\node[below,inline text,red] at ({.5*cos(-140)},{-1.5*sin(acos(1/3*cos(-140)))}) {\footnotesize\ta};
	\node[below,inline text,orange] at ({.5*cos(-40)},{-1.5*sin(acos(1/3*cos(-40)))}) {\footnotesize\tb};
	\node[above,inline text,darkgreen] at ({.5*cos(140)},{1.5*sin(acos(1/3*cos(140)))}) {\footnotesize\tc};
	\node[above,inline text,darkgreen] at ({.5*cos(40)},{1.5*sin(acos(1/3*cos(40)))}) {\footnotesize\td};
}
\newcommand{\annsspp}[4]{
	\ifthenelse{\equal{#1}{}}{}
	{
		\node[left,inline text,blue] at ($(180:.6 and .75)$() {\footnotesize#1};
		\coordinate (A) at ({.5*cos(-140)},{1*sin(140)});
		\draw[blue] ({.5*cos(-140)},{1*sin(-140)}) to[out=90+45,in=270-45] (A);
	}
	\ifthenelse{\equal{#2}{}}{}
	{
		\node[above,inline text,blue] at ($(90:.5 and .9)$() {\footnotesize#2};
		\coordinate (B) at ({.5*cos(40)},{1.7*sin(40)});
		\draw[blue] ({.5*cos(140)},{1.2*sin(140)}) to[out=45,in=270-45] (B);
	}
	\ifthenelse{\equal{#3}{}}{}
	{
		\node[right,inline text,blue] at ($(0,.3)+(0:.75 and .75)$() {\footnotesize#3};
		\coordinate (C) at ({.5*cos(40)},{2*sin(40)});
		\draw[blue] ({.5*cos(40)},{1*sin(-40)}) to[out=90-45,in=270+45] (C);
	}
	\ifthenelse{\equal{#4}{}}{}
	{
		\node[above,inline text,blue] at ($(270:.5 and 1.1)$() {\footnotesize#4};
		\coordinate (A) at ({.5*cos(-40)},{1.5*sin(-40)});
		\draw[blue] ({.5*cos(-140)},{2*sin(-140)}) to[out=45,in=270-45] (A);
	}
	\draw[red,ultra thick] ({.5*cos(-140)},{.5*sin(-140)+.1})--({.5*cos(-140)},{-1.5*sin(acos(1/3*cos(-140)))});
	\draw[red,ultra thick] ({.5*cos(-40)},{.5*sin(-40)+.1})--({.5*cos(-40)},{-1.5*sin(acos(1/3*cos(-40)))});
	\draw[darkgreen,ultra thick] ({.5*cos(140)},{.5*sin(140)-.1})--({.5*cos(140)},{1.5*sin(acos(1/3*cos(140)))});
	\draw[darkgreen,ultra thick] ({.5*cos(40)},{.5*sin(40)-.1})--({.5*cos(40)},{1.5*sin(acos(1/3*cos(40)))});
	\draw[fill=white] (0,0) ellipse (.5 and .5);
	\draw (0,0) circle (1.5 and 1.5);
	\node[below,inline text,red] at ({.5*cos(-140)},{-1.5*sin(acos(1/3*cos(-140)))}) {\footnotesize\ta};
	\node[below,inline text,red] at ({.5*cos(-40)},{-1.5*sin(acos(1/3*cos(-40)))}) {\footnotesize\tb};
	\node[above,inline text,darkgreen] at ({.5*cos(140)},{1.5*sin(acos(1/3*cos(140)))}) {\footnotesize\tc};
	\node[above,inline text,darkgreen] at ({.5*cos(40)},{1.5*sin(acos(1/3*cos(40)))}) {\footnotesize\td};
}
\newcommand{\annssss}[4]{
	\ifthenelse{\equal{#1}{}}{}
	{
		\node[left,inline text,blue] at ($(180:.6 and .75)$() {\footnotesize#1};
		\coordinate (A) at ({.5*cos(-140)},{1*sin(140)});
		\draw[blue] ({.5*cos(-140)},{1*sin(-140)}) to[out=90+45,in=270-45] (A);
	}
	\ifthenelse{\equal{#2}{}}{}
	{
		\node[above,inline text,blue] at ($(90:.5 and .9)$() {\footnotesize#2};
		\coordinate (B) at ({.5*cos(40)},{1.7*sin(40)});
		\draw[blue] ({.5*cos(140)},{1.2*sin(140)}) to[out=45,in=270-45] (B);
	}
	\ifthenelse{\equal{#3}{}}{}
	{
		\node[right,inline text,blue] at ($(0,.3)+(0:.75 and .75)$() {\footnotesize#3};
		\coordinate (C) at ({.5*cos(40)},{2*sin(40)});
		\draw[blue] ({.5*cos(40)},{1*sin(-40)}) to[out=90-45,in=270+45] (C);
	}
	\ifthenelse{\equal{#4}{}}{}
	{
		\node[above,inline text,blue] at ($(270:.5 and 1.1)$() {\footnotesize#4};
		\coordinate (A) at ({.5*cos(-40)},{1.5*sin(-40)});
		\draw[blue] ({.5*cos(-140)},{2*sin(-140)}) to[out=45,in=270-45] (A);
	}
	\draw[red,ultra thick] ({.5*cos(-140)},{.5*sin(-140)+.1})--({.5*cos(-140)},{-1.5*sin(acos(1/3*cos(-140)))});
	\draw[red,ultra thick] ({.5*cos(-40)},{.5*sin(-40)+.1})--({.5*cos(-40)},{-1.5*sin(acos(1/3*cos(-40)))});
	\draw[red,ultra thick] ({.5*cos(140)},{.5*sin(140)-.1})--({.5*cos(140)},{1.5*sin(acos(1/3*cos(140)))});
	\draw[red,ultra thick] ({.5*cos(40)},{.5*sin(40)-.1})--({.5*cos(40)},{1.5*sin(acos(1/3*cos(40)))});
	\draw[fill=white] (0,0) ellipse (.5 and .5);
	\draw (0,0) circle (1.5 and 1.5);
	\node[below,inline text,red] at ({.5*cos(-140)},{-1.5*sin(acos(1/3*cos(-140)))}) {\footnotesize\ta};
	\node[below,inline text,red] at ({.5*cos(-40)},{-1.5*sin(acos(1/3*cos(-40)))}) {\footnotesize\tb};
	\node[above,inline text,red] at ({.5*cos(140)},{1.5*sin(acos(1/3*cos(140)))}) {\footnotesize\tc};
	\node[above,inline text,red] at ({.5*cos(40)},{1.5*sin(acos(1/3*cos(40)))}) {\footnotesize\td};
}

\newcommand{\generalpantsstpq}[8]
{
	\pgfmathsetmacro{\ra}{.25};
	\pgfmathsetmacro{\sa}{.75};
	\pgfmathsetmacro{\rb}{2};
	\pgfmathsetmacro{\rc}{1.5};
	\coordinate (A) at ({-\sa},{-\rc*sin(acos(-\sa/\rb))});
	\coordinate (B) at ({\sa},{-\rc*sin(acos(-\sa/\rb))});
	\coordinate (C) at ({-\sa},{\rc*sin(acos(\sa/\rb))});
	\coordinate (D) at ({\sa},{\rc*sin(acos(\sa/\rb))});
		\ifthenelse{\equal{#1}{}}{}
		{
			\coordinate (A1) at ($(A)!.9!(-\sa,-\ra)$);
			\coordinate (B1) at ({-\sa-1.5*\ra},{0});
			\coordinate (C1) at ($(C)!.9!(-\sa,\ra)$);
			\node[left,inline text,blue] at (A1) {\footnotesize#1};
			\draw[blue] (A1) to[out=90+45,in=270] (B1) to[out=90,in=180+45] (C1);
		}
		\ifthenelse{\equal{#2}{}}{}
		{
			\coordinate (A1) at ($(A)!.6!(-\sa,-\ra)$);
			\coordinate (B1) at ({-\sa+2*\ra},{0});
			\coordinate (C1) at ($(C)!.75!(-\sa,\ra)$);
			\node[right,inline text,blue] at (A1) {\footnotesize#2};
			\draw[blue] (A1) to[out=90-45,in=270] (B1) to[out=90,in=0-45] (C1);
		}
		\ifthenelse{\equal{#3}{}}{}
		{
			\coordinate (A1) at ($(B)!.9!(\sa,-\ra)$);
			\coordinate (B1) at ({\sa-2*\ra},{0});
			\coordinate (C1) at ($(D)!.9!(\sa,\ra)$);
			\node[left,inline text,blue] at (A1) {\footnotesize#3};
			\draw[blue] (A1) to[out=90+45,in=270] (B1) to[out=90,in=180+45] (C1);
		}
		\ifthenelse{\equal{#4}{}}{}
		{
			\coordinate (A1) at ($(B)!.75!(\sa,-\ra)$);
			\coordinate (B1) at ({\sa+1.5*\ra},{0});
			\coordinate (C1) at ($(D)!.75!(\sa,\ra)$);
			\node[right,inline text,blue] at (A1) {\footnotesize#4};
			\draw[blue] (A1) to[out=90-45,in=270] (B1) to[out=90,in=0-45] (C1);
		}
		\ifthenelse{\equal{#5}{}}{}
		{
			\coordinate (A1) at ($(A)!.3!(-\sa,-\ra)$);
			\coordinate (B1) at ({-\sa-3*\ra},{0});
			\coordinate (C1) at ($(C)!.5!(-\sa,\ra)$);
			\node[left,inline text,blue] at (A1) {\footnotesize#5};
			\draw[blue] (A1) to[out=90+45,in=270] (B1) to[out=90,in=180+45] (C1);
		}
		\ifthenelse{\equal{#6}{}}{}
		{
			\coordinate (A1) at ($(C)!.25!(-\sa,-\ra)$);
			\coordinate (C1) at ($(D)!.25!(\sa,\ra)$);
			\coordinate (B1) at ($(A1)!.5!(C1)$);
			\node[below,inline text,blue] at (B1) {\footnotesize#6};
			\draw[blue] (A1) to[out=45,in=180+45] (C1);
		}
		\ifthenelse{\equal{#7}{}}{}
		{
			\coordinate (A1) at ($(B)!.3!(\sa,-\ra)$);
			\coordinate (B1) at ({\sa+3*\ra},{0});
			\coordinate (C1) at ($(D)!.1!(\sa,\ra)$);
			\node[right,inline text,blue] at (A1) {\footnotesize#7};
			\draw[blue] (A1) to[out=90-45,in=270] (B1) to[out=90,in=-45] (C1);
		}
		\ifthenelse{\equal{#8}{}}{}
		{
			\coordinate (A1) at ($(A)!.15!(-\sa,-\ra)$);
			\coordinate (C1) at ($(B)!.15!(\sa,\ra)$);
			\coordinate (B1) at ($(A1)!.5!(C1)$);
			\node[above,inline text,blue] at (B1) {\footnotesize#8};
			\draw[blue] (A1) to[out=45,in=180+45] (C1);
		}
\begin{scope}[even odd rule]
	\clip (0,0) ellipse [x radius=\rb,y radius=\rc] (-\sa,0) circle (\ra) (\sa,0) circle (\ra);
	\draw [red,ultra thick] ({-\sa},{0})--({-\sa},{-2*\rb});
	\draw [orange,ultra thick] ({\sa},{0})--({\sa},{-2*\rb});
	\draw [darkgreen,ultra thick] ({-\sa},{0})--({-\sa},{2*\rb});
	\draw [darkred,ultra thick] ({\sa},{0})--({\sa},{2*\rb});
\end{scope}
	\draw (-\sa,0) circle (\ra);
	\draw (\sa,0) circle (\ra);
	\draw (0,0) ellipse [x radius=\rb,y radius=\rc];
	\node[red,inline text,below] at (A) {\footnotesize\ta};
	\node[orange,inline text,below] at (B) {\footnotesize\tb};
	\node[darkgreen,inline text,above] at (C) {\footnotesize\tc};
	\node[darkred,inline text,above] at (D) {\footnotesize\td};
}

\newcommand{\tube}[3]{
	\ifthenelse{\equal{#2}{}}{}
	{
		\draw[domain=-90:0,smooth,variable=\x,blue] plot ({cos(\x)},{.1*sin(\x)+(\x)/360-.25});
		\draw[domain=0:180,smooth,variable=\x,blue,dashed] plot ({cos(\x)},{.1*sin(\x)+(\x)/360-.25});
		\draw[domain=180:270,smooth,variable=\x,blue] plot ({cos(\x)},{.1*sin(\x)+(\x)/360-.25});
	}
	\node[below,inline text,red] at ($(0, -1) + (270:1 and .1)$() {\footnotesize#1};
	\node[above,inline text,red] at ($(0, 1) + (-270:1 and .1)$() {\footnotesize#3};
	\draw (0,1) ellipse (1 and .1);
	\draw (-1,1)--(-1,-1) (1,1)--(1,-1);
	\draw[dashed] ($(0, -1) + (180:1 and .1)$() arc (180:0:1 and .1);
	\draw ($(0, -1) + (180:1 and .1)$() arc (180:360:1 and .1);
	\ifthenelse{\equal{#2}{}}{}
	{
		\node[inline text,blue] at (.5,-.3) {\footnotesize#2};
	}
	\draw [red,thick] ($(0, -1) + (270:1 and .1)$()--($(0, 1) + (-90:1 and .1)$();
}

\newcommand{\ttube}[6]{
	\draw [red,thick,dashed] ($(0, -1) + (70:1 and .1)$()--($(0, 1) + (70:1 and .1)$();
	\ifthenelse{\equal{#4}{}}{}
	{
		\draw[domain=-110:0,smooth,variable=\x,blue] plot ({cos(\x)},{.1*sin(\x)+(\x)/360-.25});
		\draw[domain=0:70,smooth,variable=\x,blue,dashed] plot ({cos(\x)},{.1*sin(\x)+(\x)/360-.25});
	}
	\ifthenelse{\equal{#3}{}}{}
	{
		\draw[domain=70:180,smooth,variable=\x,blue,dashed] plot ({cos(\x)},{.1*sin(\x)+(\x)/360});
		\draw[domain=180:250,smooth,variable=\x,blue] plot ({cos(\x)},{.1*sin(\x)+(\x)/360});
	}
	\node[below,inline text,red] at ($(0, -1) + (250:1 and .1)$() {\footnotesize#1};
	\node[below,inline text,red] at ($(0, -1) + (-70:1 and .1)$() {\footnotesize#2};
	\node[above,inline text,red] at ($(0, 1) + (-250:1 and .1)$() {\footnotesize#5};
	\node[above,inline text,red] at ($(0, 1) + (70:1 and .1)$() {\footnotesize#6};
	\draw (0,1) ellipse (1 and .1);
	\draw (-1,1)--(-1,-1) (1,1)--(1,-1);
	\draw[dashed] ($(0, -1) + (180:1 and .1)$() arc (180:0:1 and .1);
	\draw ($(0, -1) + (180:1 and .1)$() arc (180:360:1 and .1);
	\ifthenelse{\equal{#4}{}}{}
	{
		\node[right,inline text,blue] at ({cos(0)},{.1*sin(0)+(0)/360-.25}) {\footnotesize#4};
	}
	\ifthenelse{\equal{#3}{}}{}
	{
		\node[left,inline text,blue] at ({cos(180)},{.1*sin(180)+(180)/360}) {\footnotesize#3};
	}
	\draw [red,thick] ($(0, -1) + (250:1 and .1)$()--($(0, 1) + (250:1 and .1)$();
}

\newcommand{\ttttube}[4]{
	\draw [red,thick,dashed] ($(0, -1) + (100:1 and .1)$()--($(0, 1) + (100:1 and .1)$();
	\draw [red,thick,dashed] ($(0, -1) + (40:1 and .1)$()--($(0, 1) + (40:1 and .1)$();
	\draw (0,1) ellipse (1 and .1);
	\draw (-1,1)--(-1,-1) (1,1)--(1,-1);
	\draw[dashed] ($(0, -1) + (180:1 and .1)$() arc (180:0:1 and .1);
	\draw ($(0, -1) + (180:1 and .1)$() arc (180:360:1 and .1);
	\ifthenelse{\equal{#1}{}}{}
	{
		\draw[domain=230-360:280-360,smooth,variable=\x,blue] plot ({cos(\x)},{.1*sin(\x)+(\x)/360-.25});
	}
	\ifthenelse{\equal{#2}{}}{}
	{
		\draw[domain=280-360:0,smooth,variable=\x,blue] plot ({cos(\x)},{.1*sin(\x)+(\x)/360-.15});
		\draw[domain=0:40,smooth,variable=\x,blue,dashed] plot ({cos(\x)},{.1*sin(\x)+(\x)/360-.15});
	}
	\ifthenelse{\equal{#3}{}}{}
	{
		\draw[domain=40:100,smooth,variable=\x,blue,dashed] plot ({cos(\x)},{.1*sin(\x)+(\x)/360-.05});
	}
	\ifthenelse{\equal{#4}{}}{}
	{
		\draw[domain=100:180,smooth,variable=\x,blue,dashed] plot ({cos(\x)},{.1*sin(\x)+(\x)/360+.05});
		\draw[domain=180:230,smooth,variable=\x,blue] plot ({cos(\x)},{.1*sin(\x)+(\x)/360+.05});
	}
	\draw [red,thick] ($(0, -1) + (230:1 and .1)$()--($(0, 1) + (230:1 and .1)$();
	\draw [red,thick] ($(0, -1) + (280:1 and .1)$()--($(0, 1) + (280:1 and .1)$();
	\node[below,inline text,red] at ($(0, -1) + (230:1 and .1)$() {\footnotesize\ta};
	\node[below,inline text,red] at ($(0, -1) + (280:1 and .1)$() {\footnotesize\tb};
	\node[below,inline text,red] at ($(0, -1) + (-100:1 and .1)$() {\footnotesize\tc};
	\node[below,inline text,red] at ($(0, -1) + (-40:1 and .1)$() {\footnotesize\td};
	\node[above,inline text,red] at ($(0, 1) + (-230:1 and .1)$() {\footnotesize\tap};
	\node[above,inline text,red] at ($(0, 1) + (-280:1 and .1)$() {\footnotesize\tbp};
	\node[above,inline text,red] at ($(0, 1) + (100:1 and .1)$() {\footnotesize\tcp};
	\node[above,inline text,red] at ($(0, 1) + (40:1 and .1)$() {\footnotesize\tdp};
	\ifthenelse{\equal{#1}{}}{}
	{
		\node[above,inline text,blue] at ({cos(-90)},{.1*sin(-90)+(-90)/360-.25}) {\footnotesize#1};
	}
	\ifthenelse{\equal{#2}{}}{}
	{
		\node[right,inline text,blue] at ({cos(0)},{.1*sin(0)+(0)/360-.15}) {\footnotesize#2};
	}
	\ifthenelse{\equal{#3}{}}{}
	{
		\node[above,inline text,blue] at ({cos(60)},{.1*sin(60)+(60)/360-.05}) {\footnotesize#3};
	}
	\ifthenelse{\equal{#4}{}}{}
	{
		\node[left,inline text,blue] at ({cos(180)},{.1*sin(180)+(180)/360+.05}) {\footnotesize#4};
	}
}

\usetikzlibrary{intersections}

\tikzset{
	dot/.style={circle,inner sep=1pt,fill},
}

\newcommand{\pants}[8]{
	\draw (0,2) ellipse (1.5 and .1);
	\draw[name path=left edge] (-2.5,-2)to[out=90,in=-90](-1.5,2);
	\draw[dashed] ($(-1.5, -2) + (180:1 and .1)$() arc (180:0:1 and .1);
	\draw ($(-1.5, -2) + (180:1 and .1)$() arc (180:360:1 and .1);
	\draw[name path=right edge] (2.5,-2)to[out=90,in=-90](1.5,2);
	\draw[dashed] ($(1.5, -2) + (180:1 and .1)$() arc (180:0:1 and .1);
	\draw ($(1.5, -2) + (180:1 and .1)$() arc (180:360:1 and .1);
	\draw[name path=centre] (-.5,-2)to[out=90,in=180](0,0)to[out=0,in=90](.5,-2);
	\draw[thick,red,dashed,name path=C] ($(-1.5, -2) + (70:1 and .1)$()to[out=90,in=-90]($(0, 2) + (100:1.5 and .1)$();
	\draw[thick,red,dashed,name path=D] ($(1.5, -2) + (70:1 and .1)$()to[out=90,in=-90]($(0, 2) + (40:1.5 and .1)$();
	\path[thick,red,name path=A] ($(-1.5, -2) + (250:1 and .1)$()to[out=90,in=-90]($(0, 2) + (230:1.5 and .1)$();
	\path[thick,red,name path=B] ($(1.5, -2) + (250:1 and .1)$()to[out=90,in=-90]($(0, 2) + (280:1.5 and .1)$();
	\ifthenelse{\equal{#1}{}}{}
	{
		\begin{scope}
			\path[name path=X] (-.5,-.25)--(-2.5,0);
			\path [name intersections={of=C and X,by={c1}}];
			\path [name intersections={of=X and left edge,by={c2}}];
			\draw[blue,dashed] (c1)to[out=140,in=70] (c2);
			\path[name path=Y] (c2) -- (0,-2);
			\path [name intersections={of=Y and A,by={d1}}];
			\draw[blue] (c2) to[out=-110,in=220]  (d1);
			\node[left,inline text,blue] at (c2) {\footnotesize#1};
		\end{scope}
	}
	\ifthenelse{\equal{#2}{}}{}
	{
		\begin{scope}
			\path[name path=X] (-2.5,-2)--(0,-.5);
			\path [name intersections={of=A and X,by={c1}}];
			\path [name intersections={of=X and centre,by={c2}}];
			\draw[blue] (c1) to[out=45,in=260]  (c2);
			\path[name path=Y] (c2) -- (-2.5,2);
			\path [name intersections={of=Y and C,by={d1}}];
			\draw[blue,dashed] (c2) to[out=95,in=-40]  (d1);
			\node[right,inline text,blue] at (c2) {\footnotesize#2};
		\end{scope}
	}
	\ifthenelse{\equal{#3}{}}{}
	{
		\begin{scope}
			\path[name path=X] (0,-.5)--(2.5,-.5);
			\path[name intersections={of=D and X,by={c1}}];
			\path[name intersections={of=X and centre,by={c2}}];
			\draw[blue,dashed] (c1)to[out=140,in=100] (c2);
			\path[name path=Y] (c2) -- (2,-2);
			\path [name intersections={of=Y and B,by={d1}}];
			\draw[blue] (c2) to[out=-80,in=220]  (d1);
			\node[left,inline text,blue] at (c2) {\footnotesize#3};
		\end{scope}
	}
	\ifthenelse{\equal{#4}{}}{}
	{
		\begin{scope}
			\path[name path=X] (0,-2.5)--(2.5,-1);
			\path [name intersections={of=B and X,by={c1}}];
			\path [name intersections={of=X and right edge,by={c2}}];
			\draw[blue] (c1) to[out=45,in=280]  (c2);
			\path[name path=Y] (c2) -- (0,.5);
			\path [name intersections={of=Y and D,by={d1}}];
			\draw[blue,dashed] (c2) to[out=110,in=-40]  (d1);
			\node[right,inline text,blue] at (c2) {\footnotesize#4};
		\end{scope}
	}
	\ifthenelse{\equal{#5}{}}{}
	{
		\begin{scope}
			\path[name path=X] (-2.5,2)--(0,.5);
			\path [name intersections={of=left edge and X,by={c1}}];
			\path [name intersections={of=X and A,by={c2}}];
			\draw[blue] (c1) to[out=-110,in=250]  (c2);
			\path[name path=Y] (c1) -- (0,1.5);
			\path [name intersections={of=Y and C,by={d1}}];
			\draw[blue,dashed] (c1) to[out=80,in=110]  (d1);
			\node[left,inline text,blue] at (c1) {\footnotesize#5};
		\end{scope}
	}
	\ifthenelse{\equal{#6}{}}{}
	{
		\begin{scope}
			\path[name path=X] (-2.5,.5)--(2.5,.5);
			\path [name intersections={of=A and X,by={c1}}];
			\path [name intersections={of=X and B,by={c2}}];
			\draw[blue] (c1) to[out=70,in=250]  (c2);
			\node[left,inline text,blue] at ($ (c2) + (0,.1) $) {\footnotesize#6};
		\end{scope}
	}
	\ifthenelse{\equal{#7}{}}{}
	{
		\begin{scope}
			\path[name path=X] (0,.5)--(2.5,1);
			\path [name intersections={of=B and X,by={c1}}];
			\path [name intersections={of=X and right edge,by={c2}}];
			\draw[blue] (c1) to[out=70,in=280]  (c2);
			\path[name path=Y] (c2) -- (0,2.2);
			\path [name intersections={of=Y and D,by={d1}}];
			\draw[blue,dashed] (c2) to[out=110,in=280]  (d1);
			\node[right,inline text,blue] at (c2) {\footnotesize#7};
		\end{scope}
	}
	\ifthenelse{\equal{#8}{}}{}
	{
		\begin{scope}
			\path[name path=X] (-2.5,1.3)--(2.5,1.3);
			\path [name intersections={of=D and X,by={c1}}];
			\path [name intersections={of=X and C,by={c2}}];
			\draw[blue,dashed] (c1) to[out=120,in=300]  (c2);
			\node[left,inline text,blue] at (c1) {\footnotesize#8};
		\end{scope}
	}
	\draw[thick,red] ($(-1.5, -2) + (250:1 and .1)$()to[out=90,in=-90]($(0, 2) + (230:1.5 and .1)$();
	\draw[thick,red] ($(1.5, -2) + (250:1 and .1)$()to[out=90,in=-90]($(0, 2) + (280:1.5 and .1)$();
	\node[below,inline text,red] at ($(-1.5, -2) + (250:1 and .1)$() {\footnotesize\ta};
	\node[below,inline text,red] at ($(1.5, -2) + (250:1 and .1)$() {\footnotesize\tb};
	\node[below,inline text,red] at ($(-1.5, -2) + (-70:1 and .1)$() {\footnotesize\tc};
	\node[below,inline text,red] at ($(1.5, -2) + (-70:1 and .1)$() {\footnotesize\td};
	\node[above,inline text,red] at ($(0, 2) + (-230:1.5 and .1)$() {\footnotesize\tap};
	\node[above,inline text,red] at ($(0, 2) + (-280:1.5 and .1)$() {\footnotesize\tbp};
	\node[above,inline text,red] at ($(0, 2) + (100:1.5 and .1)$() {\footnotesize\tcp};
	\node[above,inline text,red] at ($(0, 2) + (40:1.5 and .1)$() {\footnotesize\tdp};
}

\def\ladder[#1][#2][#3][#4][#5][#6]{
	\path(#1);
	\pgfgetlastxy{\XCoord}{\YCoord};
	\begin{scope}[xshift=\XCoord,yshift=\YCoord]
		\draw (-.6,-.5)--(-.6,.5);
		\draw (.6,-.5)--(.6,.5);
		\ifthenelse{\equal{#4}{}}{}
			{
				\draw (-.6,-.25)--(.6,.25);
			};
		\node[below,inline text] at (-.6,-.5) {\footnotesize#2};
		\node[below,inline text] at (.6,-.5) {\footnotesize#3};
		\node[below,inline text] at (0,0) {\footnotesize#4};
		\node[above,inline text] at (-.6,.5) {\footnotesize#5};
		\node[above,inline text] at (.6,.5) {\footnotesize#6};
	\end{scope}
}

\def\Lladder[#1][#2][#3][#4][#5][#6][#7]{
	\path(#1);
	\pgfgetlastxy{\XCoord}{\YCoord};
	\begin{scope}[xshift=\XCoord,yshift=\YCoord]
		\draw (-1.8,-.5)--(-1.8,.5);
		\draw (-.6,-.5)--(-.6,.5);
		\draw (.6,-.5)--(.6,.5);
		\ifthenelse{\equal{#4}{}}{}
		{
			\draw (-.6,-.25)--(.6,.25);
		};
		\node[below,inline text] at (-.6,-.5) {\footnotesize#2};
		\node[below,inline text] at (.6,-.5) {\footnotesize#3};
		\node[below,inline text] at (0,0) {\footnotesize#4};
		\node[above,inline text] at (-.6,.5) {\footnotesize#5};
		\node[above,inline text] at (.6,.5) {\footnotesize#6};
		\node[below,inline text] at (-1.8,-.5) {\footnotesize#7};
	\end{scope}
}

\def\Lmorphism[#1][#2][#3][#4][#5][#6][#7]{
	\path(#1);
	\pgfgetlastxy{\XCoord}{\YCoord};
	\begin{scope}[xshift=\XCoord,yshift=\YCoord]
		\draw (-1.8,-.5)to[out=90,in=190](-.6,.25);
		\draw (-.6,-.5)--(-.6,.5);
		\draw (.6,-.5)--(.6,.5);
		\ifthenelse{\equal{#4}{}}{}
		{
			\draw (-.6,-.25)--(.6,.25);
		};
		\node[below,inline text] at (-.6,-.5) {\footnotesize#2};
		\node[below,inline text] at (.6,-.5) {\footnotesize#3};
		\node[below,inline text] at (0,0) {\footnotesize#4};
		\node[above,inline text] at (-.6,.5) {\footnotesize#5};
		\node[above,inline text] at (.6,.5) {\footnotesize#6};
		\node[below,inline text] at (-1.8,-.5) {\footnotesize#7};
	\end{scope}
}

\def\Rladder[#1][#2][#3][#4][#5][#6][#7]{
	\path(#1);
	\pgfgetlastxy{\XCoord}{\YCoord};
	\begin{scope}[xshift=\XCoord,yshift=\YCoord]
		\draw (1.8,-.5)--(1.8,.5);
		\draw (-.6,-.5)--(-.6,.5);
		\draw (.6,-.5)--(.6,.5);
		\ifthenelse{\equal{#4}{}}{}
		{
			\draw (-.6,-.25)--(.6,.25);
		};
		\node[below,inline text] at (-.6,-.5) {\footnotesize#2};
		\node[below,inline text] at (.6,-.5) {\footnotesize#3};
		\node[below,inline text] at (0,0) {\footnotesize#4};
		\node[above,inline text] at (-.6,.5) {\footnotesize#5};
		\node[above,inline text] at (.6,.5) {\footnotesize#6};
		\node[below,inline text] at (1.8,-.5) {\footnotesize#7};
	\end{scope}
}

\def\Rmorphism[#1][#2][#3][#4][#5][#6][#7]{
	\path(#1);
	\pgfgetlastxy{\XCoord}{\YCoord};
	\begin{scope}[xshift=\XCoord,yshift=\YCoord]
		\draw (1.8,-.5)to[out=90,in=-10](.6,.3);
		\draw (-.6,-.5)--(-.6,.5);
		\draw (.6,-.5)--(.6,.5);
		\ifthenelse{\equal{#4}{}}{}
		{
			\draw (-.6,-.35)--(.6,.15);
		};
		\node[below,inline text] at (-.6,-.5) {\footnotesize#2};
		\node[below,inline text] at (.6,-.5) {\footnotesize#3};
		\node[below,inline text] at (0,-.1) {\footnotesize#4};
		\node[above,inline text] at (-.6,.5) {\footnotesize#5};
		\node[above,inline text] at (.6,.5) {\footnotesize#6};
		\node[below,inline text] at (1.8,-.5) {\footnotesize#7};
	\end{scope}
}

\def\Laction[#1][#2][#3]{
		\draw (0,-.5)--(0,.5);
		\draw (-.6,-.5)to[out=90,in=210](0,0);
		\node[below,inline text] at (-.6,-.5) {\footnotesize#1};
		\node[below,inline text] at (0,-.5) {\footnotesize#2};
		\node[above,inline text]  at (0,.5) {\footnotesize#3};
}
\def\Raction[#1][#2][#3]{
	\draw (0,-.5)--(0,.5);
	\draw (.6,-.5)to[out=90,in=-30](0,0);
	\node[below,inline text] at (.6,-.5) {\footnotesize#1};
	\node[below,inline text] at (0,-.5) {\footnotesize#2};
	\node[above,inline text] at (0,.5) {\footnotesize#3};
}

\def\Lassociator[#1][#2][#3][#4]{
	\draw (0,-.5)--(0,.5);
	\draw (-.6,-.5)to[out=90,in=210](0,0);
	\draw (.6,-.5)to[out=90,in=-30](0,.25);
	\node[below,inline text] at (-.6,-.5) {\footnotesize#1};
	\node[below,inline text] at (.6,-.5) {\footnotesize#2};
	\node[below,inline text] at (0,-.5) {\footnotesize#3};
	\node[above,inline text] at (0,.5) {\footnotesize#4};
}
\def\Rassociator[#1][#2][#3][#4]{
	\draw (0,-.5)--(0,.5);
	\draw (-.6,-.5)to[out=90,in=210](0,.25);
	\draw (.6,-.5)to[out=90,in=-30](0,0);
	\node[below,inline text] at (-.6,-.5) {\footnotesize#1};
	\node[below,inline text] at (.6,-.5) {\footnotesize#2};
	\node[below,inline text] at (0,-.5) {\footnotesize#3};
	\node[above,inline text] at (0,.5) {\footnotesize#4};
}

\usepackage{hhline}
\usepackage{etoolbox}

\let\originalleft\left
\let\originalright\right
\renewcommand{\left}{\mathopen{}\mathclose\bgroup\originalleft}
\renewcommand{\right}{\aftergroup\egroup\originalright}

\newcommand{\vvec}[1]
{
	\operatorname{\bf Vec}
	\ifstrequal{#1}{}
	{}
	{\left(#1\right)}
}

\newcommand{\rrep}[1]
{
	\operatorname{\bf Rep}
	\ifstrequal{#1}{}
	{}
	{\left(#1\right)}
}

\newcommand{\vvectwist}[2]
{
	\operatorname{\bf Vec}^{#2}
	\ifstrequal{#1}{}
	{}
	{\left(#1\right)}
}

\newcommand{\lad}[1]
{
	\operatorname{\bf Lad}
	\ifstrequal{#1}{}
	{}
	{\left(#1\right)}
}

\newcommand{\ann}[3]
{
	\operatorname{\bf Ann}_{\mathcal{#1},\mathcal{#2}}
	\ifstrequal{#3}{}
	{}
	{\left(\mathcal{#3}\right)}
}

\newcommand{\kar}[1]
{
	\operatorname{\bf Kar}
	\ifstrequal{#1}{}
	{}
	{\left(#1\right)}
}

\newcommand{\bpr}[1]
{
	\operatorname{\bf BPR}
	\ifstrequal{#1}{}
	{}
	{\left(#1\right)}
}

\newcommand{\dih}[1]
{
	\operatorname{Dih}
	\ifstrequal{#1}{}
	{}
	{_{#1}}
}

\newcommand{\defect}[5]{
	\ifthenelse{\equal{#3}{}}
	{
		\begin{smallmatrix} #2\hfill	\\ #1\hfill \end{smallmatrix}\!\bigr\vert^{#5}
	}
	{
		\ifthenelse{\equal{#4}{}}
		{
			\begin{smallmatrix} #2\hfill	\\ #1\hfill \end{smallmatrix}\!\bigr\vert_{#3}^{#5}
		}
		{
			\begin{smallmatrix} #2\hfill	\\ #1\hfill \end{smallmatrix}\!\bigr\vert_{(#3,#4)}^{#5}
		}
	}
}

\newcommand{\trivalentvertex}[4]{
   \draw (0,0)--(0,1);
   \node[above,inline text] at (0,1) {#1};
   \draw (0,0)--(-0.707,-0.707);
   \node[below,inline text] at (-0.707,-0.707) {#2};
   \draw (0,0)--(0.707,-0.707);
   \node[below,inline text] at (0.707,-0.707) {#3};
   \node[left,inline text] at (0,0) {#4};
}

\newcommand{\vgeneralpantsstp}[6]
{
	\ifthenelse{\equal{#1}{}}{}
	{
		\coordinate (A1) at (0,-1);
		\coordinate (B1) at (-.3,-.6);
		\coordinate (C1) at (0,-.2);
		\node[left,inline text,blue] at (C1) {\footnotesize#1};
		\draw[blue] (A1) to[out=90+45,in=270] (B1) to[out=90,in=180+45] (C1);
	}
	\ifthenelse{\equal{#2}{}}{}
	{
		\coordinate (A1) at (0,-1.1);
		\coordinate (B1) at (.3,-.6);
		\coordinate (C1) at (0,-.1);
		\node[right,inline text,blue] at (C1) {\footnotesize#2};
		\draw[blue] (A1) to[out=90-45,in=270] (B1) to[out=90,in=0-45] (C1);
	}
	\ifthenelse{\equal{#3}{}}{}
	{
		\coordinate (A1) at (0,.2);
		\coordinate (B1) at (-.3,.6);
		\coordinate (C1) at (0,1);
		\node[left,inline text,blue] at (A1) {\footnotesize#3};
		\draw[blue] (A1) to[out=90+45,in=270] (B1) to[out=90,in=180+45] (C1);
	}
	\ifthenelse{\equal{#4}{}}{}
	{
		\coordinate (A1) at (0,.1);
		\coordinate (B1) at (.3,.6);
		\coordinate (C1) at (0,1.1);
		\node[right,inline text,blue] at (A1) {\footnotesize#4};
		\draw[blue] (A1) to[out=90-45,in=270] (B1) to[out=90,in=0-45] (C1);
	}
	\ifthenelse{\equal{#5}{}}{}
	{
		\coordinate (A1) at (0,-1.3);
		\coordinate (B1) at (-.9,0);
		\coordinate (C1) at (0,1.3);
		\node[left,inline text,blue] at (B1) {\footnotesize#5};
		\draw[blue] (A1) to[out=90+45,in=270] (B1) to[out=90,in=180+45] (C1);
	}
	\ifthenelse{\equal{#6}{}}{}
	{
		\coordinate (A1) at (0,-1.4);
		\coordinate (B1) at (.9,0);
		\coordinate (C1) at (0,1.4);
		\node[right,inline text,blue] at (B1) {\footnotesize#6};
		\draw[blue] (A1) to[out=90-45,in=270] (B1) to[out=90,in=0-45] (C1);
	}
	\begin{scope}[even odd rule]
		\clip (0,0) ellipse [x radius=1,y radius=1.5] (0,-.6) circle (.2) (0,.6) circle (.2);
		\draw [red,ultra thick] (0,-1.5)--(0,-.6);
		\draw [orange,ultra thick] (0,-.6)--(0,.6);
		\draw [darkgreen,ultra thick] (0,.6)--(0,1.5);
	\end{scope}
	\draw (0,-.6) circle (.2);
	\draw (0,.6) circle (.2);
	\draw (0,0) ellipse [x radius=1.5,y radius=1.5];
	\node[red,inline text,below] at (0,-1.5) {\footnotesize\ta};
	\node[orange,inline text,above] at (0,.4) {\footnotesize\tb};
	\node[darkgreen,inline text,above] at (0,1.5) {\footnotesize\tc};
}

\newcommand{\vpantsstp}[4]
{
	\ifthenelse{\equal{#1}{}}{}
	{
		\coordinate (A1) at (0,-1);
		\coordinate (B1) at (-.3,-.6);
		\coordinate (C1) at (0,-.2);
		\node[left,inline text,blue] at (B1) {\footnotesize#1};
		\draw[blue] (A1) to[out=90+45,in=270] (B1) to[out=90,in=180+45] (C1);
	}
	\ifthenelse{\equal{#2}{}}{}
	{
		\coordinate (A1) at (0,-1.1);
		\coordinate (B1) at (.3,-.6);
		\coordinate (C1) at (0,-.1);
		\node[right,inline text,blue] at (B1) {\footnotesize#2};
		\draw[blue] (A1) to[out=90-45,in=270] (B1) to[out=90,in=0-45] (C1);
	}
	\ifthenelse{\equal{#3}{}}{}
	{
		\coordinate (A1) at (0,.2);
		\coordinate (B1) at (-.3,.6);
		\coordinate (C1) at (0,1);
		\node[left,inline text,blue] at (B1) {\footnotesize#3};
		\draw[blue] (A1) to[out=90+45,in=270] (B1) to[out=90,in=180+45] (C1);
	}
	\ifthenelse{\equal{#4}{}}{}
	{
		\coordinate (A1) at (0,.1);
		\coordinate (B1) at (.3,.6);
		\coordinate (C1) at (0,1.1);
		\node[right,inline text,blue] at (B1) {\footnotesize#4};
		\draw[blue] (A1) to[out=90-45,in=270] (B1) to[out=90,in=0-45] (C1);
	}
	\begin{scope}[even odd rule]
		\clip (0,0) ellipse [x radius=1,y radius=1.5] (0,-.6) circle (.2) (0,.6) circle (.2);
		\draw [red,ultra thick] (0,-1.5)--(0,-.6);
		\draw [orange,ultra thick] (0,-.6)--(0,.6);
		\draw [darkgreen,ultra thick] (0,.6)--(0,1.5);
	\end{scope}
	\draw (0,-.6) circle (.2);
	\draw (0,.6) circle (.2);
	\draw (0,0) ellipse [x radius=1.5,y radius=1.5];
	\node[red,inline text,below] at (0,-1.5) {\footnotesize\ta};
	\node[orange,inline text,above] at (0,.4) {\footnotesize\tb};
	\node[darkgreen,inline text,above] at (0,1.5) {\footnotesize\tc};
}

\newlength{\tabwidth}
\newlength{\tabheight}

\newcommand{\cat}[1]{\ensuremath{\mathcal{#1}}}

\DeclareMathOperator{\id}{id}

\newcommand{\doibase}{https://doi.org/}

\allowdisplaybreaks
\begin{document}
\title{Computing data for Levin-Wen with defects}
\author{Jacob C.\ Bridgeman}

\email{jcbridgeman1@gmail.com}
\affiliation{Perimeter Institute for Theoretical Physics, Waterloo, Ontario, Canada}
\orcid{0000-0002-5638-6681}
\homepage{https://jcbridgeman.bitbucket.io}

\author{Daniel Barter}
\email{danielbarter@gmail.com}
\affiliation{Mathematical Sciences Institute, Australian National University, Canberra, Australia}
\orcid{0000-0002-6423-117X}
\homepage{https://danielbarter.github.io}

\date{\today}

\begin{abstract}
	We demonstrate how to do many computations for doubled topological phases with defects. These defects may be 1-dimensional domain walls or 0-dimensional point defects.

  Using $\vvec{S_3}$ as a guiding example, we demonstrate how domain wall fusion and associators can be computed using generalized tube algebra techniques. These domain walls can be both between distinct or identical phases.
	Additionally, we show how to compute all possible point defects, and the fusion and associator data of these. Worked examples, tabulated data and Mathematica code are provided.
\end{abstract}

\maketitle

\section{Introduction}

Topological phases are of great interest in quantum information theory. By leveraging their topological nature, quantum information can be stored and manipulated in a manner which is protected against any local noise\cite{MR1951039,Dennis2002,Brown2014,Terhal2015}. For such encodings to be useful in our attempts to build a quantum computer, it is necessary to find ways of implementing \emph{universal sets} of logic gates.

Unfortunately, many of the simplest topological codes only allow limited gate sets to be implemented in a fault tolerant manner. On the other hand, some topological codes admit universal gate sets, but are incredibly challenging to engineer. Recently, a hybrid approach has been proposed\cite{1811.06738}. This primarily relies on the (relatively) straightforward code for storage and manipulation, but included small `islands' of a more powerful phase to leverage their computational power. 

Hybrid schemes such as that described above, require information to be exchanged between codes. One way to achieve this directly in the material is to introduce \emph{gapped domain walls} between the two phases in question. Once domain walls are introduced, either between identical or distinct codes, one can also introduce point-like defects on these domain walls. These defects can also be used to increase the computational power of the underlying codes\cite{Bombin2007a,Bombin2010,Yoshida2015a,Brown2016,IrisCong1,IrisCong2,Yoshida2017,Bridgeman2017,Brown2018,1903.11634}. 
The most famous example of this is the introduction of Majorana twists into the surface code\cite{Bombin2010,Brown2016,1511.05153}.
The mathematical theory of symmetry defects is developed in \onlinecites{Barkeshli2013,Barkeshli2014,1811.02143}. In this work, we provide a computational framework for obtaining data associated to defects. This data is required when designing quantum computational schemes.

In this paper, we study the renormalization invariant properties of (2+1)D, long range entangled, doubled topological phases with defects. In particular, we study phases constructed from the symmetric group $S_3$ and its subgroups $\{0\}$, $\ZZ{2}$ and $\ZZ{3}$. These phases are often called {\em Kitaev quantum double phases} and were defined on the lattice in \onlinecite{MR1951039}. The corresponding quantum field theories were initially introduced by Dijkgraaf and Witten in \onlinecite{MR1048699}. Although our methods generalize, we focus on these models because our techniques are best illustrated by example, rather than a more abstract approach. As the smallest nonabelian group, $S_3$ is a natural example which realizes many of the complexities of the general case.

Much work has been done on the Kitaev quantum double phases corresponding to $\ZZ{2}$ and $S_3$, for example \onlinecites{0901.1345,Beigi2011,1401.7096,PhysRevB.96.195129,Bultinck2015,SETPaper,1811.06738,Shen}. The $\ZZ{2}$-phase is central to quantum information theory and quantum computing. Realizing the $\ZZ{2}$ phase is currently a major experimental effort because it can be used to build quantum memories\cite{Dennis2002,Brown2014,Terhal2015}. When realized on the torus, it has 4 ground states which are topologically protected. The $S_3$ Kitaev quantum double phase is the simplest gaped phase which supports non-abelian anyons, hence is of central interest in quantum computing. Experimentally realizing non-abelian anyons is one of the central problems in topological quantum computing\cite{0707.1889,nickreedbro,2010Natur.464..187S,chow2014implementing,Gambetta1}.

Topological phases (of the kind we discuss) are parameterized by unitary modular tensor categories (UMTCs). Our work concerns UMTCs that arise as the \emph{Drinfeld center} of a unitary fusion category $\cat{C}$.
Although not all UMTCs are Drinfeld centers of fusion categories, for example, the (chiral) Fibonacci or (chiral) semion modes, there are many interesting UMTCs which are Drinfeld centers. For example, the category describing the low energy excitations of any Levin-Wen model\cite{Levin2005} is a Drinfeld center.
The fusion categories that arise in this paper are $\vvec{}$, $\vvec{\ZZ{2}}$, $\vvec{\ZZ{3}}$ and $\vvec{S_3}$ (defined below). 

\subsection{What is being computed in this work}

This paper is a continuation of the work begun in \onlinecites{1806.01279,1810.09469,1901.08069}. In the current work, we generalize the methods developed there for $\vvec{\ZZ{p}}$ to more general input (unitary) fusion categories, and provide code to compute essentially complete data on the defect theories. We focus on the example $\vvec{S_3}$, both to illustrate the methods, and to make more complete data available to researchers interested in universal quantum computation using `small' topological phases. In particular, we show how the following data is computed. Let $\cat{A}\curvearrowright\cat{M}\curvearrowleft\cat{B}$ and $\cat{B}\curvearrowright\cat{N}\curvearrowleft\cat{C}$ be bimodule categories (domain walls). Then we compute:
\begin{itemize}
	\item The Brauer-Picard fusion tables for such bimodules. 
	\item The codimension 2 defects lying at the interface of $\cat{M}$ and $\cat{N}$.
	\item The vertical fusion rules of these defects, including ${\rm End}(\cat{M})$ as a special case.
	\item The associator for this fusion.
	\item The horizontal fusion of these defects.
	\item The `$O_3$' bimodule associator.
\end{itemize}
From a mathematical perspective, we are computing decomposition rules for relative tensor product and composition of bimodule functors. In an upcoming paper \onlinecite{BBJInPrep}, we will provide a rigorous proof of this fact using a robust theory of skeletalization of fusion categories and their bimodules.
\subsection{Key results}
The key mathematical results of this work are as follows:
\begin{itemize}

\item Given a set of domain walls, how do they fuse (compute the Brauer-Picard tables). 

\item Given a set of domain walls, it is natural to ask if their twists can be treated as anyons. The obstruction to this is the $O_3$ and $O_4$ cohomology classes defined in \onlinecite{MR2677836}.
\begin{proposition}\label{prop:O3}
  Let $I$ and $G_1$ be the invertible $\vvec{S_3} \lvert \vvec{S_3}$ domain walls. Then the $O_3$ and $O_4$ obstructions vanishes for $\{ I, G_1\}$.
\end{proposition}
This result seems to be known in \onlinecite{Barkeshli2014}, but we have been unable to find the computation in the literature.

\item If $\cat{C} \curvearrowright \cat{M}$ is a boundary to vacuum, then the binary interface defects on $\cat{M}$ form a fusion category under vertical fusion called ${\rm End}(\cat{M})$. If $\cat{M}$ is indecomposable, then $\cat{C}$ and ${\rm End}(\cat{M})$ are Morita equivalent\cite{MR3242743}. If we let $\cat{C} = \vvec{\mathbb{Z}/3\mathbb{Z} \times S_3}$ and $\cat{M} = Q_8$, then ${\rm End}(Q_8)$ is a $\mathbb{Z}/3\mathbb{Z} \times \mathbb{Z}/3\mathbb{Z}$ Tambara-Yamagami category\cite{TY}. This establishes the following:
\begin{proposition}\label{prop:TY}$\vvec{\mathbb{Z}/3\mathbb{Z} \times S_3}$ is Morita equivalent to the Tambara-Yamagami category $\cat{TY}(\mathbb{Z}/3\mathbb{Z} \times \mathbb{Z}/3\mathbb{Z},\chi,1)$, where $\chi((g_0,h_0),(g_1,h_1))=\omega^{g_0h_1+h_0g_1}$.
\end{proposition}
To the best of our knowledge, previously this TY category was only known to be group-theoretical\cite{GNN} (the associator on $\vvec{G}$ was unknown)\footnote{After this preprint was released, we were made aware that this is a special case of Cor.~5.15 in \onlinecite{MB}}.
\end{itemize}

\subsection{Structure of the paper}

This paper is structured as follows. In Section~\ref{sec:prelim} we provide some definitions and mathematical preliminaries that are required for the remainder of the paper. 
In Section~\ref{sec:S3setup}, we define all the domain walls which appear in this paper. 
In Section~\ref{sec:examples} we discuss some explicit examples of computations. 
In Section~\ref{sec:bimodule_associators}, we describe the $O_3$ and $O_4$ obstructions, and prove Proposition~\ref{prop:O3}. 
In Section~\ref{sec:remarks}, we summarize.

At its heart, this paper is about explicit computations. 
In Appendix~\ref{appendix:skein_vectors} we construct a $(2+\epsilon)$ dimensional defect TQFT, which is our main computational tool. 
In Appendix~\ref{appendix:domain_wall_definitions}, we tabulate defining data for all the domain walls which appear in this paper. 
Appendices~\ref{ap:domain_wall_fusions}-\ref{ap:BID} consist of tables of computed data. Only a small amount of what we can actually compute is included, and we have tried to only include data which hasn't previously appeared in the literature.
Perhaps the most striking computation appears in Appendix~\ref{appendix:endM}. We compute the associator for ${\rm End}(Q_8)$, thereby proving Proposition~\ref{prop:TY}. 
Code used for the computations in this paper is provided in the ancillary material\cite{anc}.


\section{Preliminaries}\label{sec:prelim}

In this paper, we make extensive and careful use of a diagrammatic formalism. Diagrams fall into two classes, those with a gray textured background and those with a white background.

Diagrams with white backgrounds do not have a direct physical interpretation. They are simply a diagrammatic calculus used to describe tensor categories and module categories. This diagrammatic calculus, first popularized by Penrose in \onlinecite{MR0281657}, is used because it is cumbersome to describe fusion categories using conventional mathematical notation. A diagram with a white background must be interpreted using the coordinate chart induced by the page on which it is drawn.

Diagrams with a gray background are to be interpreted physically. For example, compound defects (defined below) completely specify physical systems and are therefore depicted with a gray background. The ground states of a system described by such a gray diagram are the skein vectors described in Appendix~\ref{appendix:skein_vectors}.
The physical system (or state) specified by a diagram with a gray background is independent of coordinate chart. Diagrams with a gray background are closely related to the diagrammatic calculus used to describe {\em rigid} tensor categories in \onlinecite{MR3674995}.

\begin{definition}[Fusion category]
A {\em tensor category} $\cat{C}$ is a category $\cat{C}$ equipped with a functor $- \otimes - : \cat{C} \otimes \cat{C} \to \cat{C}$, a natural isomorphism $ (- \otimes -) \otimes - \cong - \otimes (- \otimes -)$ called the associator and a special object $1 \in \cat{C}$ which satisfy the pentagon equation and unit equations respectively. These can be found on Page~22 of \onlinecite{MR3242743}. If $\cat{C}$ is semi-simple, using the string diagram notation as explained in \onlinecite{MR3674995}, a vector $\alpha$ in $\cat{C}(a \otimes b, c)$ can be represented by a trivalent vertex: 
\begin{align}
\begin{array}{c}
  \includeTikz{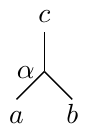}{
    \begin{tikzpicture}[scale=0.4, every node/.style={scale=0.8}]
      \trivalentvertex{$c$}{$a$}{$b$}{$\alpha$};
    \end{tikzpicture}
  }
\end{array}.
\end{align}

If we choose bases for all the vector spaces $\cat{C}(a \otimes b,c)$ (the Hom spaces), then the associator can be represented as a tensor F, where
  \begin{align}
    \begin{array}{c}
	    \includeTikz{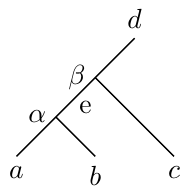}
	    {
	    	\begin{tikzpicture}[scale=0.4, every node/.style={scale=0.8}]
		    	\draw(0,0)--(3,3) node[below] at (0,0) {$a$} node[above] at (3,3) {$d$};
		    	\draw(2,0)--(1,1) node[below] at (2,0) {$b$};
		    	\draw(4,0)--(2,2) node[below] at (4,0) {$c$};
		    	\node[] at (1.75,1.25) {e};
		    	\node[left] at(1,1){$\alpha$};
		    	\node[left] at(2,2){$\beta$};
	    	\end{tikzpicture}
	    }
    \end{array}
    =\sum_{\mu\nu}
    F_{\alpha\beta}^{\mu\nu}
    \begin{array}{c}
    \includeTikz{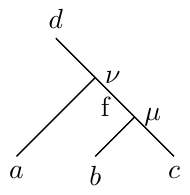}
    {
    	\begin{tikzpicture}[scale=0.4, every node/.style={scale=0.8}]
    	\draw(0,0)--(-3,3) node[below] at (0,0) {$c$} node[above] at (-3,3) {$d$};
    	\draw(-2,0)--(-1,1) node[below] at (-2,0) {$b$};
    	\draw(-4,0)--(-2,2) node[below] at (-4,0) {$a$};
    	\node at (-1.75,1.25) {f};
    	\node[right] at(-1,1){$\mu$};
    	\node[right] at(-2,2){$\nu$};
    	\end{tikzpicture}
    }
    \end{array}.
  \end{align}
  where $\alpha \in \cat{C}(a \otimes b,e), \beta \in \cat{C}(e \otimes c,d), \mu \in \cat{C}(b \otimes c,f)$ and $\nu \in \cat{C}(a \otimes f,d)$ are basis vectors.

A {\em fusion category} is a semi-simple rigid tensor category $\cat{C}$ with a finite number of simple objects and a simple unit. An object is {\em simple} if there are no non-trivial sub-objects. {\em Semi-simple} means that every object in $\cat{C}$ is a direct sum of simple objects. {\em Rigid} is a technical condition defined on Page~40 of \onlinecite{MR3242743} which implies that objects in $\cat{C}$ have duals and they behave like duals in the category of vector spaces. 
\end{definition}
In \onlinecite{Levin2005}, Levin and Wen defined a topological phase associated to any unitary fusion category $\cat{C}$. These are the phases of interest in this paper. All fusion categories considered in this work are assumed to be unitary.

\begin{definition}[$\vvec{G}$]
	Let $G$ be a finite group and $V$ a vector space. A $G$-grading on $V$ is a direct sum decomposition of $V=\oplus_{g\in G}V_g$. The fusion category $\vvec{G}$ is the category of $G$-graded (finite dimensional) vector spaces. The morphisms are linear maps which preserve the grading. The tensor product is define by
	\begin{align}
		(V\otimes W)_k=\oplus_{gh=k}V_g\otimes W_h,
	\end{align}
	and the associator is trivial. The (isomorphism classes of) simple objects in $\vvec{G}$ are parameterized by $g\in G$, where $g$ is synonymous with the 1 dimensional vector space in degree $g$. Using this notation, the tensor product can be written
	\begin{align}
		g\otimes h=gh
	\end{align}
	without confusion. We can choose trivalent vertices
  \begin{align}
    \begin{array}{c}
      \includeTikz{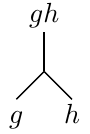}{
        \begin{tikzpicture}[scale=0.4, every node/.style={scale=0.8}]
          \trivalentvertex{$gh$}{$g$}{$h$}{};
        \end{tikzpicture}
  }
    \end{array}
  \end{align}
  such that
\begin{align}
    \begin{array}{c}
	    \includeTikz{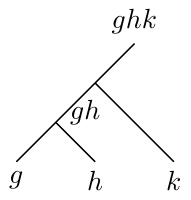}
	    {
	    	\begin{tikzpicture}[scale=0.4, every node/.style={scale=0.8}]
		    	\draw(0,0)--(3,3) node[below] at (0,0) {$g$} node[above] at (3,3) {$ghk$};
		    	\draw(2,0)--(1,1) node[below] at (2,0) {$h$};
		    	\draw(4,0)--(2,2) node[below] at (4,0) {$k$};
		    	\node[] at (1.75,1.25) {$gh$};
		    	\node[left] at(1,1){};
		    	\node[left] at(2,2){};
	    	\end{tikzpicture}
	    }
    \end{array}
    =
    \begin{array}{c}
    \includeTikz{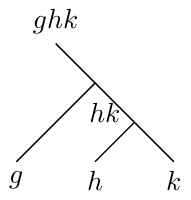}
    {
    	\begin{tikzpicture}[scale=0.4, every node/.style={scale=0.8}]
    	\draw(0,0)--(-3,3) node[below] at (0,0) {$k$} node[above] at (-3,3) {$ghk$};
    	\draw(-2,0)--(-1,1) node[below] at (-2,0) {$h$};
    	\draw(-4,0)--(-2,2) node[below] at (-4,0) {$g$};
    	\node at (-1.75,1.25) {$hk$};
    	\node[right] at(-1,1){};
    	\node[right] at(-2,2){};
    	\end{tikzpicture}
    }
    \end{array}
\end{align}
\end{definition}

\begin{definition}[Bimodule category]\label{def:bimod}
  Let $\cat{C},\,\cat{D}$ be fusion categories. A {\em bimodule category} $\cat{C} \curvearrowright \cat{M} \curvearrowleft \cat{D}$ is a semi-simple category equipped with functors $ - \vartriangleright - : \cat{C} \otimes \cat{M} \to \cat{M}$ and $ - \vartriangleleft - : \cat{M} \otimes \cat{D} \to \cat{M}$ and three natural isomorphisms
  \begin{subequations}
  \begin{align}
    &- \vartriangleright (- \vartriangleright -) \cong (- \otimes - ) \vartriangleright - \\
    & (- \vartriangleleft -) \vartriangleleft - \cong - \vartriangleleft (- \otimes -) \\
    & - \vartriangleright (- \vartriangleleft -) \cong (- \vartriangleright -) \vartriangleleft -.
  \end{align}
  \end{subequations}
  If we choose bases for the Hom spaces $\cat{M}(c \vartriangleright m, n)$ and $\cat{M}(m \vartriangleleft d,n)$, then these natural isomorphisms can be represented as tensors: 
  \begin{subequations}
  \begin{align}
  \begin{array}{c}
	  \includeTikz{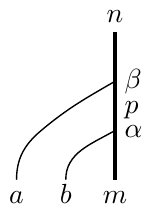}
	  {
	  	\begin{tikzpicture}[scale=0.5, every node/.style={scale=0.8}]
	  	\draw[very thick](0,0)--(0,3);
	  	\draw(-1,0)to[out=90,in=210](0,1);
	  	\draw(-2,0)to[out=90,in=220](-1.5,1)to[out=40,in=210](0,2);
	  	\node[below,inline text] at (-2,0) {$a$};
	  	\node[below,inline text] at (-1,0) {$b$};
	  	\node[below,inline text] at (0,0) {$m$};
	  	\node[above,inline text] at (0,3) {$n$};
	  	\node[right,inline text] at (0,1) {$\alpha$};
	  	\node[right,inline text] at (0,1.5) {$p$};
	  	\node[right,inline text] at (0,2) {$\beta$};
	  	\end{tikzpicture}
	  }
  \end{array}
  &=\sum_{\mu\nu}
  L_{\alpha\beta}^{\mu\nu}
  \begin{array}{c}
  \includeTikz{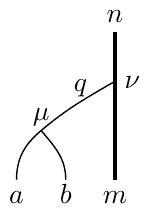}
  {
  	\begin{tikzpicture}[scale=0.5, every node/.style={scale=0.8}]
	  	\draw[very thick](0,0)--(0,3);
	  	\draw(-1,0)to[out=90,in=-50](-1.5,1);
	  	\draw(-2,0)to[out=90,in=220](-1.5,1)to[out=40,in=210](0,2);
	  	\node[below,inline text] at (-2,0) {$a$};
	  	\node[below,inline text] at (-1,0) {$b$};
	  	\node[below,inline text] at (0,0) {$m$};
	  	\node[above,inline text] at (0,3) {$n$};
	  	\node[above,inline text] at (-1.5,1) {$\mu$};
	  	\node[above,inline text] at (-.7,1.6) {$q$};
	  	\node[right,inline text] at (0,2) {$\nu$};
  	\end{tikzpicture}
  }
  \end{array}.\label{eqn:leftassociatorA} \\
  \begin{array}{c}
  \includeTikz{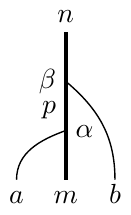}
  {
  	\begin{tikzpicture}[scale=0.5, every node/.style={scale=0.8}]
  	\draw[very thick](0,0)--(0,3);
  	\draw(1,0)to[out=90,in=-40](0,2);
  	\draw(-1,0)to[out=90,in=200](0,1);
  	\node[below,inline text] at (-1,0) {$a$};
  	\node[below,inline text] at (1,0) {$b$};
  	\node[below,inline text] at (0,0) {$m$};
  	\node[above,inline text] at (0,3) {$n$};
  	\node[right,inline text] at (0,1) {$\alpha$};
  	\node[left,inline text] at (0,1.5) {$p$};
  	\node[left,inline text] at (0,2) {$\beta$};
  	\end{tikzpicture}
  }
  \end{array}
  &=\sum_{\mu\nu} C_{\alpha\beta}^{\mu\nu}
  \begin{array}{c}
  \includeTikz{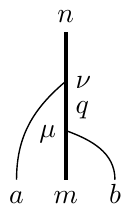}
  {
  	\begin{tikzpicture}[scale=0.5, every node/.style={scale=0.8}]
  	\draw[very thick](0,0)--(0,3);
  	\draw(1,0)to[out=90,in=-20](0,1);
  	\draw(-1,0)to[out=90,in=220](0,2);
  	\node[below,inline text] at (-1,0) {$a$};
  	\node[below,inline text] at (1,0) {$b$};
  	\node[below,inline text] at (0,0) {$m$};
  	\node[above,inline text] at (0,3) {$n$};
  	\node[left,inline text] at (0,1) {$\mu$};
  	\node[right,inline text] at (0,1.5) {$q$};
  	\node[right,inline text] at (0,2) {$\nu$};
  	\end{tikzpicture}
  }
  \end{array},\\
  \begin{array}{c}
  \includeTikz{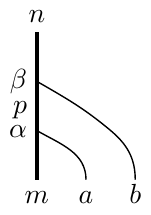}
  {
  	\begin{tikzpicture}[xscale=-0.5,yscale=.5, every node/.style={scale=0.8}]
  	\draw[very thick](0,0)--(0,3);
  	\draw(-1,0)to[out=90,in=210](0,1);
  	\draw(-2,0)to[out=90,in=220](-1.5,1)to[out=40,in=210](0,2);
  	\node[below,inline text] at (-2,0) {$b$};
  	\node[below,inline text] at (-1,0) {$a$};
  	\node[below,inline text] at (0,0) {$m$};
  	\node[above,inline text] at (0,3) {$n$};
  	\node[left,inline text] at (0,1) {$\alpha$};
  	\node[left,inline text] at (0,1.5) {$p$};
  	\node[left,inline text] at (0,2) {$\beta$};
  	\end{tikzpicture}
  }
  \end{array}
  &=\sum_{\mu\nu}
  R_{\alpha\beta}^{\mu\nu}
  \begin{array}{c}
  \includeTikz{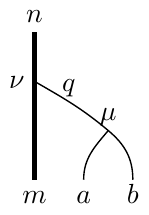}
  {
  	\begin{tikzpicture}[xscale=-0.5,yscale=.5, every node/.style={scale=0.8}]
  	\draw[very thick](0,0)--(0,3);
  	\draw(-1,0)to[out=90,in=-50](-1.5,1);
  	\draw(-2,0)to[out=90,in=220](-1.5,1)to[out=40,in=210](0,2);
  	\node[below,inline text] at (-2,0) {$b$};
  	\node[below,inline text] at (-1,0) {$a$};
  	\node[below,inline text] at (0,0) {$m$};
  	\node[above,inline text] at (0,3) {$n$};
  	\node[above,inline text] at (-1.5,1) {$\mu$};
  	\node[above,inline text] at (-.7,1.6) {$q$};
  	\node[left,inline text] at (0,2) {$\nu$};
  	\end{tikzpicture}
  }
  \end{array}.
  \end{align}
  \end{subequations}
\end{definition}
Bimodule categories parameterize domain walls between the Levin-Wen phases\cite{FUCHS2002353,MR2942952,MR3370609}. We will therefore use the words \emph{domain wall} and \emph{bimodule} interchangeably. 
\begin{theorem} \label{thm:defect_tqft} There exists a $(2+\epsilon)D$ defect TQFT which captures the renormalization invariant properties of all long range entangled, doubled 2D topological phases and their defects.
\end{theorem}
Theorem \ref{thm:defect_tqft} is the main technical tool which facilitates the computations of interest in this paper. The TQFT is constructed in Appendix~\ref{appendix:skein_vectors}. It is built from the tensors $F,L,C,R$ and some other data related to rigid structures on fusion categories. It is a $(2+\epsilon)$D theory because we can compute path integrals on cylinders, but not all 3-manifolds. 
\begin{definition}[Domain wall structure] \label{def:domain_wall_structure_prelim}
  Let $\Sigma$ be an oriented surface and $D \subseteq \Sigma$ an embedded oriented 1-manifold transverse to the boundary of $\Sigma$ such that $\partial D \subseteq \partial \Sigma$. 
Label each connected component of $\Sigma \backslash D$ with a unitary fusion category. 
Label each component of $D$ with a bimodule between the adjacent fusion categories.
  We call the pair $D \subseteq \Sigma$ together with the face and edge labels a {\em domain wall structure}. 
  A small open neighborhood in a domain wall structure looks something like
  \begin{align}
    \begin{array}{c}
      \includeTikz{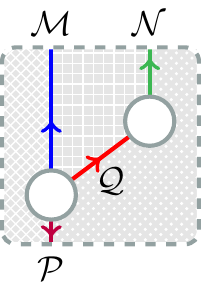}{
        \begin{tikzpicture}[scale=0.5]
          \path[fill=bulkcolor,postaction={pattern=crosshatch,pattern color=white}] (-1,2) -- (-1,-2) -- (-2,-2) -- (-2,2) -- cycle;
          \filldraw[fill=bulkcolor,postaction={pattern=grid,pattern color=white}] (-1,2) -- (1,2) -- (1,0.5) -- (-1,-1) -- cycle;
          \filldraw[fill=bulkcolor,postaction={pattern=crosshatch dots,pattern color=white}] (1,2) -- (2,2) -- (2,-2) -- (-1,-2) -- (-1,-1) -- (1,0.5) -- cycle;
\begin{scope}[very thick,decoration={markings,mark=at position 0.5 with {\arrow{>}}}]
          \draw[postaction={decorate},red] (-1,-1) -- (1,0.5);
          \draw[postaction={decorate},blue] (-1,-1) -- (-1,2);
\end{scope}
\begin{scope}[very thick,decoration={markings,mark=at position 0.85 with {\arrow{>}}}]
          \draw[postaction={decorate},nicegreen] (1,0.5) -- (1,2);
          \draw[postaction={decorate},purple] (-1,-1) -- (-1,-2);
          \end{scope}
          \draw[boundarycolor,fill=white,very thick] (-1,-1) circle (0.5);
          \draw[boundarycolor,fill=white,very thick] (1,0.5) circle (0.5);
          \draw[very thick,white] (-2,-2) rectangle (2,2);
          \draw[very thick,white,rounded corners] (-2,-2) rectangle (2,2);
          \draw[very thick,rounded corners,boundarycolor,dashed] (-2,-2) rectangle (2,2);
          \node[above] at (-1,2) {$\cat{M}$};
          \node[above] at (1,2) {$\cat{N}$};
          \node[below] at (-1,-2) {$\cat{P}$};
          \node[below] at (0.2,-0.2) {$\cat{Q}$};
        \end{tikzpicture}
        }
    \end{array}.
  \end{align}
It is important to notice that the boundary components of $\Sigma$ are not labeled with boundaries to vacuum. Therefore, a domain wall structure does not completely specify a physical system (it has a mixed gray/white background). In order to get a physical system, the boundary components of $\Sigma$ must be filled in with point defects. We stress that a system with a boundary would involve a \emph{filled} region labeled by $\vvec{}$.
  The term domain wall structure first appeared in \onlinecite{1901.08069}. Surfaces decorated with domain walls play a central role in the theory of defect TQFTs. The survey \onlinecite{1607.05747} by Carqueville is a great introduction to this field.
\end{definition}
\begin{definition}[Point defect] \label{def:point_defect}
  In order to define a physical system from a domain wall structure $\Sigma$, the holes in $\Sigma$ must be resolved. {\em Point defects} are the ``particles'' used to fill these holes. As is often the case in quantum field theory, the particle types are parameterized by representations of a local observable algebra. We define the local observable algebra in Definition~\ref{def:annular_category_appendix} in Appendix~\ref{appendix:skein_vectors}. Some of the point defects of interest in this paper will be familiar. For example anyons, which parameterize the low energy excitations for a long range entangled topological phase, and twists which terminate invertible domain walls are both examples of point defects.
  \begin{align}
    \begin{array}{c}
      \includeTikz{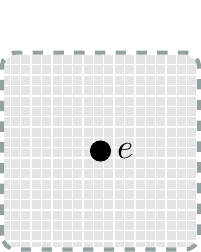}{
        \begin{tikzpicture}
          \filldraw[fill=bulkcolor,postaction={pattern=grid,pattern color=white}] (-1,-1) -- (-1,1) -- (1,1) -- (1,-1) -- cycle;
          \draw[very thick,white] (-1,-1) rectangle (1,1);
          \draw[very thick,white,rounded corners] (-1,-1) rectangle (1,1);
          \draw[very thick,rounded corners,boundarycolor,dashed] (-1,-1) rectangle (1,1);
          \filldraw[pointdefectcolor] (0,0) circle (0.1);
          \node at (0.25,0) {$e$};
          \node[above,white] at (0,1) {$F_1$};
        \end{tikzpicture}
        }
    \end{array}
\begin{array}{c}
      \includeTikz{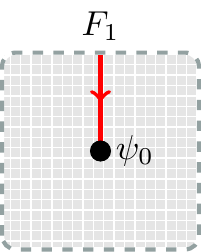}{
        \begin{tikzpicture}
          \filldraw[fill=bulkcolor,postaction={pattern=grid,pattern color=white}] (-1,-1) -- (-1,1) -- (1,1) -- (1,-1) -- cycle;
\begin{scope}[very thick,decoration={markings,mark=at position 0.5 with {\arrow{>}}}]
  \draw[postaction={decorate},red] (0,1) -- (0,0);
  \end{scope}
          \draw[very thick,white] (-1,-1) rectangle (1,1);
          \draw[very thick,white,rounded corners] (-1,-1) rectangle (1,1);
          \draw[very thick,rounded corners,boundarycolor,dashed] (-1,-1) rectangle (1,1);
          \filldraw[pointdefectcolor] (0,0) circle (0.1);
          \node at (0.35,0) {$\psi_0$};
          \node[above] at (0,1) {$F_1$};
        \end{tikzpicture}
        }
    \end{array}
  \end{align}
  In this paper, we are also interested in several other types of point defects. {\em Binary interface defects} are the particles which interface between two domain walls
  \begin{align}
    \begin{array}{c}
      \includeTikz{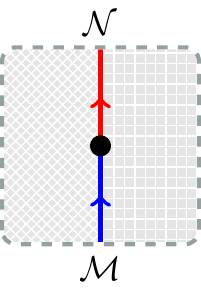}{
        \begin{tikzpicture}
          \path[fill=bulkcolor,postaction={pattern=crosshatch,pattern color=white}] (-1,-1) -- (0,-1) -- (0,1) -- (-1,1) -- cycle;
          \path[fill=bulkcolor,postaction={pattern=grid,pattern color=white}] (1,-1) -- (0,-1) -- (0,1) -- (1,1) -- cycle;
\begin{scope}[very thick,decoration={markings,mark=at position 0.5 with {\arrow{>}}}]
           \draw[postaction={decorate},blue] (0,-1) -- (0,0);
           \draw[postaction={decorate},red] (0,0) -- (0,1);
\end{scope}
           \filldraw[pointdefectcolor] (0,0) circle (0.1);
           \node[above] at (0,1) {$\cat{N}$};
           \node[below] at (0,-1) {$\cat{M}$};
           \draw[very thick,white] (-1,-1) rectangle (1,1);
           \draw[very thick,white,rounded corners] (-1,-1) rectangle (1,1);
           \draw[very thick,rounded corners,boundarycolor,dashed] (-1,-1) rectangle (1,1);
        \end{tikzpicture}
        }
      \end{array}.\label{eqn:BID}
  \end{align}
\end{definition}
Point defects were first defined by Morrison and Walker in \onlinecite{MR2978449}, where they are called {\em sphere modules}. Point defects in the ${\bf Vec}(\mathbb{Z}/p\mathbb{Z})$ phase were studied in \onlinecites{1810.09469,1901.08069}. 
\begin{definition}[Compound defect]
A {\em Compound Defect} is a domain wall structure $\Sigma$ together with point defects to resolve all the holes in $\Sigma$.
  \begin{align}
    \begin{array}{c}
      \includeTikz{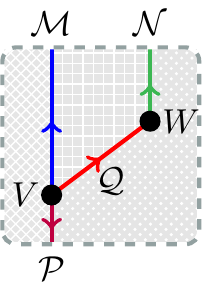}{
        \begin{tikzpicture}[scale=0.5]
          \filldraw[fill=bulkcolor,postaction={pattern=crosshatch,pattern color=white}] (-1,2) -- (-1,-2) -- (-2,-2) -- (-2,2) -- cycle;
          \filldraw[fill=bulkcolor,postaction={pattern=grid,pattern color=white}] (-1,2) -- (1,2) -- (1,0.5) -- (-1,-1) -- cycle;
          \filldraw[fill=bulkcolor,postaction={pattern=crosshatch dots,pattern color=white}] (1,2) -- (2,2) -- (2,-2) -- (-1,-2) -- (-1,-1) -- (1,0.5) -- cycle;
\begin{scope}[very thick,decoration={markings,mark=at position 0.5 with {\arrow{>}}}]
          \draw[postaction={decorate},red] (-1,-1) -- (1,0.5);
          \draw[postaction={decorate},blue] (-1,-1) -- (-1,2);
          \draw[postaction={decorate},nicegreen] (1,0.5) -- (1,2);
\end{scope}
\begin{scope}[very thick,decoration={markings,mark=at position 0.7 with {\arrow{>}}}]
          \draw[postaction={decorate},purple] (-1,-1) -- (-1,-2);
\end{scope}
          \filldraw[pointdefectcolor] (-1,-1) circle (0.2);
          \filldraw[pointdefectcolor] (1,0.5) circle (0.2);
          \draw[very thick,white,rounded corners] (-2,-2) rectangle (2,2);
          \draw[very thick,white] (-2,-2) rectangle (2,2);
          \draw[very thick,rounded corners,boundarycolor,dashed] (-2,-2) rectangle (2,2);
          \node[above] at (-1,2) {$\cat{M}$};
          \node[above] at (1,2) {$\cat{N}$};
          \node[below] at (-1,-2) {$\cat{P}$};
          \node[below] at (0.2,-0.2) {$\cat{Q}$};
          \node[left] at  (-1,-1) {$V$};
          \node[right] at (1,0.5) {$W$};
        \end{tikzpicture}
        }
    \end{array}.
  \end{align}
Compound defects are a complete specification of a physical system. The ground states $Z(\Sigma)$ of this system are spanned by the skein vectors which can be drawn on top of the system as described in Appendix~\ref{appendix:skein_vectors}. Two skein vectors are equivalent if one can be transformed into the other using local relations.
\end{definition}

\begin{definition}[Generalized Levin-Wen lattice models]
  Let $\Sigma$ be a compound defect. We can define a Levin-Wen\cite{Levin2005} type model as follows: The local degrees of freedom on the edges are simple objects in the labeling bimodule. The local degrees of freedom on the vertices are vectors in the corresponding point defect representation. Let $H$ be the global Hilbert space. The Hamiltonian is a commuting projector Hamiltonian with two types of terms, vertex terms $P_v$ and face terms $P_f$. The vertex terms project onto the states where the object labels for the vertex states match the local degrees of freedom on the edges. The face terms apply the operator
  \begin{align}
   P_f = \frac{1}{{\rm dim}(\mathcal{C})} \sum_{a\in \cat{C}} {\rm dim}(a) a
  \end{align}
  depicted in equation \eqref{eq:levin_wen_operator}. The ground states of this model are
  \begin{align}
   &\{ \psi \lvert P_v \psi = \psi, P_f \psi = \psi \}
     \cong H / \mathbb{C}\{ \psi- P_v \psi, \psi - P_f \psi \} 
     \cong Z(\Sigma) \label{eq:kitaev_ground_space} 
  \end{align}
  The projection onto the ground states in \eqref{eq:kitaev_ground_space} is
  \begin{align}
   \prod_f P_f \prod_v P_v.
  \end{align}
  For this reason, we can write a Levin-Wen ground state as
  \begin{align}
    \left\lvert \psi \right\rangle &\propto
    \left\lvert
    \begin{array}{c}
      \includeTikz{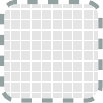}{
        \begin{tikzpicture}
          \filldraw[fill=bulkcolor,postaction={pattern=grid,pattern color=white}] (-0.5,-0.5) rectangle (0.5,0.5);
          \draw[very thick,white,rounded corners] (-0.5,-0.5) rectangle (0.5,0.5);
          \draw[very thick,white] (-0.5,-0.5) rectangle (0.5,0.5);
          \draw[very thick,boundarycolor,rounded corners,dashed] (-0.5,-0.5) rectangle (0.5,0.5);
          \end{tikzpicture}
        }
      \end{array}
    \right\rangle + {\rm dim}(a)
\left\lvert
    \begin{array}{c}
      \includeTikz{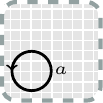}{
        \begin{tikzpicture}
          \filldraw[fill=bulkcolor,postaction={pattern=grid,pattern color=white}] (-0.5,-0.5) rectangle (0.5,0.5);
          \begin{scope}[thick,decoration={markings,mark=at position 0.5 with {\arrow[scale=0.7]{>}}}]
            \draw[postaction={decorate}] (-0.2,-0.2) circle (0.2);
          \end{scope}
          \node at (0.1,-0.2) {\tiny $a$};
          \draw[very thick,white,rounded corners] (-0.5,-0.5) rectangle (0.5,0.5);
          \draw[very thick,white] (-0.5,-0.5) rectangle (0.5,0.5);
          \draw[very thick,boundarycolor,rounded corners,dashed] (-0.5,-0.5) rectangle (0.5,0.5);
          \end{tikzpicture}
        }
      \end{array}
    \right\rangle 
+ {\rm dim}(a){\rm dim}(b)
\left\lvert
    \begin{array}{c}
      \includeTikz{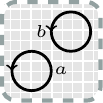}{
        \begin{tikzpicture}
          \filldraw[fill=bulkcolor,postaction={pattern=grid,pattern color=white}] (-0.5,-0.5) rectangle (0.5,0.5);
          \begin{scope}[thick,decoration={markings,mark=at position 0.5 with {\arrow[scale=0.7]{>}}}]
            \draw[postaction={decorate}] (-0.2,-0.2) circle (0.2);
            \draw[postaction={decorate}] (0.2,0.2) circle (0.2);
          \end{scope}
          \node at (0.1,-0.2) {\tiny $a$};
          \node at (-0.1,0.2) {\tiny $b$};
          \draw[very thick,white,rounded corners] (-0.5,-0.5) rectangle (0.5,0.5);
          \draw[very thick,white] (-0.5,-0.5) rectangle (0.5,0.5);
          \draw[very thick,boundarycolor,rounded corners,dashed] (-0.5,-0.5) rectangle (0.5,0.5);
          \end{tikzpicture}
        } 
      \end{array}
    \right\rangle 
+ {\rm dim}(c)
\left\lvert
    \begin{array}{c}
      \includeTikz{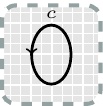}{
        \begin{tikzpicture}
          \filldraw[fill=bulkcolor,postaction={pattern=grid,pattern color=white}] (-0.5,-0.5) rectangle (0.5,0.5);
          \begin{scope}[thick,decoration={markings,mark=at position 0.5 with {\arrow[scale=0.7]{>}}}]
            \draw[postaction={decorate}] (0,0) ellipse (0.2 and 0.3);
          \end{scope}
          \node at (0,0.4) {\tiny $c$};
          \draw[very thick,white,rounded corners] (-0.5,-0.5) rectangle (0.5,0.5);
          \draw[very thick,white] (-0.5,-0.5) rectangle (0.5,0.5);
          \draw[very thick,boundarycolor,rounded corners,dashed] (-0.5,-0.5) rectangle (0.5,0.5);
          \end{tikzpicture}
        }
      \end{array}
    \right\rangle + \cdots
  \end{align}
  In this paper we shall use both the skein theoretic description of ground states and the loop superposition description of ground states interchangeably. We will always surround the loop superposition description with bra-kets.
\end{definition}

\begin{definition}[Renormalization] A compound defect can be renormalized to a point defect:
  \begin{align}
    \begin{array}{c}
      \includeTikz{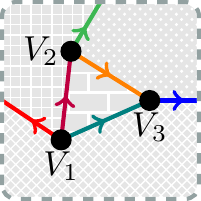}{
        \begin{tikzpicture}
          \path[fill=bulkcolor,postaction={pattern=crosshatch,pattern color=white}] (0,0) -- (0,1) -- (0.6,0.6) -- (1.5,1) -- (2,1) -- (2,0) -- cycle;
          \path[fill=bulkcolor,postaction={pattern=grid,pattern color=white}] (0,1) -- (0,2) -- (1,2) -- (0.7,1.5) -- (0.6,0.6) -- cycle;
          \path[fill=bulkcolor,postaction={pattern=crosshatch dots,pattern color=white}] (1,2) -- (2,2) -- (2,1) -- (1.5,1) -- (0.7,1.5) -- cycle;
\path[fill=bulkcolor,postaction={pattern=bricks,pattern color=white}] (0.6,0.6) -- (0.7,1.5) -- (1.5,1) -- cycle;
\begin{scope}[very thick,decoration={markings,mark=at position 0.5 with {\arrow{>}}}]
  \draw[postaction={decorate},purple] (0.6,0.6) -- (0.7,1.5);
          \draw[postaction={decorate},teal] (0.6,0.6) -- (1.5,1);
          \draw[postaction={decorate},orange] (0.7,1.5) -- (1.5,1);
          \draw[postaction={decorate},red] (0.6,0.6) -- (0,1);
          \draw[postaction={decorate},nicegreen] (0.7,1.5) -- (1,2);
\end{scope}
\begin{scope}[very thick,decoration={markings,mark=at position 0.7 with {\arrow{>}}}]
          \draw[postaction={decorate},blue] (1.5,1) -- (2,1);
          \end{scope}
          \filldraw (0.6,0.6) circle (0.1);
          \node[below] at (0.6,0.6) {$V_1$};
          \filldraw (0.7,1.5) circle (0.1);
          \node[left] at (0.7,1.5) {$V_2$};
          \filldraw (1.5,1) circle (0.1);
          \node[below] at (1.5,1) {$V_3$};
          \draw[very thick,white,rounded corners] (0,0) rectangle (2,2);
          \draw[very thick,white] (0,0) rectangle (2,2);
          \draw[very thick,boundarycolor,rounded corners,dashed] (0,0) rectangle (2,2);
        \end{tikzpicture}
        }
    \end{array} \cong \bigoplus_i
    \begin{array}{c}
      \includeTikz{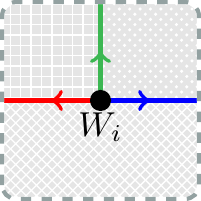}{
        \begin{tikzpicture}
          \path[fill=bulkcolor,postaction={pattern=grid,pattern color=white}] (0,1) -- (1,1) -- (1,2) -- (0,2) -- cycle;
          \path[fill=bulkcolor,postaction={pattern=crosshatch dots,pattern color=white}] (1,1) -- (2,1) -- (2,2) -- (1,2) -- cycle;
          \path[fill=bulkcolor,postaction={pattern=crosshatch,pattern color=white}] (0,1) -- (2,1) -- (2,0) -- (0,0) -- cycle;
\begin{scope}[very thick,decoration={markings,mark=at position 0.5 with {\arrow{>}}}]
          \draw[postaction={decorate},red] (1,1) -- (0,1);
          \draw[postaction={decorate},nicegreen] (1,1) -- (1,2);
          \draw[postaction={decorate},blue] (1,1) -- (2,1);
          \end{scope}
          \filldraw (1,1) circle (0.1);
          \node[below] at (1,1) {$W_i$};
          \draw[very thick,white,rounded corners] (0,0) rectangle (2,2);
          \draw[very thick,white] (0,0) rectangle (2,2);
          \draw[very thick,boundarycolor,rounded corners,dashed] (0,0) rectangle (2,2);
        \end{tikzpicture}
        }
    \end{array}
  \end{align}
The point defects $W_i$ are determined by taking the space of skein vectors supported on the left hand compound defect and looking how they transform with respect to the local observable algebra.
  \end{definition}
The following four examples of renormalization are the main focus of this paper:
\begin{definition}[Domain wall fusion]
	  We can horizontally fuse domain walls by horizontally concatenating and then renormalizing
	\begin{align}
	\begin{array}{c}
	\includeTikz{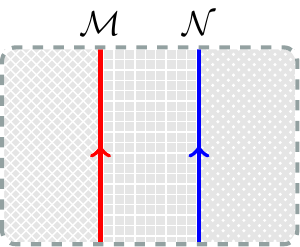}{
		\begin{tikzpicture}
		\draw[fill=bulkcolor,postaction={pattern=crosshatch,pattern color=white}] (-1.5,-1) rectangle (-0.5,1);
		\draw[fill=bulkcolor,postaction={pattern=grid,pattern color=white}] (-0.5,-1) rectangle (0.5,1);
		\draw[fill=bulkcolor,postaction={pattern=crosshatch dots,pattern color=white}] (0.5,-1) rectangle (1.5,1);
		\begin{scope}[very thick,decoration={markings,mark=at position 0.5 with {\arrow{>}}}]
		\draw[postaction={decorate},red] (-0.5,-1) -- (-0.5,1);
		\draw[postaction={decorate},blue] (0.5,-1) -- (0.5,1);
		\end{scope}
		\draw[white,very thick] (-1.5,-1) rectangle (1.5,1);
		\draw[rounded corners,white,very thick] (-1.5,-1) rectangle (1.5,1);
		\draw[rounded corners,dashed,boundarycolor,very thick] (-1.5,-1) rectangle (1.5,1);
		\node[above] at (-0.5,1) {$\cat{M}$};
		\node[above] at (0.5,1) {$\cat{N}$};
		\end{tikzpicture}
	}
	\end{array}
	\cong \bigoplus_i 
	\begin{array}{c}
	\includeTikz{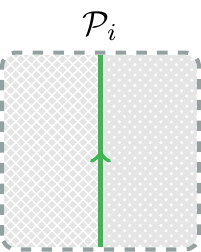}{
		\begin{tikzpicture}
		\draw[fill=bulkcolor,postaction={pattern=crosshatch,pattern color=white}] (-1,-1) rectangle (0,1);
		\draw[fill=bulkcolor,postaction={pattern=crosshatch dots,pattern color=white}] (0,-1) rectangle (1,1);
		\begin{scope}[very thick,decoration={markings,mark=at position 0.5 with {\arrow{>}}}]
		\draw[postaction={decorate},nicegreen] (0,-1) -- (0,1);
		\end{scope}
		\draw[white,very thick] (-1,-1) rectangle (1,1);
		\draw[rounded corners,white,very thick] (-1,-1) rectangle (1,1);
		\draw[rounded corners,dashed,boundarycolor,very thick] (-1,-1) rectangle (1,1);
		\node[above] at (0,1) {$\cat{P}_i$};
		\end{tikzpicture}
	}
	\end{array}.
	\end{align}
	We call the fusion ring for domain walls the \emph{Brauer-Picard ring} (BPR). Computing the BPR for $\vvec{\ZZ{p}}$ was the subject of \onlinecite{1806.01279}.

	Using the inflation trick from Definition~\ref{def:inflation_trick} in Appendix~\ref{appendix:skein_vectors}, we can encode these domain wall fusions as point defects
	\begin{align}
	\begin{array}{c}
	\includeTikz{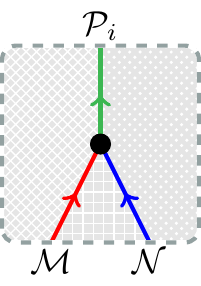}{
		\begin{tikzpicture}
		\path[fill=bulkcolor,postaction={pattern=crosshatch,pattern color=white}] (-1,-1) -- (-0.5,-1) -- (0,0) -- (0,1) -- (-1,1) -- cycle;
		\path[fill=bulkcolor,postaction={pattern=crosshatch dots,pattern color=white}] (1,-1) -- (0.5,-1) -- (0,0) -- (0,1) -- (1,1) -- cycle;
		\path[fill=bulkcolor,postaction={pattern=grid,pattern color=white}] (-0.5,-1) -- (0,0) -- (0.5,-1) -- cycle;
		\begin{scope}[very thick,decoration={markings,mark=at position 0.5 with {\arrow{>}}}]
		\draw[postaction={decorate},red]  (-0.5,-1) -- (0,0);
		\draw[postaction={decorate},blue]  (0.5,-1)-- (0,0);
		\draw[postaction={decorate},nicegreen] (0,0) -- (0,1);
		\end{scope}
		\filldraw (0,0) circle (0.1);
		\draw[white,very thick] (-1,-1) rectangle (1,1);
		\draw[rounded corners,white,very thick] (-1,-1) rectangle (1,1);
		\draw[rounded corners,dashed,boundarycolor,very thick] (-1,-1) rectangle (1,1);
		\node at (0,1.2) {$\cat{P}_i$};
		\node at (-0.5,-1.2) {$\cat{M}$};
		\node at (0.5,-1.2) {$\cat{N}$};
		\end{tikzpicture}
	}
	\end{array}
	\begin{array}{c}
	\includeTikz{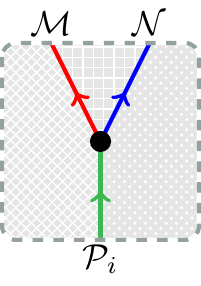}{
		\begin{tikzpicture}[yscale=-1]
		\path[fill=bulkcolor,postaction={pattern=crosshatch,pattern color=white}] (-1,-1) -- (-0.5,-1) -- (0,0) -- (0,1) -- (-1,1) -- cycle;
		\path[fill=bulkcolor,postaction={pattern=crosshatch dots,pattern color=white}] (1,-1) -- (0.5,-1) -- (0,0) -- (0,1) -- (1,1) -- cycle;
		\path[fill=bulkcolor,postaction={pattern=grid,pattern color=white}] (-0.5,-1) -- (0,0) -- (0.5,-1) -- cycle;
		\begin{scope}[very thick,decoration={markings,mark=at position 0.5 with {\arrow{>}}}]
		\draw[postaction={decorate},red] (0,0) -- (-0.5,-1);
		\draw[postaction={decorate},blue] (0,0) -- (0.5,-1);
		\draw[postaction={decorate},nicegreen] (0,1) -- (0,0);
		\end{scope}
		\filldraw (0,0) circle (0.1);
		\draw[white,very thick] (-1,-1) rectangle (1,1);
		\draw[rounded corners,white,very thick] (-1,-1) rectangle (1,1);
		\draw[rounded corners,dashed,boundarycolor,very thick] (-1,-1) rectangle (1,1);
		\node at (0,1.2) {$\cat{P}_i$};
		\node at (-0.5,-1.2) {$\cat{M}$};
		\node at (0.5,-1.2) {$\cat{N}$};
		\end{tikzpicture}
	}
	\end{array}.
	\end{align}
\end{definition}
\begin{definition}[Vertical fusion]
  \begin{align} \label{eq:vertical_fusion_decomp}
    \begin{array}{c}
      \includeTikz{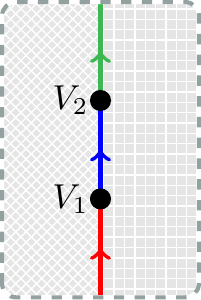}{
        \begin{tikzpicture}
          \draw[fill=bulkcolor,postaction={pattern=crosshatch,pattern color=white}] (0,0) rectangle (1,3);
          \draw[fill=bulkcolor,postaction={pattern=grid,pattern color=white}] (1,0) rectangle (2,3);
\begin{scope}[very thick,decoration={markings,mark=at position 0.5 with {\arrow{>}}}]
          \draw[postaction={decorate},red] (1,0) -- (1,1);
          \draw[postaction={decorate},blue] (1,1) -- (1,2);
          \draw[postaction={decorate},nicegreen] (1,2) -- (1,3);
          \end{scope}
          \filldraw (1,1) circle (0.1);
          \node[left] at (1,1) {$V_1$};
          \node[left] at (1,2) {$V_2$};
          \filldraw (1,2) circle (0.1);
          \draw[very thick,white,rounded corners] (0,0) rectangle (2,3);
          \draw[very thick,white] (0,0) rectangle (2,3);
          \draw[very thick,boundarycolor,rounded corners,dashed] (0,0) rectangle (2,3);
        \end{tikzpicture}
        }
    \end{array} \cong \bigoplus_i
    \begin{array}{c}
      \includeTikz{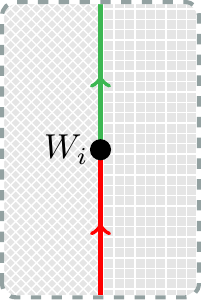}{
        \begin{tikzpicture}
          \draw[fill=bulkcolor,postaction={pattern=crosshatch,pattern color=white}] (0,0) rectangle (1,3);
          \draw[fill=bulkcolor,postaction={pattern=grid,pattern color=white}] (1,0) rectangle (2,3);
\begin{scope}[very thick,decoration={markings,mark=at position 0.5 with {\arrow{>}}}]
          \draw[postaction={decorate},red] (1,0) -- (1,1.5);
          \draw[postaction={decorate},nicegreen] (1,1.5) -- (1,3);
\end{scope}
          \filldraw (1,1.5) circle (0.1);
          \node[left] at (1,1.5) {$W_i$};
          \draw[very thick,white,rounded corners] (0,0) rectangle (2,3);
          \draw[very thick,white] (0,0) rectangle (2,3);
          \draw[very thick,boundarycolor,rounded corners,dashed] (0,0) rectangle (2,3);
        \end{tikzpicture}
        }
      \end{array}
  \end{align}
The defects interfacing between the domain wall $\cat{M}$ and itself form a fusion category ${\rm End}(\cat{M})$. The components of the isomorphism \eqref{eq:vertical_fusion_decomp} are trivalent vertices in ${\rm End}(\cat{M})$ and we can compute the associator for ${\rm End}(\cat{M})$ as described in Definition~\ref{def:vertical_defect_fusion_associator}. 
\end{definition}

\begin{definition}[Horizontal fusion]
  \begin{align}
    \begin{array}{c}
      \includeTikz{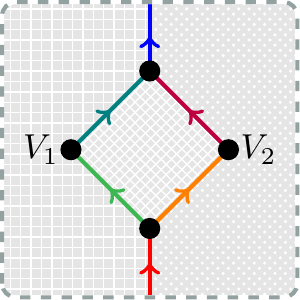}{
        \begin{tikzpicture}
          \path[fill=bulkcolor,postaction={pattern=crosshatch,pattern color=white}]
          (1.5,0.7) -- (0.7,1.5) -- (1.5,2.3) -- (2.3,1.5) -- cycle;
          \path[fill=bulkcolor,postaction={pattern=grid,pattern color=white}] (1.5,0) -- (0,0) -- (0,3) -- (1.5,3) -- (1.5,2.3) -- (0.7,1.5) -- (1.5,0.7) -- cycle;
          \path[fill=bulkcolor,postaction={pattern=crosshatch dots,pattern color=white}] (1.5,0) -- (3,0) -- (3,3) -- (1.5,3) -- (1.5,2.3) -- (2.3,1.5) -- (1.5,0.7) -- cycle;
          \begin{scope}[very thick,decoration={markings,mark=at position 0.5 with {\arrow{>}}}]
            \draw[postaction={decorate},red] (1.5,0) -- (1.5,0.7);
            \draw[postaction={decorate},nicegreen] (1.5,0.7) -- (0.7,1.5);
            \draw[postaction={decorate},orange] (1.5,0.7) --  (2.3,1.5);
            \draw[postaction={decorate},teal] (0.7,1.5) -- (1.5,2.3);
            \draw[postaction={decorate},purple] (2.3,1.5) -- (1.5,2.3);
            \draw[postaction={decorate},blue] (1.5,2.3) -- (1.5,3);
          \end{scope}
          \node[left] at (0.7,1.5) {$V_1$};
          \node[right] at (2.3,1.5) {$V_2$};
          \filldraw (1.5,0.7) circle (0.1);
          \filldraw (0.7,1.5) circle (0.1);
          \filldraw (2.3,1.5) circle (0.1);
          \filldraw (1.5,2.3) circle (0.1);
          \draw[very thick,white,rounded corners] (0,0) rectangle (3,3);
          \draw[very thick,white] (0,0) rectangle (3,3);
          \draw[very thick,boundarycolor,rounded corners,dashed] (0,0) rectangle (3,3);
          \end{tikzpicture}
        }
    \end{array} \cong \bigoplus_i
    \begin{array}{c}
      \includeTikz{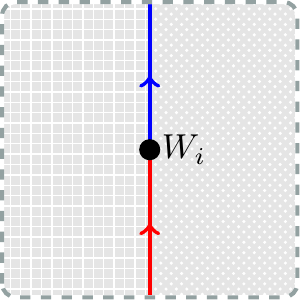}{
        \begin{tikzpicture}
          \draw[fill=bulkcolor,postaction={pattern=grid,pattern color=white}] (0,0) rectangle (1.5,3);
\draw[fill=bulkcolor,postaction={pattern=crosshatch dots,pattern color=white}] (1.5,0) rectangle (3,3);
          \begin{scope}[very thick,decoration={markings,mark=at position 0.5 with {\arrow{>}}}]
            \draw[postaction={decorate},red] (1.5,0) -- (1.5,1.5);
            \draw[postaction={decorate},blue] (1.5,1.5) -- (1.5,3);
            \end{scope}
          \filldraw (1.5,1.5) circle (0.1);
          \node[right] at (1.5,1.5) {$W_i$};
          \draw[very thick,white,rounded corners] (0,0) rectangle (3,3);
          \draw[very thick,white] (0,0) rectangle (3,3);
          \draw[very thick,boundarycolor,rounded corners,dashed] (0,0) rectangle (3,3);
          \end{tikzpicture}
        }
    \end{array}
  \end{align}
  where the trivalent point defects are those which encode the domain wall fusions as described in Definition~\ref{def:point_defect}.
\end{definition}

\begin{definition}[Associators]
  \begin{align}
    \begin{array}{c}
      \includeTikz{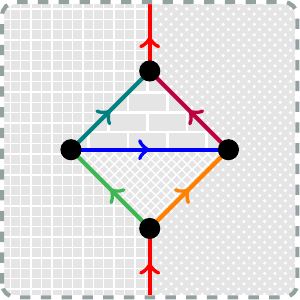}{
        \begin{tikzpicture}
          \path[fill=bulkcolor,postaction={pattern=crosshatch,pattern color=white}] (1.5,0.7) -- (0.7,1.5) -- (2.3,1.5) -- cycle;
\path[fill=bulkcolor,postaction={pattern=bricks,pattern color=white}]
          (1.5,2.3) -- (0.7,1.5) -- (2.3,1.5) -- cycle;
          \path[fill=bulkcolor,postaction={pattern=grid,pattern color=white}] (1.5,0) -- (0,0) -- (0,3) -- (1.5,3) -- (1.5,2.3) -- (0.7,1.5) -- (1.5,0.7) -- cycle;
          \path[fill=bulkcolor,postaction={pattern=crosshatch dots,pattern color=white}] (1.5,0) -- (3,0) -- (3,3) -- (1.5,3) -- (1.5,2.3) -- (2.3,1.5) -- (1.5,0.7) -- cycle;
          \begin{scope}[very thick,decoration={markings,mark=at position 0.5 with {\arrow{>}}}]
            \draw[postaction={decorate},red] (1.5,0) -- (1.5,0.7);
            \draw[postaction={decorate},nicegreen] (1.5,0.7) -- (0.7,1.5);
            \draw[postaction={decorate},orange] (1.5,0.7) --  (2.3,1.5);
            \draw[postaction={decorate},teal] (0.7,1.5) -- (1.5,2.3);
            \draw[postaction={decorate},purple] (2.3,1.5) -- (1.5,2.3);
            \draw[postaction={decorate},red] (1.5,2.3) -- (1.5,3);
            \draw[postaction={decorate},blue] (0.7,1.5) -- (2.3,1.5);
          \end{scope}
          \filldraw (1.5,0.7) circle (0.1);
          \filldraw (0.7,1.5) circle (0.1);
          \filldraw (2.3,1.5) circle (0.1);
          \filldraw (1.5,2.3) circle (0.1);
          \draw[very thick,white,rounded corners] (0,0) rectangle (3,3);
          \draw[very thick,white] (0,0) rectangle (3,3);
          \draw[very thick,boundarycolor,rounded corners,dashed] (0,0) rectangle (3,3);
          \end{tikzpicture}
        }
    \end{array} \cong \bigoplus_i
   \begin{array}{c}
      \includeTikz{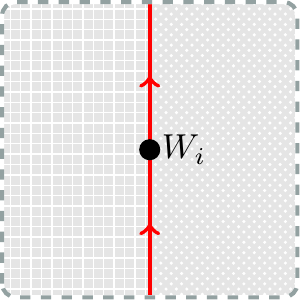}{
        \begin{tikzpicture}
          \draw[fill=bulkcolor,postaction={pattern=grid,pattern color=white}] (0,0) rectangle (1.5,3);
\draw[fill=bulkcolor,postaction={pattern=crosshatch dots,pattern color=white}] (1.5,0) rectangle (3,3);
          \begin{scope}[very thick,decoration={markings,mark=at position 0.5 with {\arrow{>}}}]
            \draw[postaction={decorate},red] (1.5,0) -- (1.5,1.5);
            \draw[postaction={decorate},red] (1.5,1.5) -- (1.5,3);
            \end{scope}
          \filldraw (1.5,1.5) circle (0.1);
          \node[right] at (1.5,1.5) {$W_i$};
          \draw[very thick,white,rounded corners] (0,0) rectangle (3,3);
          \draw[very thick,white] (0,0) rectangle (3,3);
          \draw[very thick,boundarycolor,rounded corners,dashed] (0,0) rectangle (3,3);
          \end{tikzpicture}
        }
    \end{array}
  \end{align}
  All the trivalent point defects are encoding domain wall fusions.
\end{definition}

\begin{definition}[$S_3$ as a semi-direct product]
  It is well known that $S_3 \cong \mathbb{Z}/2\mathbb{Z} \ltimes \mathbb{Z}/3\mathbb{Z}$ where the involution on $\mathbb{Z}/3\mathbb{Z}$ is multiplication by $2$. Throughout this paper, $S_3$ plays an important role and we use the semi-direct product presentation of $S_3$ because it is more amenable to computer implementation. In this representation, multiplication is given by
\begin{align}
  (a_0,b_0) \cdot (a_1,b_1) = (a_0 + a_1, (1+a_1)b_0 + b_1),
\end{align}
where the first entry is taken modulo 2 and the second modulo 3.
One caveat of this notation is that sometimes you need to be careful about applying the ``obvious'' rules of arithmetic like associativity and distributivity.
  \end{definition}


\section{Domain Wall Definitions}\label{sec:S3setup}

In this section, we briefly discuss the domain walls (bimodule categories) that occur in this paper.
Recall from Chapter 7 of \onlinecite{MR3242743} that $\vvec{G_1}|\vvec{G_2}$ bimodules are classified by a subgroup $M\leqslant G_1\times G_2^{\rm op}$ and a 2-cocycle
\begin{align}
\omega\in& H^2(M,U(1)) \label{eq:associator_spaceA}\\
 &\cong H^2(G_1 \times G_2^{\rm op},{\rm maps}(G_1 \times G_2^{\rm op}/M,U(1))) \label{eq:associator_spaceB}
\end{align}
up to conjugation.
This defining data is provided in Appendix \ref{appendix:domain_wall_definitions}. Where these domain walls have occurred in the literature, we have attempted to be consistent with already established names. In many cases, we have named corresponding $\cat{C}|\cat{D}$ and $\cat{D}|\cat{C}$ bimodules with the same label. Which version of the bimodule is being discussed can be easily inferred from context. 

Simple objects in the bimodule category are labeled by cosets of $M$. The cocycle $\omega$ in Eqn.~\ref{eq:associator_spaceA} does not directly give the associators of the fusion category action.
One way to see this is that the associator is defined for a triple $(g_1,g_2,m)\in G_1\times G_2^{\rm op}\times (G_1 \times G_2^{\rm op} / M)$, whereas Eqn.~\ref{eq:associator_spaceA} has domain $M\times M$. The associator is obtained via the isomorphism to Eqn.~\ref{eq:associator_spaceB}. 
An implementation of this isomorphism is given by Tyler Lawson in \onlinecite{288304}. For all the bimodules considered in this paper, a gauge can be chosen such that the left and right associators are trivial. We have used the following equations for $\vvec{S_3} \lvert \vvec{S_3}$ domain walls
\begin{subequations}
\begin{align}
A_{2,1}:\Omega((a,b),m,(c,d))&=(-1)^{ac}\\
A_{3,1}:\Omega((a,b),m,(c,d))&=\omega^{(1+c)bd(-1)^{\lfloor m/2\rfloor}},\, m\in\{1,2,3,4\}\\
D_{R,1}:\Omega((a,b),m,(c,d))&=(-1)^{ac}\\
D_{L,1}:\Omega((a,b),m,(c,d))&=(-1)^{ac}\\
G_{1}:\Omega((a,b),m,(c,d))&=\omega^{(1+c)bd(m+1)},\, m\in\{0,1\}\\
F_{1}:\Omega((a,b),*,(c,d))&=(-1)^{ac},
\end{align}
\end{subequations}
with all other associators being trivial. The gauge choices for other bimodules can be found in the code provided in the ancillary material\cite{anc}.


\section{Examples}\label{sec:examples}

In this section, we discuss some interesting example computations. This serves to both show how calculations are done, and to provide some physical interpretation for the results. 

The computations performed in this paper are very similar to those in \onlinecites{1806.01279,1810.09469,1901.08069}, and we refer the reader to those papers for more examples. Due to the increased complexity of $\vvec{S_3}$, the computations here require computer assistance\cite{anc}.

\subsection{Domain wall fusion}

To compute the result of domain wall fusion, we use the algorithm described in Definition~\ref{def:laddercat}. Extensive examples are provided in \onlinecite{1806.01279} for the $\vvec{\ZZ{p}}$ case. We therefore only provide two examples here, and refer the reader to the previous work for more details.

\subsubsection{$G_1\otimes_{\vvec{S_3}}G_1$}
The first step in computing the decomposition of $G_1\otimes_{\vvec{S_3}}G_1$ is to construct the Karoubi envelope of the ladder category (defined in Def.~\ref{def:laddercat}). The trivalent vertices for $G_1$ are 
\begin{align}
	\begin{array}{c}
	\includeTikz{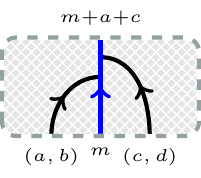}{
		\begin{tikzpicture}
		\fill[fill=bulkcolor,postaction={pattern=crosshatch,pattern color=white}] (-1,-0.5) rectangle (0,0.5);
		\fill[fill=bulkcolor,postaction={pattern=crosshatch,pattern color=white}] (0,-0.5) rectangle (1,0.5);
		\begin{scope}[very thick,decoration={markings,mark=at position 0.5 with {\arrow{>}}}]
		\draw[postaction={decorate}] (0.5,-0.5) to[out=90,in=0] (0,0.3);
		\draw[postaction={decorate}] (-0.5,-0.5) to[out=90,in=180] (0,0.1);
		\draw[postaction={decorate},blue] (0,-0.5) -- (0,0.5);
		\end{scope}
		\node[below] at (0,-0.5) {\tiny $m$};
		\node[below] at (-0.5,-0.5) {\tiny$(a,b)$};
		\node[below] at (0.5,-0.5) {\tiny$(c,d)$};
		\node[above] at (0,0.5) {\tiny $m{+}a{+}c$};
		\draw[very thick,white] (-1,-0.5) rectangle (1,0.5);
		\draw[very thick,white,rounded corners] (-1,-0.5) rectangle (1,0.5);
		\draw[very thick,boundarycolor,rounded corners,dashed] (-1,-0.5) rectangle (1,0.5);
		\end{tikzpicture}
	}
	\end{array}
	&=\omega^{(1+c)bd(1+m)}
	\begin{array}{c}
	\includeTikz{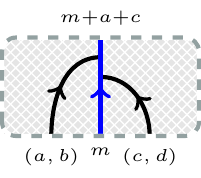}{
		\begin{tikzpicture}
		\fill[fill=bulkcolor,postaction={pattern=crosshatch,pattern color=white}] (-1,-0.5) rectangle (0,0.5);
		\fill[fill=bulkcolor,postaction={pattern=crosshatch,pattern color=white}] (0,-0.5) rectangle (1,0.5);
		\begin{scope}[very thick,decoration={markings,mark=at position 0.5 with {\arrow{>}}}]
		\draw[postaction={decorate}] (0.5,-0.5) to[out=90,in=0] (0,0.1);
		\draw[postaction={decorate}] (-0.5,-0.5) to[out=90,in=180] (0,0.3);
		\draw[postaction={decorate},blue] (0,-0.5) -- (0,0.5);
		\end{scope}
		\node[below] at (0,-0.5) {\tiny $m$};
		\node[below] at (-0.5,-0.5) {\tiny$(a,b)$};
		\node[below] at (0.5,-0.5) {\tiny$(c,d)$};
		\node[above] at (0,0.5) {\tiny $m{+}a{+}c$};
		\draw[very thick,white] (-1,-0.5) rectangle (1,0.5);
		\draw[very thick,white,rounded corners] (-1,-0.5) rectangle (1,0.5);
		\draw[very thick,boundarycolor,rounded corners,dashed] (-1,-0.5) rectangle (1,0.5);
		\end{tikzpicture}
	}
	\end{array},
\end{align}
with $m\in \{0,1\}$. To construct the Karoubi envelope, we need to construct idempotent ladder diagrams
\begin{align}
\begin{array}{c}
\includeTikz{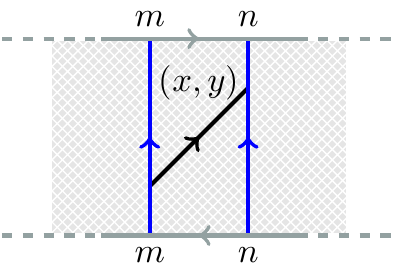}{
	\begin{tikzpicture}
	\draw[white,fill=bulkcolor,postaction={pattern=crosshatch,pattern color=white}] (-1.5,-1) rectangle (-0.5,1);
	\draw[white,fill=bulkcolor,postaction={pattern=crosshatch,pattern color=white}] (-0.5,-1) rectangle (0.5,1);
	\draw[white,fill=bulkcolor,postaction={pattern=crosshatch,pattern color=white}] (0.5,-1) rectangle (1.5,1);
	\begin{scope}[very thick,decoration={markings,mark=at position 0.5 with {\arrow{>}}}]
	\draw[postaction={decorate}] (-0.5,-0.5) -- (0.5,0.5);
	\draw[blue,postaction={decorate}] (-0.5,-1) -- (-0.5,1);
	\draw[blue,postaction={decorate}] (0.5,-1) -- (0.5,1);
	\draw[boundarycolor,postaction={decorate}] (-1,1) -- (1,1);
	\draw[white] (1,1) -- (2,1);
	\draw[white] (-2,1) -- (-1,1);
	\draw[white] (1,-1) -- (2,-1);
	\draw[white] (-2,-1) -- (-1,-1);
	\draw[boundarycolor,dashed] (1,1) -- (2,1);
	\draw[boundarycolor,dashed] (-2,1) -- (-1,1);
	\draw[boundarycolor,dashed] (1,-1) -- (2,-1);
	\draw[boundarycolor,dashed] (-2,-1) -- (-1,-1);
	\draw[boundarycolor,postaction={decorate}] (1,-1) -- (-1,-1);
	\end{scope}
	\node[below] at (-0.5,-1) {$m$};
	\node[below] at (0.5,-1) {$n$};
	\node[above] at (-0.5,1) {$m$};
	\node[above] at (0.5,1) {$n$};
	\node[above] at (0,.25) {$(x,y)$};
	\end{tikzpicture}
}
\end{array}.
\end{align}
For such a diagram to be an idempotent, it must be the case that $x=0$. Diagrams which can be related by multiplying by a general ladder diagram are isomorphic in the Ladder category\cite{1806.01279}, so it is sufficient to consider diagrams of the form 
\begin{align}
\begin{array}{c}
\includeTikz{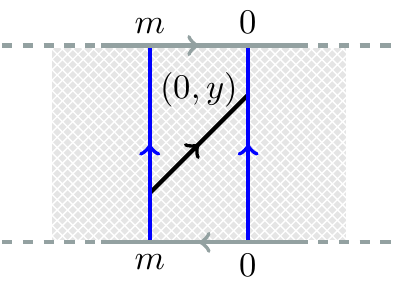}{
	\begin{tikzpicture}
	\draw[white,fill=bulkcolor,postaction={pattern=crosshatch,pattern color=white}] (-1.5,-1) rectangle (-0.5,1);
	\draw[white,fill=bulkcolor,postaction={pattern=crosshatch,pattern color=white}] (-0.5,-1) rectangle (0.5,1);
	\draw[white,fill=bulkcolor,postaction={pattern=crosshatch,pattern color=white}] (0.5,-1) rectangle (1.5,1);
	\begin{scope}[very thick,decoration={markings,mark=at position 0.5 with {\arrow{>}}}]
	\draw[postaction={decorate}] (-0.5,-0.5) -- (0.5,0.5);
	\draw[blue,postaction={decorate}] (-0.5,-1) -- (-0.5,1);
	\draw[blue,postaction={decorate}] (0.5,-1) -- (0.5,1);
	\draw[boundarycolor,postaction={decorate}] (-1,1) -- (1,1);
	\draw[white] (1,1) -- (2,1);
	\draw[white] (-2,1) -- (-1,1);
	\draw[white] (1,-1) -- (2,-1);
	\draw[white] (-2,-1) -- (-1,-1);
	\draw[boundarycolor,dashed] (1,1) -- (2,1);
	\draw[boundarycolor,dashed] (-2,1) -- (-1,1);
	\draw[boundarycolor,dashed] (1,-1) -- (2,-1);
	\draw[boundarycolor,dashed] (-2,-1) -- (-1,-1);
	\draw[boundarycolor,postaction={decorate}] (1,-1) -- (-1,-1);
	\end{scope}
	\node[below] at (-0.5,-1) {$m$};
	\node[below] at (0.5,-1) {$0$};
	\node[above] at (-0.5,1) {$m$};
	\node[above] at (0.5,1) {$0$};
	\node[above] at (0,.25) {$(0,y)$};
	\end{tikzpicture}
}
\end{array}.
\end{align}
These diagrams form the group algebra $\mathbb{C}[\ZZ{3}]$, so the primitive idempotents are
\begin{align}
X_{m,\alpha}:=\frac{1}{3}\sum_y \omega^{\alpha y}
\begin{array}{c}
\includeTikz{ladder_category_morphism_G1G1_2}{
	\begin{tikzpicture}
	\draw[white,fill=bulkcolor,postaction={pattern=crosshatch,pattern color=white}] (-1.5,-1) rectangle (-0.5,1);
	\draw[white,fill=bulkcolor,postaction={pattern=crosshatch,pattern color=white}] (-0.5,-1) rectangle (0.5,1);
	\draw[white,fill=bulkcolor,postaction={pattern=crosshatch,pattern color=white}] (0.5,-1) rectangle (1.5,1);
	\begin{scope}[very thick,decoration={markings,mark=at position 0.5 with {\arrow{>}}}]
	\draw[postaction={decorate}] (-0.5,-0.5) -- (0.5,0.5);
	\draw[blue,postaction={decorate}] (-0.5,-1) -- (-0.5,1);
	\draw[blue,postaction={decorate}] (0.5,-1) -- (0.5,1);
	\draw[boundarycolor,postaction={decorate}] (-1,1) -- (1,1);
	\draw[white] (1,1) -- (2,1);
	\draw[white] (-2,1) -- (-1,1);
	\draw[white] (1,-1) -- (2,-1);
	\draw[white] (-2,-1) -- (-1,-1);
	\draw[boundarycolor,dashed] (1,1) -- (2,1);
	\draw[boundarycolor,dashed] (-2,1) -- (-1,1);
	\draw[boundarycolor,dashed] (1,-1) -- (2,-1);
	\draw[boundarycolor,dashed] (-2,-1) -- (-1,-1);
	\draw[boundarycolor,postaction={decorate}] (1,-1) -- (-1,-1);
	\end{scope}
	\node[below] at (-0.5,-1) {$m$};
	\node[below] at (0.5,-1) {$0$};
	\node[above] at (-0.5,1) {$m$};
	\node[above] at (0.5,1) {$0$};
	\node[above] at (0,.25) {$(0,y)$};
	\end{tikzpicture}
}
\end{array}.
\end{align}
The left and right trivalent vertices are skein vectors that absorb these idempotents on the top and bottom. Being careful with the associator, one can check that in this case the vertices are

\begin{align}
\begin{array}{c}
\includeTikz{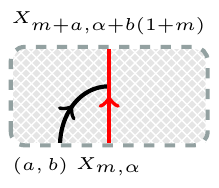}{
	\begin{tikzpicture}
	\fill[fill=bulkcolor,postaction={pattern=crosshatch,pattern color=white}] (-1,-0.5) rectangle (0,0.5);
	\fill[fill=bulkcolor,postaction={pattern=crosshatch,pattern color=white}] (0,-0.5) rectangle (1,0.5);
	\begin{scope}[very thick,decoration={markings,mark=at position 0.5 with {\arrow{>}}}]
	\draw[postaction={decorate}] (-0.5,-0.5) to[out=90,in=180] (0,0.1);
	\draw[postaction={decorate},red] (0,-0.5) -- (0,0.5);
	\end{scope}
	\node[below] at (0,-0.5) {\tiny $X_{m,\alpha}$};
	\node[below] at (-0.7,-0.5) {\tiny$(a,b)$};
	\node[above] at (0,0.5) {\tiny $X_{m+a,\alpha+b(1+m)}$};
	\draw[very thick,white] (-1,-0.5) rectangle (1,0.5);
	\draw[very thick,white,rounded corners] (-1,-0.5) rectangle (1,0.5);
	\draw[very thick,boundarycolor,rounded corners,dashed] (-1,-0.5) rectangle (1,0.5);
	\end{tikzpicture}
}
\end{array}
\mapsto
\sum_y \frac{\omega^{\alpha y}}{3}
\begin{array}{c}
\includeTikz{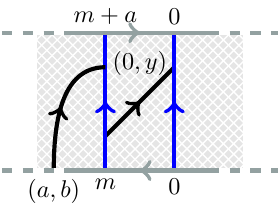}{
	\begin{tikzpicture}[scale=.7,every node/.style={scale=.7}]
	\draw[white,fill=bulkcolor,postaction={pattern=crosshatch,pattern color=white}] (-1.5,-1) rectangle (-0.5,1);
	\draw[white,fill=bulkcolor,postaction={pattern=crosshatch,pattern color=white}] (-0.5,-1) rectangle (0.5,1);
	\draw[white,fill=bulkcolor,postaction={pattern=crosshatch,pattern color=white}] (0.5,-1) rectangle (1.5,1);
	\begin{scope}[very thick,decoration={markings,mark=at position 0.5 with {\arrow{>}}}]
	\draw[postaction={decorate}] (-0.5,-0.5) -- (0.5,0.5);
	\draw[blue,postaction={decorate}] (-0.5,-1) -- (-0.5,1);
	\draw[blue,postaction={decorate}] (0.5,-1) -- (0.5,1);
	\draw[boundarycolor,postaction={decorate}] (-1,1) -- (1,1);
	\draw[postaction={decorate}] (-1.25,-1) to[out=90,in=180] (-0.5,0.5);
	\draw[white] (1,1) -- (2,1);
	\draw[white] (-2,1) -- (-1,1);
	\draw[white] (1,-1) -- (2,-1);
	\draw[white] (-2,-1) -- (-1,-1);
	\draw[boundarycolor,dashed] (1,1) -- (2,1);
	\draw[boundarycolor,dashed] (-2,1) -- (-1,1);
	\draw[boundarycolor,dashed] (1,-1) -- (2,-1);
	\draw[boundarycolor,dashed] (-2,-1) -- (-1,-1);
	\draw[boundarycolor,postaction={decorate}] (1,-1) -- (-1,-1);
	\end{scope}
	\node[below] at (-1.25,-1) {$(a,b)$};
	\node[below] at (-0.5,-1) {$m$};
	\node[below] at (0.5,-1) {$0$};
	\node[above] at (-0.5,1) {$m+a$};
	\node[above] at (0.5,1) {$0$};
	\node[above] at (0,.25) {$(0,y)$};
	\end{tikzpicture}
}
\end{array},\\
\begin{array}{c}
\includeTikz{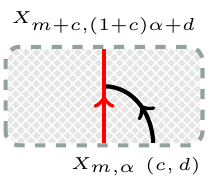}{
	\begin{tikzpicture}
	\fill[fill=bulkcolor,postaction={pattern=crosshatch,pattern color=white}] (-1,-0.5) rectangle (0,0.5);
	\fill[fill=bulkcolor,postaction={pattern=crosshatch,pattern color=white}] (0,-0.5) rectangle (1,0.5);
	\begin{scope}[very thick,decoration={markings,mark=at position 0.5 with {\arrow{>}}}]
	\draw[postaction={decorate}] (0.5,-0.5) to[out=90,in=0] (0,0.1);
	\draw[postaction={decorate},red] (0,-0.5) -- (0,0.5);
	\end{scope}
	\node[below] at (0,-0.5) {\tiny $X_{m,\alpha}$};
	\node[below] at (0.7,-0.5) {\tiny$(c,d)$};
	\node[above] at (0,0.5) {\tiny $X_{m+c,(1+c)\alpha+d}$};
	\draw[very thick,white] (-1,-0.5) rectangle (1,0.5);
	\draw[very thick,white,rounded corners] (-1,-0.5) rectangle (1,0.5);
	\draw[very thick,boundarycolor,rounded corners,dashed] (-1,-0.5) rectangle (1,0.5);
	\end{tikzpicture}
}
\end{array}
\mapsto
\sum_y \frac{\omega^{\alpha y}}{3}
\begin{array}{c}
\includeTikz{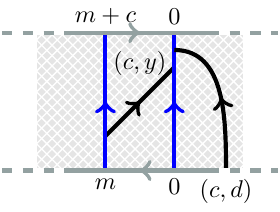}{
	\begin{tikzpicture}[scale=.7,every node/.style={scale=.7}]
	\draw[white,fill=bulkcolor,postaction={pattern=crosshatch,pattern color=white}] (-1.5,-1) rectangle (-0.5,1);
	\draw[white,fill=bulkcolor,postaction={pattern=crosshatch,pattern color=white}] (-0.5,-1) rectangle (0.5,1);
	\draw[white,fill=bulkcolor,postaction={pattern=crosshatch,pattern color=white}] (0.5,-1) rectangle (1.5,1);
	\begin{scope}[very thick,decoration={markings,mark=at position 0.5 with {\arrow{>}}}]
	\draw[postaction={decorate}] (-0.5,-0.5) -- (0.5,0.5);
	\draw[blue,postaction={decorate}] (-0.5,-1) -- (-0.5,1);
	\draw[blue,postaction={decorate}] (0.5,-1) -- (0.5,1);
	\draw[boundarycolor,postaction={decorate}] (-1,1) -- (1,1);
	\draw[postaction={decorate}] (1.25,-1) to[out=90,in=0] (0.5,0.75);
	\draw[white] (1,1) -- (2,1);
	\draw[white] (-2,1) -- (-1,1);
	\draw[white] (1,-1) -- (2,-1);
	\draw[white] (-2,-1) -- (-1,-1);
	\draw[boundarycolor,dashed] (1,1) -- (2,1);
	\draw[boundarycolor,dashed] (-2,1) -- (-1,1);
	\draw[boundarycolor,dashed] (1,-1) -- (2,-1);
	\draw[boundarycolor,dashed] (-2,-1) -- (-1,-1);
	\draw[boundarycolor,postaction={decorate}] (1,-1) -- (-1,-1);
	\end{scope}
	\node[below] at (1.25,-1) {$(c,d)$};
	\node[below] at (-0.5,-1) {$m$};
	\node[below] at (0.5,-1) {$0$};
	\node[above] at (-0.5,1) {$m+c$};
	\node[above] at (0.5,1) {$0$};
	\node[above] at (0,.25) {$(c,y)$};
	\end{tikzpicture}
}
\end{array}.
\end{align}
It is straightforward to verify that the associator is trivial. From this, we obtain 
\begin{align}
G_1\otimes_{\vvec{S_3}}G_1\cong I.
\end{align}

\subsubsection{Interpreting $\cat{M}\otimes_{\cat{C}}\cat{N}=\cat{P}+\cat{Q}$}
In the Brauer-Picard tables of Appendix~\ref{ap:domain_wall_fusions}, there are instances of domain wall fusions of the kind
\begin{align}
	\cat{M}\otimes_{\cat{C}}\cat{N}=\cat{P}+\cat{Q}.
\end{align}
We now provide a physical interpretation of this result.
For concreteness, we will consider the fusion of $K_4$ walls. The fusion for these domain walls is
\begin{align}
	K_4\otimes_{\vvec{\ZZ{2}}} K_4&=I_2\\
	K_4\otimes_{\vvec{S_3}}K_4&=T^{\ZZ{2}}+X_1^{\ZZ{2}}.
\end{align}
We define three domain wall structures on the cylinder $\mathbb{R} \times S^1$:
\begin{subequations}
\begin{align} \Sigma_{X_1} &= 
\begin{array}{c}
		\includeTikz{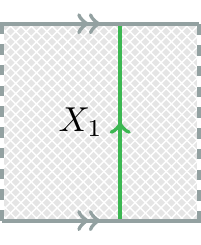}{
			\begin{tikzpicture}
			\node[above] at (0,1) {};
			\node[below] at (0,-1) {};
			\fill[fill=bulkcolor,postaction={pattern=crosshatch,pattern color=white}] (-1,-1) rectangle (1,1);
			\draw[very thick,white] (-1,-1) rectangle (1,1);
			\begin{scope}[very thick,decoration={markings,mark=at position 0.5 with {\arrow{>>}}}]
			\draw[postaction={decorate},boundarycolor] (-1,1) -- (1,1);
			\draw[postaction={decorate},boundarycolor] (-1,-1) -- (1,-1);
			\end{scope}
			\begin{scope}[very thick,decoration={markings,mark=at position 0.5 with {\arrow{>}}}]
			\draw[postaction={decorate},nicegreen] (0.2,-1) -- (0.2,1);
			\end{scope}
			\node at (-0.2,0) {$X_1$};
			\draw[very thick,boundarycolor,dashed] (-1,-1) rectangle (1,1);
			\end{tikzpicture}
		}
	\end{array} \\
	\Sigma_{T} &= 
	\begin{array}{c}
		\includeTikz{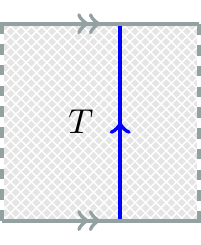}{
			\begin{tikzpicture}
			\node[above] at (0,1) {};
			\node[below] at (0,-1) {};
			\fill[fill=bulkcolor,postaction={pattern=crosshatch,pattern color=white}] (-1,-1) rectangle (1,1);
			\draw[very thick,white] (-1,-1) rectangle (1,1);
			\begin{scope}[very thick,decoration={markings,mark=at position 0.5 with {\arrow{>>}}}]
			\draw[postaction={decorate},boundarycolor] (-1,1) -- (1,1);
			\draw[postaction={decorate},boundarycolor] (-1,-1) -- (1,-1);
			\end{scope}
			\begin{scope}[very thick,decoration={markings,mark=at position 0.5 with {\arrow{>}}}]
			\draw[postaction={decorate},blue] (0.2,-1) -- (0.2,1);
			\end{scope}
			\node at (-0.2,0) {$T$};
			\draw[very thick,boundarycolor,dashed] (-1,-1) rectangle (1,1);
			\end{tikzpicture}
		}
	\end{array} \\
	\Sigma_{K_4,K_4} &= 
	\begin{array}{c}
		\includeTikz{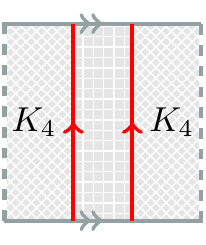}{
			\begin{tikzpicture}
			\node[above] at (0,1) {};
			\node[below] at (0,-1) {};
			\fill[fill=bulkcolor,postaction={pattern=crosshatch,pattern color=white}] (-1,-1) rectangle (-0.3,1);
			\fill[fill=bulkcolor,postaction={pattern=grid,pattern color=white}] (-0.3,-1) rectangle (0.3,1);
			\fill[fill=bulkcolor,postaction={pattern=crosshatch,pattern color=white}] (0.3,-1) rectangle (1,1);
			\draw[very thick,white] (-1,-1) rectangle (1,1);
			\begin{scope}[very thick,decoration={markings,mark=at position 0.5 with {\arrow{>>}}}]
			\draw[postaction={decorate},boundarycolor] (-1,1) -- (1,1);
			\draw[postaction={decorate},boundarycolor] (-1,-1) -- (1,-1);
			\end{scope}
			\begin{scope}[very thick,decoration={markings,mark=at position 0.5 with {\arrow{>}}}]
			\draw[postaction={decorate},red] (0.3,-1) -- (0.3,1);
			\draw[postaction={decorate},red] (-0.3,-1) -- (-0.3,1);
			\end{scope}
			\node at (-0.7,0) {$K_4$};
			\node at (0.7,0) {$K_4$};
			\draw[very thick,boundarycolor,dashed] (-1,-1) rectangle (1,1);
			\end{tikzpicture}
		}
	\end{array}
\end{align}
\end{subequations}
The ground states on $\Sigma_{X_1}$ and $\Sigma_T$ are given respectively by
\begin{subequations}
\begin{align}
	\ket{\psi_a}&=
	\bigl|\adjustbox{valign=M}{$
		\begin{array}{c}
		\includeTikz{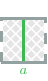}
		{
			\begin{tikzpicture}[scale=.3,every node/.style={scale=.5}]
			\fill[fill=bulkcolor,postaction={pattern=crosshatch,pattern color=white}] (-0.75,-0.75) rectangle (0.75,0.75);
			\draw[white,thick] (-.75,-.75) rectangle (.75,.75);
			\draw[nicegreen,thick] (0,-.75)--(0,.75) node [pos=0,below] {\footnotesize$a$} node [pos=1,above,white] {\footnotesize$a$};
			\draw[thick,boundarycolor,dashed] (-0.75,-0.75) rectangle (0.75,0.75);
			\draw[thick,boundarycolor] (-0.75,0.75) -- (0.75,0.75);
			\draw[thick,boundarycolor] (-0.75,-0.75) -- (0.75,-0.75);
			\end{tikzpicture}
		}
		\end{array}$}\bigr\rangle
	+
	\bigl|\adjustbox{valign=M}{$
		\begin{array}{c}
		\includeTikz{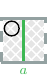}
		{
			\begin{tikzpicture}[scale=.3,every node/.style={scale=.5}]
			\fill[fill=bulkcolor,postaction={pattern=crosshatch,pattern color=white}] (-0.75,-0.75) rectangle (0.75,0.75);
			\draw[white,thick] (-.75,-.75) rectangle (.75,.75);
			\draw[nicegreen,thick] (0,-.75)--(0,.75) node [pos=0,below] {\footnotesize$a$} node [pos=1,above,white] {\footnotesize$a$};
			\draw[black] (-.4,.4) circle (.25);
			\draw[thick,boundarycolor,dashed] (-0.75,-0.75) rectangle (0.75,0.75);
			\draw[thick,boundarycolor] (-0.75,0.75) -- (0.75,0.75);
			\draw[thick,boundarycolor] (-0.75,-0.75) -- (0.75,-0.75);
			\end{tikzpicture}
		}
		\end{array}$}
	\bigr\rangle
	+
	\bigl|\adjustbox{valign=M}{$
		\begin{array}{c}
		\includeTikz{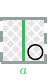}
		{
			\begin{tikzpicture}[scale=.3,every node/.style={scale=.5}]
			\fill[fill=bulkcolor,postaction={pattern=crosshatch,pattern color=white}] (-0.75,-0.75) rectangle (0.75,0.75);
			\draw[white,thick] (-.75,-.75) rectangle (.75,.75);
			\draw[nicegreen,thick] (0,-.75)--(0,.75) node [pos=0,below] {\footnotesize$a$} node [pos=1,above,white] {\footnotesize$a$};
			\draw[black] (.4,-.4) circle (.25);
			\draw[thick,boundarycolor,dashed] (-0.75,-0.75) rectangle (0.75,0.75);
			\draw[thick,boundarycolor] (-0.75,0.75) -- (0.75,0.75);
			\draw[thick,boundarycolor] (-0.75,-0.75) -- (0.75,-0.75);
			\end{tikzpicture}
		}
		\end{array}$}
	\bigr\rangle
	+
	\bigl|
	\adjustbox{valign=M}{$
		\begin{array}{c}
		\includeTikz{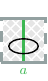}
		{
			\begin{tikzpicture}[scale=.3,every node/.style={scale=.5}]
			\fill[fill=bulkcolor,postaction={pattern=crosshatch,pattern color=white}] (-0.75,-0.75) rectangle (0.75,0.75);
			\draw[white,thick] (-.75,-.75) rectangle (.75,.75);
			\draw[nicegreen,thick] (0,-.75)--(0,.75) node [pos=0,below] {\footnotesize$a$} node [pos=1,above,white] {\footnotesize$a$};
			\draw[black] (0,-0.2) ellipse (.5 and .25);
			\draw[thick,boundarycolor,dashed] (-0.75,-0.75) rectangle (0.75,0.75);
			\draw[thick,boundarycolor] (-0.75,0.75) -- (0.75,0.75);
			\draw[thick,boundarycolor] (-0.75,-0.75) -- (0.75,-0.75);
			\end{tikzpicture}
		}
		\end{array}$}
	\bigr\rangle
	+
	\bigl|\adjustbox{valign=M}{$
		\begin{array}{c}
		\includeTikz{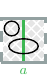}
		{
			\begin{tikzpicture}[scale=.3,every node/.style={scale=.5}]
			\fill[fill=bulkcolor,postaction={pattern=crosshatch,pattern color=white}] (-0.75,-0.75) rectangle (0.75,0.75);
			\draw[white,thick] (-.75,-.75) rectangle (.75,.75);
			\draw[nicegreen,thick] (0,-.75)--(0,.75) node [pos=0,below] {\footnotesize$a$} node [pos=1,above,white] {\footnotesize$a$};
			\draw[black] (-.4,.4) circle (.25);
			\draw[black] (0,-0.2) ellipse (.5 and .25);
			\draw[thick,boundarycolor,dashed] (-0.75,-0.75) rectangle (0.75,0.75);
			\draw[thick,boundarycolor] (-0.75,0.75) -- (0.75,0.75);
			\draw[thick,boundarycolor] (-0.75,-0.75) -- (0.75,-0.75);
			\end{tikzpicture}
		}
		\end{array}$}\bigr\rangle
	+\ldots \\
	\ket{\psi_{a,b}}&=
	\bigl|\adjustbox{valign=M}{$
		\begin{array}{c}
		\includeTikz{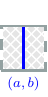}
		{
			\begin{tikzpicture}[scale=.3,every node/.style={scale=.5}]
			\fill[fill=bulkcolor,postaction={pattern=crosshatch,pattern color=white}] (-0.75,-0.75) rectangle (0.75,0.75);
			\draw[thick,white] (-.75,-.75) rectangle (.75,.75);
			\draw[blue,thick] (0,-.75)--(0,.75)  node [pos=0,below] {\footnotesize$(a,b)$} node [pos=1,above,white] {\footnotesize$(a,b)$};
			\draw[thick,boundarycolor,dashed] (-0.75,-0.75) rectangle (0.75,0.75);
			\draw[thick,boundarycolor] (-0.75,0.75) -- (0.75,0.75);
			\draw[thick,boundarycolor] (-0.75,-0.75) -- (0.75,-0.75);
			\end{tikzpicture}
		}
		\end{array}$}\bigr\rangle
	+
	\bigl|\adjustbox{valign=M}{$
		\begin{array}{c}
		\includeTikz{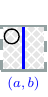}
		{
			\begin{tikzpicture}[scale=.3,every node/.style={scale=.5}]
			\fill[fill=bulkcolor,postaction={pattern=crosshatch,pattern color=white}] (-0.75,-0.75) rectangle (0.75,0.75);
			\draw[thick,white] (-.75,-.75) rectangle (.75,.75);
			\draw[blue,thick] (0,-.75)--(0,.75)  node [pos=0,below] {\footnotesize$(a,b)$} node [pos=1,above,white] {\footnotesize$(a,b)$};
			\draw[black] (-.4,.4) circle (.25);
			\draw[thick,boundarycolor,dashed] (-0.75,-0.75) rectangle (0.75,0.75);
			\draw[thick,boundarycolor] (-0.75,0.75) -- (0.75,0.75);
			\draw[thick,boundarycolor] (-0.75,-0.75) -- (0.75,-0.75);
			\end{tikzpicture}
		}
		\end{array}$}
	\bigr\rangle
	+
	\bigl|\adjustbox{valign=M}{$
		\begin{array}{c}
		\includeTikz{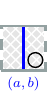}
		{
			\begin{tikzpicture}[scale=.3,every node/.style={scale=.5}]
			\fill[fill=bulkcolor,postaction={pattern=crosshatch,pattern color=white}] (-0.75,-0.75) rectangle (0.75,0.75);
			\draw[thick,white] (-.75,-.75) rectangle (.75,.75);
			\draw[blue,thick] (0,-.75)--(0,.75)  node [pos=0,below] {\footnotesize$(a,b)$} node [pos=1,above,white] {\footnotesize$(a,b)$};
			\draw[black] (.4,-.4) circle (.25);
			\draw[thick,boundarycolor,dashed] (-0.75,-0.75) rectangle (0.75,0.75);
			\draw[thick,boundarycolor] (-0.75,0.75) -- (0.75,0.75);
			\draw[thick,boundarycolor] (-0.75,-0.75) -- (0.75,-0.75);
			\end{tikzpicture}
		}
		\end{array}$}
	\bigr\rangle
	+
	\bigl|
	\adjustbox{valign=M}{$
		\begin{array}{c}
		\includeTikz{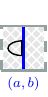}
		{
			\begin{tikzpicture}[scale=.3,every node/.style={scale=.5}]
			\fill[fill=bulkcolor,postaction={pattern=crosshatch,pattern color=white}] (-0.75,-0.75) rectangle (0.75,0.75);
			\draw[thick,white] (-.75,-.75) rectangle (.75,.75);
			\draw[blue,thick] (0,-.75)--(0,.75)  node [pos=0,below] {\footnotesize$(a,b)$} node [pos=1,above,white] {\footnotesize$(a,b)$};
			\begin{scope}[shift={(-.8pt,0)}]
			\clip (-.55,-.3) rectangle (0,.3);
			\draw[black] (0,0) ellipse (.5 and .25);
			\draw[black,thick] (0,-.25)--(0,.25);
			\end{scope}
			\draw[thick,boundarycolor,dashed] (-0.75,-0.75) rectangle (0.75,0.75);
			\draw[thick,boundarycolor] (-0.75,0.75) -- (0.75,0.75);
			\draw[thick,boundarycolor] (-0.75,-0.75) -- (0.75,-0.75);
			\end{tikzpicture}
		}
		\end{array}$}
	\bigr\rangle
	+
	\bigl|\adjustbox{valign=M}{$
		\begin{array}{c}
		\includeTikz{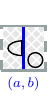}
		{
			\begin{tikzpicture}[scale=.3,every node/.style={scale=.5}]
			\fill[fill=bulkcolor,postaction={pattern=crosshatch,pattern color=white}] (-0.75,-0.75) rectangle (0.75,0.75);
			\draw[thick,white] (-.75,-.75) rectangle (.75,.75);
			\draw[blue,thick] (0,-.75)--(0,.75)  node [pos=0,below] {\footnotesize$(a,b)$} node [pos=1,above,white] {\footnotesize$(a,b)$};
			\draw[black] (.4,-.4) circle (.25);
			\begin{scope}[shift={(-.8pt,0)}]
			\clip (-.55,-.3) rectangle (0,.3);
			\draw[black] (0,0) ellipse (.5 and .25);
			\draw[black,thick] (0,-.25)--(0,.25);
			\end{scope}
			\draw[thick,boundarycolor,dashed] (-0.75,-0.75) rectangle (0.75,0.75);
			\draw[thick,boundarycolor] (-0.75,0.75) -- (0.75,0.75);
			\draw[thick,boundarycolor] (-0.75,-0.75) -- (0.75,-0.75);
			\end{tikzpicture}
		}
		\end{array}$}\bigr\rangle
	+\ldots
\end{align}
\end{subequations}
The logical operator for both $\Sigma_{X_1}$ and $\Sigma_{T}$ is wrapping a pair of $m$ anyons around the cylinder. In the $X_1$ case, this flips the index. In the $T$ case, performing the operation to the left (resp right) of the domain wall flips the left (right) index. We have an isomorphism of ground spaces
\begin{align}
	Z(\Sigma_{X_1}) \oplus Z(\Sigma_T) \cong Z(\Sigma_{K_4,K_4}).
\end{align}
The isomorphism is given locally by the following six rules:
\begin{subequations}
\begin{align}
	\begin{array}{c}
		\includeTikz{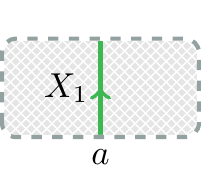}{
			\begin{tikzpicture}
			\fill[fill=bulkcolor,postaction={pattern=crosshatch,pattern color=white}] (-1,-0.5) rectangle (0,0.5);
			\fill[fill=bulkcolor,postaction={pattern=crosshatch,pattern color=white}] (0,-0.5) rectangle (1,0.5);
			\begin{scope}[very thick,decoration={markings,mark=at position 0.5 with {\arrow{>}}}]
			\draw[postaction={decorate},nicegreen] (0,-0.5) -- (0,0.5);
			\end{scope}
			\node[below] at (0,-0.5) {$a$};
			\node[above,white] at (0,0.5) {$a$};
			\node[left] at (0,0) {$X_1$};
			\draw[very thick,white] (-1,-0.5) rectangle (1,0.5);
			\draw[very thick,white,rounded corners] (-1,-0.5) rectangle (1,0.5);
			\draw[very thick,boundarycolor,rounded corners,dashed] (-1,-0.5) rectangle (1,0.5);
			\end{tikzpicture}
		}
	\end{array} &\mapsto
	\begin{array}{c}
		\includeTikz{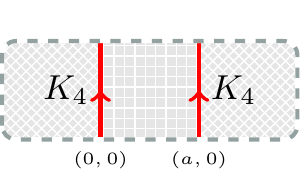}{
			\begin{tikzpicture}
			\fill[fill=bulkcolor,postaction={pattern=crosshatch,pattern color=white}] (-1.5,-0.5) rectangle (-0.5,0.5);
			\fill[fill=bulkcolor,postaction={pattern=grid,pattern color=white}] (-0.5,-0.5) rectangle (0.5,0.5);
			\fill[fill=bulkcolor,postaction={pattern=crosshatch,pattern color=white}] (0.5,-0.5) rectangle (1.5,0.5);
			\begin{scope}[very thick,decoration={markings,mark=at position 0.5 with {\arrow{>}}}]
			\draw[postaction={decorate},red] (-0.5,-0.5) -- (-0.5,0.5);
			\draw[postaction={decorate},red] (0.5,-0.5) -- (0.5,0.5);
			\end{scope}
			\node[below] at (-0.5,-0.5) {\tiny $(0,0)$};
			\node[below] at (0.5,-0.5) {\tiny $(a,0)$};
			\node[above,white] at (-0.5,0.5) {\tiny $(0,0)$};
			\node[above,white] at (0.5,0.5) {\tiny $(a,0)$};
			\node[left] at (-0.5,0) {$K_4$};
			\node[right] at (0.5,0) {$K_4$};
			\draw[very thick,white] (-1.5,-0.5) rectangle (1.5,0.5);
			\draw[very thick,white,rounded corners] (-1.5,-0.5) rectangle (1.5,0.5);
			\draw[very thick,boundarycolor,rounded corners,dashed] (-1.5,-0.5) rectangle (1.5,0.5);
			\end{tikzpicture}
		}
	\end{array} \label{eqn:K4K4X1} \\
	\begin{array}{c}
		\includeTikz{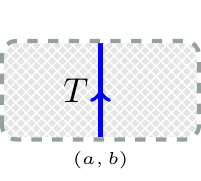}{
			\begin{tikzpicture}
			\fill[fill=bulkcolor,postaction={pattern=crosshatch,pattern color=white}] (-1,-0.5) rectangle (0,0.5);
			\fill[fill=bulkcolor,postaction={pattern=crosshatch,pattern color=white}] (0,-0.5) rectangle (1,0.5);
			\begin{scope}[very thick,decoration={markings,mark=at position 0.5 with {\arrow{>}}}]
			\draw[postaction={decorate},blue] (0,-0.5) -- (0,0.5);
			\end{scope}
			\node[below] at (0,-0.5) {\tiny $(a,b)$};
			\node[above,white] at (0,0.5) {\tiny $(0,0)$};
			\node[left] at (0,0) {$T$};
			\draw[very thick,white] (-1,-0.5) rectangle (1,0.5);
			\draw[very thick,white,rounded corners] (-1,-0.5) rectangle (1,0.5);
			\draw[very thick,boundarycolor,rounded corners,dashed] (-1,-0.5) rectangle (1,0.5);
			\end{tikzpicture}
		}
	\end{array} &\mapsto
	\begin{array}{c}
		\includeTikz{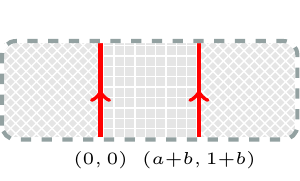}{
			\begin{tikzpicture}
			\fill[fill=bulkcolor,postaction={pattern=crosshatch,pattern color=white}] (-1.5,-0.5) rectangle (-0.5,0.5);
			\fill[fill=bulkcolor,postaction={pattern=grid,pattern color=white}] (-0.5,-0.5) rectangle (0.5,0.5);
			\fill[fill=bulkcolor,postaction={pattern=crosshatch,pattern color=white}] (0.5,-0.5) rectangle (1.5,0.5);
			\begin{scope}[very thick,decoration={markings,mark=at position 0.5 with {\arrow{>}}}]
			\draw[postaction={decorate},red] (-0.5,-0.5) -- (-0.5,0.5);
			\draw[postaction={decorate},red] (0.5,-0.5) -- (0.5,0.5);
			\end{scope}
			\node[below] at (-0.5,-0.5) {\tiny $(0,0)$};
			\node[below] at (0.5,-0.5) {\tiny $(a{+}b,1{+}b)$};
			\node[above,white] at (-0.5,0.5) {\tiny $(0,0)$};
			\node[above,white] at (0.5,0.5) {\tiny $(a,0)$};
			\draw[very thick,white] (-1.5,-0.5) rectangle (1.5,0.5);
			\draw[very thick,white,rounded corners] (-1.5,-0.5) rectangle (1.5,0.5);
			\draw[very thick,boundarycolor,rounded corners,dashed] (-1.5,-0.5) rectangle (1.5,0.5);
			\end{tikzpicture}
		}
	\end{array} \label{eqn:K4K4T}\\
	\begin{array}{c}
		\includeTikz{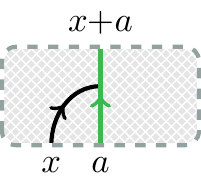}{
			\begin{tikzpicture}
			\fill[fill=bulkcolor,postaction={pattern=crosshatch,pattern color=white}] (-1,-0.5) rectangle (0,0.5);
			\fill[fill=bulkcolor,postaction={pattern=crosshatch,pattern color=white}] (0,-0.5) rectangle (1,0.5);
			\begin{scope}[very thick,decoration={markings,mark=at position 0.5 with {\arrow{>}}}]
			\draw[postaction={decorate}] (-0.5,-0.5) to[out=90,in=180] (0,0.1);
			\draw[postaction={decorate},nicegreen] (0,-0.5) -- (0,0.5);
			\end{scope}
			\node[below] at (0,-0.5) {$a$};
			\node[below] at (-0.5,-0.5) {$x$};
			\node[above] at (0,0.5) {$x{+}a$};
			\draw[very thick,white] (-1,-0.5) rectangle (1,0.5);
			\draw[very thick,white,rounded corners] (-1,-0.5) rectangle (1,0.5);
			\draw[very thick,boundarycolor,rounded corners,dashed] (-1,-0.5) rectangle (1,0.5);
			\end{tikzpicture}
		}
	\end{array} &\mapsto
	\begin{array}{c}
		\includeTikz{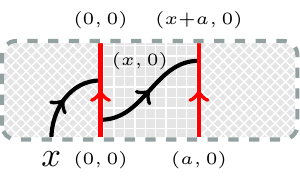}{
			\begin{tikzpicture}
			\fill[fill=bulkcolor,postaction={pattern=crosshatch,pattern color=white}] (-1.5,-0.5) rectangle (-0.5,0.5);
			\fill[fill=bulkcolor,postaction={pattern=grid,pattern color=white}] (-0.5,-0.5) rectangle (0.5,0.5);
			\fill[fill=bulkcolor,postaction={pattern=crosshatch,pattern color=white}] (0.5,-0.5) rectangle (1.5,0.5);
			\begin{scope}[very thick,decoration={markings,mark=at position 0.5 with {\arrow{>}}}]
			\draw[postaction={decorate}] (-1,-0.5) to[out=90,in=180] (-0.5,0.1);
			\draw[postaction={decorate}] (-0.5,-0.3) to[out=0,in=180] (0.5,0.3);
			\draw[postaction={decorate},red] (-0.5,-0.5) -- (-0.5,0.5);
			\draw[postaction={decorate},red] (0.5,-0.5) -- (0.5,0.5);
			\end{scope}
			\node[below] at (-0.5,-0.5) {\tiny $(0,0)$};
			\node at (-0.1,0.3) {\tiny $(x,0)$};
			\node[below] at (0.5,-0.5) {\tiny $(a,0)$};
			\node[below] at (-1,-0.5) {$x$};
			\node[above] at (-0.5,0.5) {\tiny $(0,0)$};
			\node[above] at (0.5,0.5) {\tiny $(x{+}a,0)$};
			\draw[very thick,white] (-1.5,-0.5) rectangle (1.5,0.5);
			\draw[very thick,white,rounded corners] (-1.5,-0.5) rectangle (1.5,0.5);
			\draw[very thick,boundarycolor,rounded corners,dashed] (-1.5,-0.5) rectangle (1.5,0.5);
			\end{tikzpicture}
		}
	\end{array} \\
	\begin{array}{c}
		\includeTikz{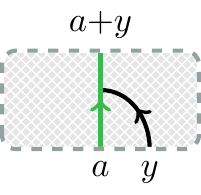}{
			\begin{tikzpicture}
			\fill[fill=bulkcolor,postaction={pattern=crosshatch,pattern color=white}] (-1,-0.5) rectangle (0,0.5);
			\fill[fill=bulkcolor,postaction={pattern=crosshatch,pattern color=white}] (0,-0.5) rectangle (1,0.5);
			\begin{scope}[very thick,decoration={markings,mark=at position 0.5 with {\arrow{>}}}]
			\draw[postaction={decorate}] (0.5,-0.5) to[out=90,in=0] (0,0.1);
			\draw[postaction={decorate},nicegreen] (0,-0.5) -- (0,0.5);
			\end{scope}
			\node[below] at (0,-0.5) {$a$};
			\node[below] at (0.5,-0.5) {$y$};
			\node[above] at (0,0.5) {$a{+}y$};
			\draw[very thick,white] (-1,-0.5) rectangle (1,0.5);
			\draw[very thick,white,rounded corners] (-1,-0.5) rectangle (1,0.5);
			\draw[very thick,boundarycolor,rounded corners,dashed] (-1,-0.5) rectangle (1,0.5);
			\end{tikzpicture}
		}
	\end{array} &\mapsto
	\begin{array}{c}
		\includeTikz{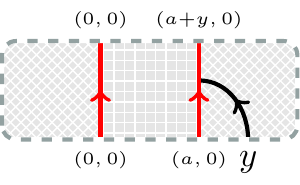}{
			\begin{tikzpicture}
			\fill[fill=bulkcolor,postaction={pattern=crosshatch,pattern color=white}] (-1.5,-0.5) rectangle (-0.5,0.5);
			\fill[fill=bulkcolor,postaction={pattern=grid,pattern color=white}] (-0.5,-0.5) rectangle (0.5,0.5);
			\fill[fill=bulkcolor,postaction={pattern=crosshatch,pattern color=white}] (0.5,-0.5) rectangle (1.5,0.5);
			\begin{scope}[very thick,decoration={markings,mark=at position 0.5 with {\arrow{>}}}]
			\draw[postaction={decorate}] (1,-0.5) to[out=90,in=0] (0.5,0.1);
			\draw[postaction={decorate},red] (-0.5,-0.5) -- (-0.5,0.5);
			\draw[postaction={decorate},red] (0.5,-0.5) -- (0.5,0.5);
			\end{scope}
			\node[below] at (1,-0.5) {$y$};
			\node[below] at (-0.5,-0.5) {\tiny $(0,0)$};
			\node[below] at (0.5,-0.5) {\tiny $(a,0)$};
			\node[above] at (-0.5,0.5) {\tiny $(0,0)$};
			\node[above] at (0.5,0.5) {\tiny $(a{+}y,0)$};
			\draw[very thick,white] (-1.5,-0.5) rectangle (1.5,0.5);
			\draw[very thick,white,rounded corners] (-1.5,-0.5) rectangle (1.5,0.5);
			\draw[very thick,boundarycolor,rounded corners,dashed] (-1.5,-0.5) rectangle (1.5,0.5);
			\end{tikzpicture}
		}
	\end{array} \\
	\begin{array}{c}
		\includeTikz{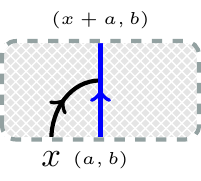}{
			\begin{tikzpicture}
			\fill[fill=bulkcolor,postaction={pattern=crosshatch,pattern color=white}] (-1,-0.5) rectangle (0,0.5);
			\fill[fill=bulkcolor,postaction={pattern=crosshatch,pattern color=white}] (0,-0.5) rectangle (1,0.5);
			\begin{scope}[very thick,decoration={markings,mark=at position 0.5 with {\arrow{>}}}]
			\draw[postaction={decorate}] (-0.5,-0.5) to[out=90,in=180] (0,0.1);
			\draw[postaction={decorate},blue] (0,-0.5) -- (0,0.5);
			\end{scope}
			\node[below] at (0,-0.5) {\tiny $(a,b)$};
			\node[above] at (0,0.5) {\tiny $(x+a,b)$};
			\node[below] at (-0.5,-0.5) {$x$};
			\draw[very thick,white] (-1,-0.5) rectangle (1,0.5);
			\draw[very thick,white,rounded corners] (-1,-0.5) rectangle (1,0.5);
			\draw[very thick,boundarycolor,rounded corners,dashed] (-1,-0.5) rectangle (1,0.5);
			\end{tikzpicture}
		}
	\end{array} &\mapsto
	\begin{array}{c}
		\includeTikz{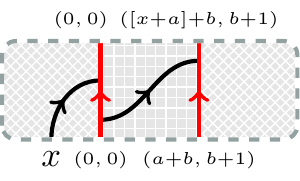}{
			\begin{tikzpicture}
			\fill[fill=bulkcolor,postaction={pattern=crosshatch,pattern color=white}] (-1.5,-0.5) rectangle (-0.5,0.5);
			\fill[fill=bulkcolor,postaction={pattern=grid,pattern color=white}] (-0.5,-0.5) rectangle (0.5,0.5);
			\fill[fill=bulkcolor,postaction={pattern=crosshatch,pattern color=white}] (0.5,-0.5) rectangle (1.5,0.5);
			\begin{scope}[very thick,decoration={markings,mark=at position 0.5 with {\arrow{>}}}]
			\draw[postaction={decorate}] (-1,-0.5) to[out=90,in=180] (-0.5,0.1);
			\draw[postaction={decorate}] (-0.5,-0.3) to[out=0,in=180] (0.5,0.3);
			\draw[postaction={decorate},red] (-0.5,-0.5) -- (-0.5,0.5);
			\draw[postaction={decorate},red] (0.5,-0.5) -- (0.5,0.5);
			\end{scope}
			\node[below] at (-0.5,-0.5) {\tiny $(0,0)$};
			\node[below] at (0.5,-0.5) {\tiny $(a{+}b,b{+}1)$};
			\node[below] at (-1,-0.5) {$x$};
			\node[above] at (-0.7,0.5) {\tiny $(0,0)$};
			\node[above] at (0.5,0.5) {\tiny $([x{+}a]{+}b,b{+}1)$};
			\draw[very thick,white] (-1.5,-0.5) rectangle (1.5,0.5);
			\draw[very thick,white,rounded corners] (-1.5,-0.5) rectangle (1.5,0.5);
			\draw[very thick,boundarycolor,rounded corners,dashed] (-1.5,-0.5) rectangle (1.5,0.5);
			\end{tikzpicture}
		}
	\end{array} \\
	\begin{array}{c}
		\includeTikz{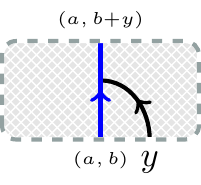}{
			\begin{tikzpicture}
			\fill[fill=bulkcolor,postaction={pattern=crosshatch,pattern color=white}] (-1,-0.5) rectangle (0,0.5);
			\fill[fill=bulkcolor,postaction={pattern=crosshatch,pattern color=white}] (0,-0.5) rectangle (1,0.5);
			\begin{scope}[very thick,decoration={markings,mark=at position 0.5 with {\arrow{>}}}]
			\draw[postaction={decorate}] (0.5,-0.5) to[out=90,in=0] (0,0.1);
			\draw[postaction={decorate},blue] (0,-0.5) -- (0,0.5);
			\end{scope}
			\node[below] at (0,-0.5) {\tiny $(a,b)$};
			\node[below] at (0.5,-0.5) {$y$};
			\node[above] at (0,0.5) {\tiny $(a,b{+}y)$};
			\draw[very thick,white] (-1,-0.5) rectangle (1,0.5);
			\draw[very thick,white,rounded corners] (-1,-0.5) rectangle (1,0.5);
			\draw[very thick,boundarycolor,rounded corners,dashed] (-1,-0.5) rectangle (1,0.5);
			\end{tikzpicture}
		}
	\end{array} &\mapsto
	\begin{array}{c}
		\includeTikz{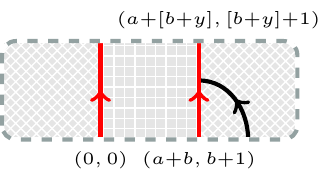}{
			\begin{tikzpicture}
			\fill[fill=bulkcolor,postaction={pattern=crosshatch,pattern color=white}] (-1.5,-0.5) rectangle (-0.5,0.5);
			\fill[fill=bulkcolor,postaction={pattern=grid,pattern color=white}] (-0.5,-0.5) rectangle (0.5,0.5);
			\fill[fill=bulkcolor,postaction={pattern=crosshatch,pattern color=white}] (0.5,-0.5) rectangle (1.5,0.5);
			\begin{scope}[very thick,decoration={markings,mark=at position 0.5 with {\arrow{>}}}]
			\draw[postaction={decorate}] (1,-0.5) to[out=90,in=0] (0.5,0.1);
			\draw[postaction={decorate},red] (-0.5,-0.5) -- (-0.5,0.5);
			\draw[postaction={decorate},red] (0.5,-0.5) -- (0.5,0.5);
			\end{scope}
			\node[below] at (-0.5,-0.5) {\tiny $(0,0)$};
			\node[below] at (0.5,-0.5) {\tiny $(a{+}b,b{+}1)$};
			\node[above] at (0.7,0.5) {\tiny $(a{+}[b{+}y],[b{+}y]{+}1)$};
			\draw[very thick,white] (-1.5,-0.5) rectangle (1.5,0.5);
			\draw[very thick,white,rounded corners] (-1.5,-0.5) rectangle (1.5,0.5);
			\draw[very thick,boundarycolor,rounded corners,dashed] (-1.5,-0.5) rectangle (1.5,0.5);
			\end{tikzpicture}
		}
	\end{array}
\end{align}
\end{subequations}
\subsubsection{Factoring domain walls}
As observed in \onlinecite{1806.01279}, many of the $\vvec{S_3}|\vvec{S_3}$ domain walls are noninvertible. From the BPR table~\ref{tab:S3BPR}, we see that there are only 2 invertible walls, with a $\ZZ{2}$ fusion rule. All other domain walls `factor through' a smaller intermediate phase. The full set of (generalized) BPR tables are presented in Appendix~\ref{ap:domain_wall_fusions}, we obtain physical interpretations for all bimodules. For example the $\vvec{S_3}|\vvec{S_3}$ wall named $T$ is actually a pair of $\mathcal{A}_1$ boundaries separated by a strip of vacuum, just like the $T^{\ZZ{2}}$ in Table~\ref{tab:Zpbimodinterp}. Similar physical interpretations can be obtained for all walls.


\subsection{Codimension 2 defects}\label{sec:defects}

\begin{table}[t]
	\resizebox{\textwidth}{!}{
	\begin{tabular}{!{\vrule width 1pt}>{\columncolor[gray]{.9}[\tabcolsep]}c!{\vrule width 2pt}c!{\color[gray]{.9}\vrule}c!{\color[gray]{.9}\vrule}c!{\color[gray]{.9}\vrule}c!{\color[gray]{.9}\vrule}c!{\color[gray]{.9}\vrule}c!{\vrule width 1pt}}
		\toprule[1pt]
		\rowcolor[gray]{.9}[\tabcolsep]$\otimes_{\vvec{S_3}}$&$ \sigma   $&$ B \sigma    $&$ C \sigma    $&$ \rho   $&$ B \rho    $&$ C \rho   $\\
		\toprule[2pt]
		$\sigma   $&$ A+G  $&$ B+G  $&$ C+F+H  $&$ E  $&$ D  $&$ D+E $\\
		$	B \sigma    $&$ B+G  $&$ A+G  $&$ C+F+H  $&$ D  $&$ E  $&$ D+E $\\
		$	C \sigma    $&$ C+F+H  $&$ C+F+H  $&$ A+B+C+F+2 G+H  $&$ D+E  $&$ D+E  $&$ 2 D+2 E $\\
		$	\rho   $&$ E  $&$ D  $&$ D+E  $&$ A+H  $&$ B+H  $&$ C+F+G $\\
		$	B \rho    $&$ D  $&$ E  $&$ D+E  $&$ B+H  $&$ A+H  $&$ C+F+G $\\
		$	C \rho    $&$ D+E  $&$ D+E  $&$ 2 D+2 E  $&$ C+F+G  $&$ C+F+G  $&$ A+B+C+F+G+2 H $\\
		\toprule[1pt]
	\end{tabular}
}
	\caption{Fusion rules for twists}\label{tab:S3twists}
\end{table}

We now discuss the excitations that live on domain walls. For the case of $\vvec{\ZZ{p}}$, more examples can be found in \onlinecites{1810.09469,1901.08069}.
	
\subsubsection{Finding point defects}
The first task is to find the possible point defects. We describe this process for the special case of binary interface defects (BIDs), as defined in Eqn.~\ref{eqn:BID}. In particular, consider the case where $\cat{M}=\cat{M}=I$ ($I$ defined in Table~\ref{tab:S3S3dws}). The BIDs are labeled by representations of the annular category (in this case, this is the \emph{tube algebra}\cite{ocneanu,Bultinck2015,SETPaper}). A basis for the morphisms $g\to hgh^{-1}$ is
\begin{align}
\mathcal{T}^g_{h}&:=
\begin{array}{c}
\includeTikz{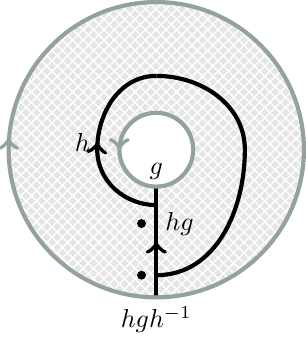}{
	\begin{tikzpicture}[scale=.75,every node/.style={scale=.75}]
	\filldraw[fill=bulkcolor,postaction={pattern=crosshatch,pattern color=white}] (0,0) circle (2);
	\begin{scope}[very thick,decoration={markings,mark=at position 0.5 with {\arrow{>}}}]
	\draw[postaction={decorate}] (0,-2) -- (0,-0.5);
	\draw[postaction={decorate}] (0,-.75) to[out=180,in=270] (-0.8,0) to[out=90,in=180] (0,1);
	\draw (0,-1.7) to[out=0,in=270] (1.2,0) to[out=90,in=0] (0,1);
	\draw[boundarycolor,postaction={decorate},fill=white] (0,0) circle (0.5);
	\end{scope}
	\begin{scope}[very thick,decoration={markings,mark=at position 0.5 with {\arrow{<}}}]
	\draw[boundarycolor,postaction={decorate}] (0,0) circle (2);
	\end{scope}
	\node at (-2,0) {};
	\node at (0,-2.3) {$hgh^{-1}$};
	\node at (-1,.1) {$h$};
	\node[right] at (0,-1) {$hg$};
	\node at (0,-0.3) {$g$};
	\draw[fill=black] (-0.2,-1.7) circle (0.05);
	\draw[fill=black] (-0.2,-1) circle (0.05);
	\end{tikzpicture}
}
\end{array},
\end{align}
which compose as
\begin{align}
\mathcal{T}^{hgh^{-1}}_{k}\mathcal{T}^{g}_{h}=\mathcal{T}^{g}_{kh}.
\end{align}

The first step in computing the representations of this category is to find the Karoubi envelope. This amounts to finding primitive idempotents in endomorphism algebras. By first identifying the $h$ for which $g=hgh^{-1}$, and then finding primitive idempotents for each matrix algebra, a simple computation shows that the isomorphism classes of simple objects in the Karoubi envelope are represented by
\begin{subequations}
\begin{align}
A=&\frac{1}{6}\left(\mathcal{T}^{(0,0)}_{(0,0)}+\mathcal{T}^{(0,0)}_{(0,1)}+\mathcal{T}^{(0,0)}_{(0,2)}+\mathcal{T}^{(0,0)}_{(1,0)}+\mathcal{T}^{(0,0)}_{(1,1)}+\mathcal{T}^{(0,0)}_{(1,2)}\right)\\
B=&\frac{1}{6}\left(\mathcal{T}^{(0,0)}_{(0,0)}+\mathcal{T}^{(0,0)}_{(0,1)}+\mathcal{T}^{(0,0)}_{(0,2)}-\mathcal{T}^{(0,0)}_{(1,0)}-\mathcal{T}^{(0,0)}_{(1,1)}-\mathcal{T}^{(0,0)}_{(1,2)}\right)\\
C=&\frac{1}{3}\left(\mathcal{T}^{(0,0)}_{(0,0)}-\mathcal{T}^{(0,0)}_{(0,1)}-\mathcal{T}^{(0,0)}_{(1,0)}+\mathcal{T}^{(0,0)}_{(1,1)}\right)\\
D=&\frac{1}{2}\left(\mathcal{T}^{(1,0)}_{(0,0)}+\mathcal{T}^{(1,0)}_{(1,0)}\right)\\
E=&\frac{1}{2}\left(\mathcal{T}^{(1,0)}_{(0,0)}-\mathcal{T}^{(1,0)}_{(1,0)}\right)\\
F=&\frac{1}{3}\left(\mathcal{T}^{(0,1)}_{(0,0)}+\mathcal{T}^{(0,1)}_{(0,1)}+\mathcal{T}^{(0,1)}_{(0,2)}\right)\\
G=&\frac{1}{3}\left(\mathcal{T}^{(0,1)}_{(0,0)}+\omega\mathcal{T}^{(0,1)}_{(0,1)}+\omega^2\mathcal{T}^{(0,1)}_{(0,2)}\right)\\
H=&\frac{1}{3}\left(\mathcal{T}^{(0,1)}_{(0,0)}+\omega^2\mathcal{T}^{(0,1)}_{(0,1)}+\omega\mathcal{T}^{(0,1)}_{(0,2)}\right),
\end{align}
\end{subequations}
where $\omega=\exp(\frac{2\pi i}{3})$.

Given these idempotents, the full representation can be constructed as described in Appendix~\ref{appendix:skein_vectors}. The bimodule $I$ is actually $\vvec{S_3}$ as a $\vvec{S_3}$ bimodule, and these representations describe the anyons of the theory.

\subsubsection{Vertical fusion}

Once we have obtained representations of point defects, we can compute more interesting properties of these excitations such as their fusion. In the case of vertical fusion, this is particularly straightforward. By examining how tubes act on the compound defect, we can obtain an explicit isomorphism to indecomposable point defects. For a triple compound defect, this isomorphism can be formed in two distinct ways as shown in Eqns.~\ref{eqn:verticalA}-\ref{eqn:verticalB}. The change of basis between these two families of isomorphisms gives the associator for ${\rm End}(\mathcal{M})$. An explicit example is given in Appendix~\ref{appendix:endM}.

\subsubsection{Fusion of twists}
One of the most studied domain wall excitations are termination points of invertible domain walls\cite{Bombin2010,Barkeshli2013,Barkeshli2014,Brown2016,Bridgeman2017,SETPaper,Brown2018}. Commonly referred to as twists, these excitations are of interest for quantum computation\cite{Bombin2010,Brown2016,1511.05153}. For the only nontrivial invertible domain wall for $\vvec{S_3}$, the horizontal fusion rules for twists is shown in Table~\ref{tab:S3twists}. From this table, we see that there are two kinds of `bare' twist that cannot be interchanged by the action of anyons (the action is shown in Table~\ref{tab:anyonontwist}).

In \onlinecite{Barkeshli2014}, it was conjectured that for a twist $t$, both $t\times t=A+G$ and $t\times t=A+H$ should occur. From Table~\ref{tab:S3twists}, we see that this is indeed the case.


\subsection{Logical operations on the toric code} 

The Toric code is one of the most famous examples in quantum information theory. Consider the Levin-Wen model for $\vvec{\mathbb{Z}/2\mathbb{Z}}$ on the torus $\Sigma$. This system has a 4 dimensional ground space:
\begin{align} \mathbb{C} \left\{
  \begin{array}{c}
    \includeTikz{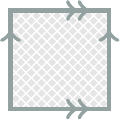}{
      \begin{tikzpicture}
\fill[fill=bulkcolor,postaction={pattern=crosshatch,pattern color=white}] (-0.5,-0.5) rectangle (0.5,0.5);
        \draw[white] (-0.6,-0.6) rectangle (0.6,0.6);
        \begin{scope}[very thick,decoration={markings,mark=at position 0.8 with {\arrow{>>}}}]
          \draw[boundarycolor,postaction={decorate}] (-0.5,-0.5) -- (0.5,-0.5);
          \draw[boundarycolor,postaction={decorate}] (-0.5,0.5) -- (0.5,0.5);
        \end{scope}
\begin{scope}[very thick,decoration={markings,mark=at position 0.8 with {\arrow{>}}}]
  \draw[boundarycolor,postaction={decorate}] (-0.5,-0.5) -- (-0.5,0.5);
          \draw[boundarycolor,postaction={decorate}] (0.5,-0.5) -- (0.5,0.5);
          \end{scope}
        \draw[very thick,boundarycolor] (-0.5,-0.5) rectangle (0.5,0.5);
      \end{tikzpicture}
      }
  \end{array},
\begin{array}{c}
    \includeTikz{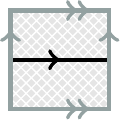}{
      \begin{tikzpicture}
\fill[fill=bulkcolor,postaction={pattern=crosshatch,pattern color=white}] (-0.5,-0.5) rectangle (0.5,0.5);
        \draw[white] (-0.6,-0.6) rectangle (0.6,0.6);
        \begin{scope}[very thick,decoration={markings,mark=at position 0.8 with {\arrow{>>}}}]
          \draw[boundarycolor,postaction={decorate}] (-0.5,-0.5) -- (0.5,-0.5);
          \draw[boundarycolor,postaction={decorate}] (-0.5,0.5) -- (0.5,0.5);
        \end{scope}
        \begin{scope}[very thick,decoration={markings,mark=at position 0.5 with {\arrow{>}}}]
          \draw[postaction={decorate}] (-0.5,0) -- (0.5,0);
          \end{scope}
\begin{scope}[very thick,decoration={markings,mark=at position 0.8 with {\arrow{>}}}]
          \draw[boundarycolor,postaction={decorate}] (-0.5,-0.5) -- (-0.5,0.5);
          \draw[boundarycolor,postaction={decorate}] (0.5,-0.5) -- (0.5,0.5);
          \end{scope}
        \draw[very thick,boundarycolor] (-0.5,-0.5) rectangle (0.5,0.5);
      \end{tikzpicture}
      }
  \end{array},
\begin{array}{c}
    \includeTikz{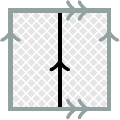}{
      \begin{tikzpicture}
\fill[fill=bulkcolor,postaction={pattern=crosshatch,pattern color=white}] (-0.5,-0.5) rectangle (0.5,0.5);
        \draw[white] (-0.6,-0.6) rectangle (0.6,0.6);
        \begin{scope}[very thick,decoration={markings,mark=at position 0.8 with {\arrow{>>}}}]
          \draw[boundarycolor,postaction={decorate}] (-0.5,-0.5) -- (0.5,-0.5);
          \draw[boundarycolor,postaction={decorate}] (-0.5,0.5) -- (0.5,0.5);
        \end{scope}
        \begin{scope}[very thick,decoration={markings,mark=at position 0.5 with {\arrow{>}}}]
          \draw[postaction={decorate}] (0,-0.5) -- (0,0.5);
          \end{scope}
\begin{scope}[very thick,decoration={markings,mark=at position 0.8 with {\arrow{>}}}]
          \draw[boundarycolor,postaction={decorate}] (-0.5,-0.5) -- (-0.5,0.5);
          \draw[boundarycolor,postaction={decorate}] (0.5,-0.5) -- (0.5,0.5);
          \end{scope}
        \draw[very thick,boundarycolor] (-0.5,-0.5) rectangle (0.5,0.5);
      \end{tikzpicture}
      }
\end{array},
\begin{array}{c}
    \includeTikz{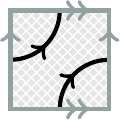}{
      \begin{tikzpicture}
\fill[fill=bulkcolor,postaction={pattern=crosshatch,pattern color=white}] (-0.5,-0.5) rectangle (0.5,0.5);
        \draw[white] (-0.6,-0.6) rectangle (0.6,0.6);
        \begin{scope}[very thick,decoration={markings,mark=at position 0.8 with {\arrow{>>}}}]
          \draw[boundarycolor,postaction={decorate}] (-0.5,-0.5) -- (0.5,-0.5);
          \draw[boundarycolor,postaction={decorate}] (-0.5,0.5) -- (0.5,0.5);
        \end{scope}
        \begin{scope}[very thick,decoration={markings,mark=at position 0.5 with {\arrow{>}}}]
          \draw[postaction={decorate}] (-0.5,0) to[out=0,in=270] (0,0.5);
          \draw[postaction={decorate}] (0,-0.5) to[out=90,in=180] (0.5,0);
          \end{scope}
\begin{scope}[very thick,decoration={markings,mark=at position 0.8 with {\arrow{>}}}]
          \draw[boundarycolor,postaction={decorate}] (-0.5,-0.5) -- (-0.5,0.5);
          \draw[boundarycolor,postaction={decorate}] (0.5,-0.5) -- (0.5,0.5);
          \end{scope}
        \draw[very thick,boundarycolor] (-0.5,-0.5) rectangle (0.5,0.5);
      \end{tikzpicture}
      }
  \end{array} \right\}
\end{align}
Two types of logical operations often discussed for this model are the Dehn twists and the anyon loop operators. Using the tools developed in this paper, we can compute how these operations interact with each other and deduce the group of logical operators which they generate. The mapping class group of the torus, ${\rm SL}(2,\mathbb{Z})$, is generated by two Dehn twists. The first cuts the torus along a horizontal loop (in our presentation), rotates a boundary circle by $2\pi$ and then glues the cylinder back into a torus. This Dehn twist acts as the matrix
\begin{align}\small
  \begin{bmatrix} 1 & 0 & 0 & 0 \\ 0 & 1 & 0 & 0 \\ 0 & 0 & 0 & 1 \\ 0 & 0 & 1 & 0
    \end{bmatrix}
\end{align}
We can think of the torus as $S^1 \times S^1$. The second Dehn twist swaps the factors and acts by the matrix
\begin{align}\small
  \begin{bmatrix} 1 & 0 & 0 & 0 \\ 0 & 0 & 1 & 0 \\ 0 & 1 & 0 & 0 \\ 0 & 0 & 0 & 1
    \end{bmatrix}
\end{align}
The reader is most likely familiar with these Dehn twists acting via the $T$ and $S$ matrices of $Z(\vvec{\mathbb{Z}/2\mathbb{Z}})$ respectively. If we conjugate by
\begin{align}\small
  \begin{bmatrix}
    1/2 & 1/2 & 0 & 0 \\ 1/2 & -1/2 & 0 & 0 \\ 0 & 0 & 1/2 & 1/2 \\ 0 & 0 & 1/2 & -1/2
  \end{bmatrix}
\end{align}
  which is the change of basis matrix from the skein basis to the basis which represents anyon flux through the center of the torus, we get
  \begin{align}
    \small
    \begin{bmatrix}
      1 & 0 & 0 & 0 \\ 0 & 1 & 0 & 0 \\ 0 & 0 & 1 & 0 \\ 0 & 0 & 0 & -1
    \end{bmatrix} \qquad
    \frac{1}{2}
    \begin{bmatrix}
      1 & 1 & 1 & 1 \\ 1 & 1 & -1 & -1 \\ 1 & -1 & 1 & -1 \\ 1 & -1 & -1 & 1
    \end{bmatrix}
  \end{align}
  respectively, as expected. To compute the anyon loop operators in the skein basis, we need to describe the anyon braidings, defined in Appendix~\ref{appendix:skein_vectors}, for $e$ and $m$. The $e$ particle doesn't leave a trail as it propagates through the material, but when it crosses a string, the state gets multiplied by $-1$. The $m$ particle leaves a trail as it propagates, but doesn't pick up a phase when it crosses a string. A horizontal $m$ loop acts via the matrix
  \begin{align}
    \small
    \begin{bmatrix}
      0 & 1 & 0 & 0 \\ 1 & 0 & 0 & 0 \\ 0 & 0 & 0 & 1 \\ 0 & 0 & 1 & 0
    \end{bmatrix}.
    \end{align}
Therefore, in the skein basis, every permutation matrix in $S_4$ is a logical operator. A horizontal $e$ loop acts via the matrix 
  \begin{align}
    \small
    \begin{bmatrix}
1 & 0 & 0 & 0 \\ 0 & 1 & 0 & 0 \\ 0 & 0 & -1 & 0 \\ 0 & 0 & 0 & -1
    \end{bmatrix}.
    \end{align}
  By conjugating with our permutation operators, we get a logical operator for each element of
  \begin{align}
   (\mathbb{Z}/2\mathbb{Z})^3  &\cong \left\{ (a,b,c,d) \in (\mathbb{Z}/2\mathbb{Z})^4 \lvert a + b + c + d = 0  \right\}
  \end{align}
  The full group of logical operators generated by Dehn twists and anyon loops is isomorphic to $S_4 \ltimes (\mathbb{Z}/2\mathbb{Z})^3$.


\subsection{Bimodule associators and obstructions} \label{sec:bimodule_associators}

We now discuss the bimodule associators. For $\vvec{\ZZ{p}}$, example computations can be found in \onlinecite{1901.08069}.

Suppose that we have a fusion category $\cat{C}$, a finite group $G$, a family of $\cat{C}{-}\cat{C}$ bimodules $\{ \cat{M}_g \}_{g \in G}$ and bimodule isomorphisms $\cat{M}_g \otimes_{\cat{C}} \cat{M}_h \cong \cat{M}_{gh}$. We present each of these bimodule isomorphisms as a pair of point defects:
  \begin{align}
    \begin{array}{c}
      \includeTikz{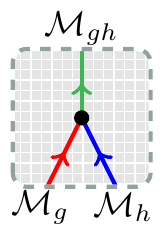}{
        \begin{tikzpicture}[scale=0.7]
          \path[fill=bulkcolor,postaction={pattern=grid,pattern color=white}] (-1,-1) -- (-0.5,-1) -- (0,0) -- (0,1) -- (-1,1) -- cycle;
          \path[fill=bulkcolor,postaction={pattern=grid,pattern color=white}] (1,-1) -- (0.5,-1) -- (0,0) -- (0,1) -- (1,1) -- cycle;
          \path[fill=bulkcolor,postaction={pattern=grid,pattern color=white}] (-0.5,-1) -- (0,0) -- (0.5,-1) -- cycle;
\begin{scope}[very thick,decoration={markings,mark=at position 0.5 with {\arrow{>}}}]
          \draw[postaction={decorate},red]  (-0.5,-1) -- (0,0);
          \draw[postaction={decorate},blue]  (0.5,-1)-- (0,0);
          \draw[postaction={decorate},nicegreen] (0,0) -- (0,1);
         \end{scope}
          \filldraw (0,0) circle (0.1);
          \draw[white,very thick] (-1,-1) rectangle (1,1);
          \draw[rounded corners,white,very thick] (-1,-1) rectangle (1,1);
          \draw[rounded corners,dashed,boundarycolor,very thick] (-1,-1) rectangle (1,1);
          \node at (0,1.3) {$\cat{M}_{gh}$};
          \node at (-0.6,-1.3) {$\cat{M}_g$};
          \node at (0.6,-1.3) {$\cat{M}_h$};
        \end{tikzpicture}
        }
    \end{array}
\begin{array}{c}
      \includeTikz{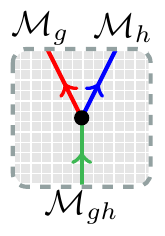}{
        \begin{tikzpicture}[yscale=-1,scale=0.7]
          \path[fill=bulkcolor,postaction={pattern=grid,pattern color=white}] (-1,-1) -- (-0.5,-1) -- (0,0) -- (0,1) -- (-1,1) -- cycle;
          \path[fill=bulkcolor,postaction={pattern=grid,pattern color=white}] (1,-1) -- (0.5,-1) -- (0,0) -- (0,1) -- (1,1) -- cycle;
          \path[fill=bulkcolor,postaction={pattern=grid,pattern color=white}] (-0.5,-1) -- (0,0) -- (0.5,-1) -- cycle;
\begin{scope}[very thick,decoration={markings,mark=at position 0.5 with {\arrow{>}}}]
          \draw[postaction={decorate},red] (0,0) -- (-0.5,-1);
          \draw[postaction={decorate},blue] (0,0) -- (0.5,-1);
          \draw[postaction={decorate},nicegreen] (0,1) -- (0,0);
\end{scope}
          \filldraw (0,0) circle (0.1);
          \draw[white,very thick] (-1,-1) rectangle (1,1);
          \draw[rounded corners,white,very thick] (-1,-1) rectangle (1,1);
          \draw[rounded corners,dashed,boundarycolor,very thick] (-1,-1) rectangle (1,1);
          \node at (0,1.3) {$\cat{M}_{gh}$};
          \node at (-0.6,-1.3) {$\cat{M}_g$};
          \node at (0.6,-1.3) {$\cat{M}_h$};
        \end{tikzpicture}
        }
      \end{array}
    \end{align}
  which satisfy the following renormalization equations:
\begin{subequations}
  \begin{align}
    \begin{array}{c}
      \includeTikz{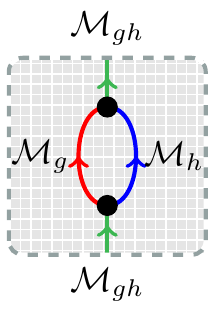}{
        \begin{tikzpicture}
          \filldraw[fill=bulkcolor,postaction={pattern=grid,pattern color=white}] (-1,-1) rectangle (1,1);
          \begin{scope}[very thick,decoration={markings,mark=at position 0.5 with {\arrow{>}}}]
            \draw[postaction={decorate},red] (0,-0.5) to[out=180,in=180] (0,0.5);
            \draw[postaction={decorate},blue] (0,-0.5) to[out=0,in=0] (0,0.5);
          \end{scope}
\begin{scope}[very thick,decoration={markings,mark=at position 0.6 with {\arrow{>}}}]
            \draw[postaction={decorate},nicegreen] (0,-1) -- (0,-0.5);
            \draw[postaction={decorate},nicegreen] (0,0.5) -- (0,1);
          \end{scope}
          \filldraw (0,-0.5) circle (0.1);
          \filldraw (0,0.5) circle (0.1);
          \draw[white,very thick] (-1,-1) rectangle (1,1);
          \draw[rounded corners,white,very thick] (-1,-1) rectangle (1,1);
          \draw[rounded corners,dashed,boundarycolor,very thick] (-1,-1) rectangle (1,1);
          \node at (0,1.3) {$\cat{M}_{gh}$};
          \node at (0,-1.3) {$\cat{M}_{gh}$};
          \node at (-0.68,0) {$\cat{M}_g$};
          \node at (0.68,0) {$\cat{M}_h$};
        \end{tikzpicture}
        }
    \end{array} &\cong
\begin{array}{c}
      \includeTikz{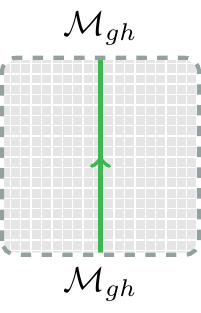}{
        \begin{tikzpicture}
          \filldraw[fill=bulkcolor,postaction={pattern=grid,pattern color=white}] (-1,-1) rectangle (1,1);
\begin{scope}[very thick,decoration={markings,mark=at position 0.5 with {\arrow{>}}}]
            \draw[postaction={decorate},nicegreen] (0,-1) -- (0,1);
          \end{scope}
          \draw[white,very thick] (-1,-1) rectangle (1,1);
          \draw[rounded corners,white,very thick] (-1,-1) rectangle (1,1);
          \draw[rounded corners,dashed,boundarycolor,very thick] (-1,-1) rectangle (1,1);
          \node at (0,1.3) {$\cat{M}_{gh}$};
          \node at (0,-1.3) {$\cat{M}_{gh}$};
        \end{tikzpicture}
        }
\end{array} \\
    \begin{array}{c}
      \includeTikz{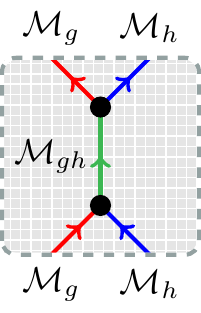}{
        \begin{tikzpicture}
          \filldraw[fill=bulkcolor,postaction={pattern=grid,pattern color=white}] (-1,-1) rectangle (1,1);
          \begin{scope}[very thick,decoration={markings,mark=at position 0.5 with {\arrow{>}}}]
            \draw[postaction={decorate},nicegreen] (0,-0.5) -- (0,0.5);
          \end{scope}
          \begin{scope}[very thick,decoration={markings,mark=at position 0.6 with {\arrow{>}}}]
            \draw[postaction={decorate},red] (-0.5,-1) -- (0,-0.5);
            \draw[postaction={decorate},blue] (0.5,-1) -- (0,-0.5);
\draw[postaction={decorate},red] (0,0.5) -- (-0.5,1);
            \draw[postaction={decorate},blue] (0,0.5) -- (0.5,1);
          \end{scope}
          \filldraw (0,-0.5) circle (0.1);
          \filldraw (0,0.5) circle (0.1);
          \draw[white,very thick] (-1,-1) rectangle (1,1);
          \draw[rounded corners,white,very thick] (-1,-1) rectangle (1,1);
          \draw[rounded corners,dashed,boundarycolor,very thick] (-1,-1) rectangle (1,1);
          \node at (-0.5,-1.3) {$\cat{M}_g$};
          \node at (0.5,-1.3) {$\cat{M}_h$};
          \node at (-0.5,1.3) {$\cat{M}_g$};
          \node at (0.5,1.3) {$\cat{M}_h$};
          \node at (-0.5,0) {$\cat{M}_{gh}$};
        \end{tikzpicture}
        }
    \end{array} &\cong
\begin{array}{c}
      \includeTikz{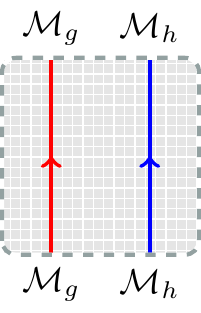}{
        \begin{tikzpicture}
          \filldraw[fill=bulkcolor,postaction={pattern=grid,pattern color=white}] (-1,-1) rectangle (1,1);
\begin{scope}[very thick,decoration={markings,mark=at position 0.5 with {\arrow{>}}}]
  \draw[postaction={decorate},red] (-0.5,-1) -- (-0.5,1);
  \draw[postaction={decorate},blue] (0.5,-1) -- (0.5,1);
          \end{scope}
          \draw[white,very thick] (-1,-1) rectangle (1,1);
          \draw[rounded corners,white,very thick] (-1,-1) rectangle (1,1);
          \draw[rounded corners,dashed,boundarycolor,very thick] (-1,-1) rectangle (1,1);
          \node at (-0.5,-1.3) {$\cat{M}_g$};
          \node at (0.5,-1.3) {$\cat{M}_h$};
          \node at (-0.5,1.3) {$\cat{M}_g$};
          \node at (0.5,1.3) {$\cat{M}_h$};
        \end{tikzpicture}
        }
      \end{array}
    \end{align}
\end{subequations}
  There are two obstructions for $\oplus_{g \in G} \cat{M}_g$ to form a tensor category. Firstly, it need not be the case that
  \begin{align} \label{eq:associator_domain_and_codomain}
    \begin{array}{c}
      \includeTikz{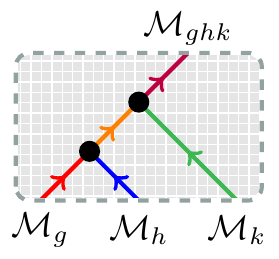}{
        \begin{tikzpicture}[scale=0.5]
          \filldraw[fill=bulkcolor,postaction={pattern=grid,pattern color=white}] (-0.5,0) rectangle (4.5,3);
          \begin{scope}[very thick,decoration={markings,mark=at position 0.5 with {\arrow{>}}}]
            \draw[red,postaction={decorate}] (0,0)--(1,1);
            \draw[orange,postaction={decorate}] (1,1) -- (2,2);
            \draw[purple,postaction={decorate}] (2,2) -- (3,3);
		    	\draw[blue,postaction={decorate}] (2,0)--(1,1);
		    	\draw[nicegreen,postaction={decorate}]  (4,0)--(2,2);
            \end{scope}
          \filldraw (1,1) circle (0.2);
          \filldraw (2,2) circle (0.2);
          \draw[white,very thick]  (-0.5,0) rectangle (4.5,3);
          \draw[rounded corners,white,very thick] (-0.5,0) rectangle (4.5,3);
          \draw[rounded corners,dashed,boundarycolor,very thick]  (-0.5,0) rectangle (4.5,3);
          \node at (0,-0.6) {$\cat{M}_g$};
          \node at (2,-0.6) {$\cat{M}_h$};
          \node at (4,-0.6) {$\cat{M}_k$};
          \node at (3,3.5) {$\cat{M}_{ghk}$};
        \end{tikzpicture}
        }
    \end{array} \; {\rm and} \;
\begin{array}{c}
      \includeTikz{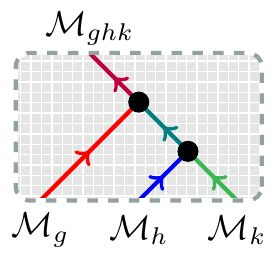}{
        \begin{tikzpicture}[scale=0.5]
          \filldraw[fill=bulkcolor,postaction={pattern=grid,pattern color=white}] (0.5,0) rectangle (-4.5,3);
          \begin{scope}[very thick,decoration={markings,mark=at position 0.5 with {\arrow{>}}}]
            \draw[nicegreen,postaction={decorate}] (0,0)--(-1,1);
            \draw[teal,postaction={decorate}]  (-1,1) -- (-2,2);
            \draw[purple,postaction={decorate}] (-2,2) -- (-3,3);
    	      \draw[blue,postaction={decorate}] (-2,0)--(-1,1);
    	      \draw[red,postaction={decorate}] (-4,0)--(-2,2);
            \end{scope}
          \filldraw (-1,1) circle (0.2);
          \filldraw (-2,2) circle (0.2);
          \draw[white,very thick]  (0.5,0) rectangle (-4.5,3);
          \draw[rounded corners,white,very thick] (0.5,0) rectangle (-4.5,3);
          \draw[rounded corners,dashed,boundarycolor,very thick]  (0.5,0) rectangle (-4.5,3);
          \node at (-4,-0.6) {$\cat{M}_g$};
          \node at (-2,-0.6) {$\cat{M}_h$};
          \node at (0,-0.6) {$\cat{M}_k$};
          \node at (-3,3.5) {$\cat{M}_{ghk}$};
        \end{tikzpicture}
        }
\end{array}
  \end{align}
  are isomorphic representations of the 4-bimodule annular category. The obstruction to such an isomorphism existing is the following defect
  \begin{align}
    \begin{array}{c}
      \includeTikz{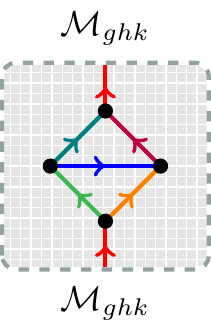}{
\begin{tikzpicture}[scale=0.7]
          \path[fill=bulkcolor,postaction={pattern=grid,pattern color=white}] (1.5,0.7) -- (0.7,1.5) -- (2.3,1.5) -- cycle;
\path[fill=bulkcolor,postaction={pattern=grid,pattern color=white}]
          (1.5,2.3) -- (0.7,1.5) -- (2.3,1.5) -- cycle;
          \path[fill=bulkcolor,postaction={pattern=grid,pattern color=white}] (1.5,0) -- (0,0) -- (0,3) -- (1.5,3) -- (1.5,2.3) -- (0.7,1.5) -- (1.5,0.7) -- cycle;
          \path[fill=bulkcolor,postaction={pattern=grid,pattern color=white}] (1.5,0) -- (3,0) -- (3,3) -- (1.5,3) -- (1.5,2.3) -- (2.3,1.5) -- (1.5,0.7) -- cycle;
          \begin{scope}[very thick,decoration={markings,mark=at position 0.5 with {\arrow{>}}}]
            \draw[postaction={decorate},red] (1.5,0) -- (1.5,0.7);
            \draw[postaction={decorate},nicegreen] (1.5,0.7) -- (0.7,1.5);
            \draw[postaction={decorate},orange] (1.5,0.7) --  (2.3,1.5);
            \draw[postaction={decorate},teal] (0.7,1.5) -- (1.5,2.3);
            \draw[postaction={decorate},purple] (2.3,1.5) -- (1.5,2.3);
            \draw[postaction={decorate},red] (1.5,2.3) -- (1.5,3);
            \draw[postaction={decorate},blue] (0.7,1.5) -- (2.3,1.5);
          \end{scope}
          \filldraw (1.5,0.7) circle (0.1);
          \filldraw (0.7,1.5) circle (0.1);
          \filldraw (2.3,1.5) circle (0.1);
          \filldraw (1.5,2.3) circle (0.1);
          \draw[very thick,white,rounded corners] (0,0) rectangle (3,3);
          \draw[very thick,white] (0,0) rectangle (3,3);
          \draw[very thick,boundarycolor,rounded corners,dashed] (0,0) rectangle (3,3);
          \node at (1.5,-0.5) {$\cat{M}_{ghk}$};
          \node at (1.5,3.5) {$\cat{M}_{ghk}$};
          \end{tikzpicture}
        }
    \end{array} \cong
\begin{array}{c}
      \includeTikz{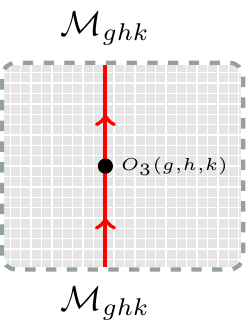}{
\begin{tikzpicture}[scale=0.7]
          \filldraw[fill=bulkcolor,postaction={pattern=grid,pattern color=white}] (0,0) rectangle (3.5,3);
          \begin{scope}[very thick,decoration={markings,mark=at position 0.5 with {\arrow{>}}}]
            \draw[postaction={decorate},red] (1.5,0) -- (1.5,1.5);
            \draw[postaction={decorate},red] (1.5,1.5) -- (1.5,3);
          \end{scope}
          \filldraw (1.5,1.5) circle (0.1);
          \draw[very thick,white,rounded corners] (0,0) rectangle (3.5,3);
          \draw[very thick,white] (0,0) rectangle (3.5,3);
          \draw[very thick,boundarycolor,rounded corners,dashed] (0,0) rectangle (3.5,3);
          \node at (1.5,-0.5) {$\cat{M}_{ghk}$};
          \node at (1.5,3.5) {$\cat{M}_{ghk}$};
          \node at (2.5,1.5) {\tiny$O_3(g{,}h{,}k)$};
          \end{tikzpicture}
        }
    \end{array}
  \end{align}
  as explained in \onlinecite{MR2677836}, $O_3(g,h,k)$ is a cohomology class in $H^3(G,Z)$ where $Z$ is the group of invertible anyons in $Z(\cat{C})$. If $O_3$ vanishes, then we can choose an isomorphism $\alpha_{g,h,k}$ between the two annular category representations in equation \eqref{eq:associator_domain_and_codomain}. Secondly, there is no reason that the associator $\alpha_{g,h,k}$ satisfies the pentagon equation. The obstruction to this is called $O_4$ and as explained in \onlinecite{MR2677836}, lives in $H^4(G,U(1))$. Since the loop space of $U(1)$ is homotopy equivalent to $\mathbb{Z}$, we have $H^4(G,U(1)) \cong H^5(G,\mathbb{Z})$.

  In this paper, we are primarily interested in the case where $\cat{C} = \vvec{S_3}$, $G = \mathbb{Z}/2\mathbb{Z}$, and $\cat{M}_1 = G_1$. The group of invertible anyons in the $\vvec{S_3}$ phase is $\mathbb{Z}/2\mathbb{Z}$, so we have
  \begin{align}
    O_3 &\in H^3(\mathbb{Z}/2\mathbb{Z},\mathbb{Z}/2\mathbb{Z}) \cong \mathbb{Z}/2\mathbb{Z} \\
    O_4 &\in H^4(\mathbb{Z}/2\mathbb{Z},U(1)) \cong H^5(\mathbb{Z}/2\mathbb{Z},\mathbb{Z}) = 0.
  \end{align}
Therefore, it is not obvious that the $O_3$ obstruction vanishes. We have explicitly computed the $O_3$ obstruction in this case and it is zero. The $O_4$ obstruction vanishes trivially.


\section{Remarks}\label{sec:remarks}

In this work, we have studied doubled topological phases and their defects. We have shown how various data, including fusion and associators can be computed by evaluating \emph{compound defects} in topological phases. We have utilized these techniques (in a fully automated manner) to compute this data for $\vvec{\ZZ{p}}$ and $\vvec{S_3}$, and interfaces between these.

Although we have specialized to $\vvec{G}$ in this work, the techniques described work more generally. Apart from the computational complexity, there is no obstruction to applying these algorithms to general fusion categories. The main challenge in that case is finding the bimodule categories, which we take as input data. It would be valuable to develop algorithms to find the data associated to these bimodules, beyond brute force solving of coherence equations.

From a quantum computational perspective, one of the most useful applications of this work is to study the `color code' (the phase described by the fusion category $\vvec{\ZZ{2}\times\ZZ{2}}$)\cite{ColorCode,Yoshida2015a}. This category has 270 bimodules, so a computational approach, such as the one described here, is necessary. Since the color code supports transversal Clifford gates, generalizations of the universal hybrid scheme described in \onlinecite{1811.06738} to this code may prove important for topological quantum computation.

Although the cohomology group is nontrivial, we have shown the $O_3$ obstruction for $\vvec{S_3}$ vanishes. It would be interesting to study a case where this obstruction is nontrivial. The explicit nature of the computations performed here may shed light on when this is expected to occur.

\acknowledgments
This work is supported by the Australian Research Council (ARC) via Discovery Project ``Subfactors and symmetries'' DP140100732 and Discovery Project ``Low dimensional categories'' DP16010347. This research was supported in part by Perimeter Institute for Theoretical Physics. Research at Perimeter Institute is supported in part by the Government of Canada through the Department of Innovation, Science and Economic Development Canada and by the Province of Ontario through the Ministry of Economic Development, Job Creation and Trade. Many of the ideas presented here were first suggested to us by Corey Jones. This paper wouldn't be possible without him. We thank Christopher Chubb for his comments on the manuscript.

\bibliographystyle{apsrev_jacob}
\bibliography{refs}

\appendix
\clearpage

\clearpage

\section{Skein vectors, annular categories and local relations}
\label{appendix:skein_vectors}
In this appendix, we define a $(2+\epsilon)$-dimensional defect TQFT, $Z$, which encodes the renormalization invariant properties of all long range entangled, doubled 2D topological phases and their defects. This TQFT first appeared in \onlinecite{MR2978449}, but in a somewhat disguised form. A more detailed construction which works in all dimensions is going to appear in the forthcoming paper \emph{Algebraic completions of higher categories} by Morrison and Walker. 
It is a $(2+\epsilon)$ theory because we explain how to evaluate path integrals on cylinders, but not all 3-manifolds. As explained by Morrison and Walker in \onlinecite{MR2978449}, for a topological theory, path integrals on a cylinder are equivalent to local relations. We have two classes of local relations: those which are required to define $Z(\Sigma)$ where $\Sigma$ is a domain wall structure or compound defect, and renormalization. Renormalization can modify $\Sigma$, which is why we treat it separately.

\begin{definition}[Domain wall structure] \label{def:domain_wall_structure_appendix}
Let $\Sigma$ be an oriented surface (possibly with boundary) and $D \subseteq \Sigma$ an embedded 1-dimensional oriented manifold (possibly with boundary) transverse to the boundary of $\Sigma$ with $\partial D \subseteq \partial \Sigma$. Since $D$ is a manifold, it is diffeomorphic to a disjoint union of oriented circles and oriented intervals connecting boundary components of $\Sigma$. In particular, there is no neighborhood of $D$ which is homeomorphic to a trivalent vertex. Label the connected components of $\Sigma \backslash D$ with fusion categories. Let $D'$ be a connected component of $D$. Since $D'$ and $\Sigma$ are oriented, we can unambiguously define the left face of $D'$ and the right face of $D'$. If we let $e_2$ be a tangent vector which agrees with the orientation of $D'$, and $e_1,e_2$ agrees with the orientation of $\Sigma$, then $e_1$ points towards the right face. Let $\cat{C}_L$ and $\cat{C}_R$ be the fusion categories labeling the left and right faces of $D'$ respectively. We label $D'$ with a bimodule category $\cat{C}_L \curvearrowright \cat{M} \curvearrowleft \cat{C}_R$. The pair $D \subseteq \Sigma$ equipped with fusion category and bimodule labels is called a {\em domain wall structure}.
\end{definition}

\begin{definition}[Skein vector] Let $D \subseteq \Sigma$ be a domain wall structure. A {\em skein vector} consists of the following data:
\begin{itemize}
\item[] {\bf graph:} A graph $\Gamma \subseteq \Sigma$ transverse to the boundary of $\Sigma$. Edges of $\Gamma$ are allowed to terminate on the boundary of $\Sigma$. Edges of $\Gamma$ are allowed to terminate on $D$, but they aren't allowed to cross $D$. Let $E(\Gamma)$ be the set of edges in $\Gamma$ and $E(D)$ be the set of edges of $D$. 
\item[] {\bf edge orientations:} An orientation for each edge in $E(\Gamma)$. The edges in $E(D)$ already have a fixed orientation induced from $D$.
\item[] {\bf edge labels:} Each edge $e \in E(\Gamma)$ lives in a connected component of $\Sigma \backslash D$ which has an associated fusion category $\cat{C}$. We label $e$ with a simple object $l(e)$ in $\cat{C}$. Each edge $f \in E(D)$ has an associated bimodule $\cat{M}$. We label $f$ with a simple object $l(f)$ in $\cat{M}$.
\item[] {\bf dots:} Let $v$ be a vertex internal to $\Sigma \backslash D$. A distinguished base point for $E_v$, the set of edges adjacent to the vertex $v$, which we call a {\em dot}. The orientation of $\Sigma$ induces a cyclic ordering on $E_v$, so the base point gives us a linear ordering of $E_v$, say $e_1,\dots,e_d$, where $e_1$ is the distinguished edge. Define $a_i = l(e_i)$. We denote this base point using a dot in one of the sectors around $v$. The distinguished edge is the first edge encountered starting from the dot following the orientation of the surface:
  \begin{align} \label{vertex_picture}
    \begin{array}{c}
      \includeTikz{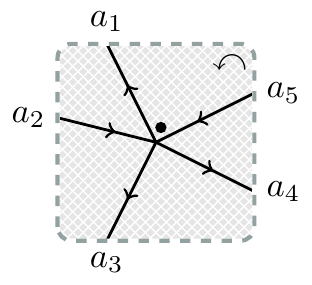}{
        \begin{tikzpicture}[scale=0.5]
        \filldraw[fill=bulkcolor,postaction={pattern=crosshatch,pattern color=white}] (-2,-2) rectangle (2,2);
          \begin{scope}[thick,decoration={markings,mark=at position 0.5 with {\arrow{<}}}]
          \draw[postaction={decorate}] (-1,2) to (0,0);
          \draw[postaction={decorate}] (0,0) to (-2,0.5);
          \draw[postaction={decorate}] (-1,-2) to (0,0);
          \draw[postaction={decorate}] (0,0) to (2,1);
          \draw[postaction={decorate}] (2,-1) to (0,0);
          \end{scope}
          \node[above] at (-1,2) {$a_1$};
          \node[left] at (-2,0.5) {$a_2$};
          \node[below] at (-1,-2) {$a_3$};
          \node[right] at (2,-1) {$a_4$};
          \node[right] at (2,1) {$a_5$};
          \node at (1.5,1.6) {$\curvearrowleft$};
          \draw[fill=black] (0.1,0.3) circle (0.1);
          \draw[white,very thick] (-2,-2) rectangle (2,2);
          \draw[rounded corners,white,very thick] (-2,-2) rectangle (2,2);
          \draw[rounded corners,boundarycolor,dashed,very thick] (-2,-2) rectangle (2,2);
        \end{tikzpicture}
        }
      \end{array}
  \end{align}
Vertices which live on $D$ are not given dots. Intuitively, this is because the rotational symmetry is already broken by the orientation of $D$. The dot method for breaking rotational symmetry is borrowed from \onlinecite{MR3624901}.
\item[] {\bf vertex vectors:} Let $v \in \Sigma \backslash D$ be a vertex of $\Gamma$ in a component with fusion category label $\cat{C}$. Let $e_1,\dots,e_d$ be the adjacent edges starting with the distinguished one. Define
  \begin{align}
  o_v(e) = \begin{cases} 1 & \text{$e$ is pointing towards $v$} \\ * & \text{$e$ is oriented away from $v$}. \end{cases}
  \end{align}
  We label $v$ with a morphism
  \begin{align}
  l(e_1)^{o_v(e_1)} l(e_2)^{o_v(e_2)} \cdots l(e_d)^{o_v(e_d)} \to 1
  \end{align}
  in $\cat{C}$. For example, the vertex in \eqref{vertex_picture} is labeled with a morphism $a_1^* a_2 a_3^* a_4^* a_5 \to 1$. There are four types of vertices which live on $D$. If $M$ is the bimodule which labels the relevant component of $D$, the vertices are labeled with morphisms in $\cat{M}$ as follows:
\begin{subequations}
\begin{align}
\begin{array}{c}
\includeTikz{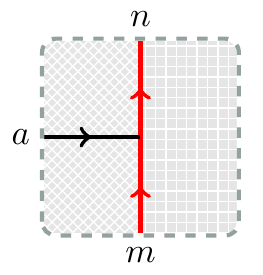}{
\begin{tikzpicture}
\filldraw[fill=bulkcolor,postaction={pattern=crosshatch,pattern color=white}] (0,0) rectangle (1,2);
\filldraw[fill=bulkcolor,postaction={pattern=grid,pattern color=white}] (1,0) rectangle (2,2);
\begin{scope}[very thick,decoration={markings,mark=at position 0.5 with {\arrow{>}}}]
\draw[red,postaction={decorate}] (1,0) -- (1,1);
\draw[red,postaction={decorate}] (1,1) -- (1,2);
\draw[postaction={decorate}] (0,1) -- (1,1);
\end{scope}
\node[left] at (0,1) {$a$};
\node[right] at (2,1) {};
\node[above] at (1,2) {$n$};
\node[below] at (1,0) {$m$};
\draw[very thick,white] (0,0) rectangle (2,2);
\draw[very thick,white,rounded corners] (0,0) rectangle (2,2);
\draw[very thick,boundarycolor,rounded corners,dashed] (0,0) rectangle (2,2);
\end{tikzpicture}
}
\end{array} :  am \to n \qquad
\begin{array}{c}
\includeTikz{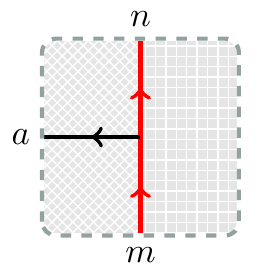}{
\begin{tikzpicture}
\filldraw[fill=bulkcolor,postaction={pattern=crosshatch,pattern color=white}] (0,0) rectangle (1,2);
\filldraw[fill=bulkcolor,postaction={pattern=grid,pattern color=white}] (1,0) rectangle (2,2);
\begin{scope}[very thick,decoration={markings,mark=at position 0.5 with {\arrow{>}}}]
\draw[red,postaction={decorate}] (1,0) -- (1,1);
\draw[red,postaction={decorate}] (1,1) -- (1,2);
\draw[postaction={decorate}] (1,1) -- (0,1);
\end{scope}
\node[right] at (2,1) {};
\node[left] at (0,1) {$a$};
\node[above] at (1,2) {$n$};
\node[below] at (1,0) {$m$};
\draw[very thick,white] (0,0) rectangle (2,2);
\draw[very thick,white,rounded corners] (0,0) rectangle (2,2);
\draw[very thick,boundarycolor,rounded corners,dashed] (0,0) rectangle (2,2);
\end{tikzpicture}
}
\end{array} : a^*m \to n \\
\begin{array}{c}
\includeTikz{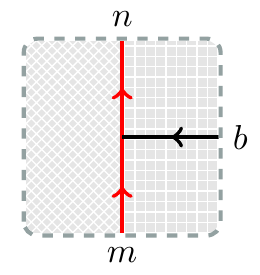}{
\begin{tikzpicture}
\filldraw[fill=bulkcolor,postaction={pattern=crosshatch,pattern color=white}] (0,0) rectangle (1,2);
\filldraw[fill=bulkcolor,postaction={pattern=grid,pattern color=white}] (1,0) rectangle (2,2);
\begin{scope}[very thick,decoration={markings,mark=at position 0.5 with {\arrow{>}}}]
\draw[red,postaction={decorate}] (1,0) -- (1,1);
\draw[red,postaction={decorate}] (1,1) -- (1,2);
\draw[postaction={decorate}] (2,1) -- (1,1);
\end{scope}
\node[left] at (0,1) {};
\node[right] at (2,1) {$b$};
\node[above] at (1,2) {$n$};
\node[below] at (1,0) {$m$};
\draw[very thick,white] (0,0) rectangle (2,2);
\draw[very thick,white,rounded corners] (0,0) rectangle (2,2);
\draw[very thick,boundarycolor,rounded corners,dashed] (0,0) rectangle (2,2);
\end{tikzpicture}
}
\end{array} : mb \to n
\qquad
\begin{array}{c}
\includeTikz{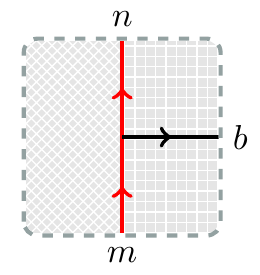}{
\begin{tikzpicture}
\filldraw[fill=bulkcolor,postaction={pattern=crosshatch,pattern color=white}] (0,0) rectangle (1,2);
\filldraw[fill=bulkcolor,postaction={pattern=grid,pattern color=white}] (1,0) rectangle (2,2);
\begin{scope}[very thick,decoration={markings,mark=at position 0.5 with {\arrow{>}}}]
\draw[red,postaction={decorate}] (1,0) -- (1,1);
\draw[red,postaction={decorate}] (1,1) -- (1,2);
\draw[postaction={decorate}] (1,1) -- (2,1);
\end{scope}
\node[left] at (0,1) {};
\node[right] at (2,1) {$b$};
\node[above] at (1,2) {$n$};
\node[below] at (1,0) {$m$};
\draw[very thick,white] (0,0) rectangle (2,2);
\draw[very thick,white,rounded corners] (0,0) rectangle (2,2);
\draw[very thick,boundarycolor,rounded corners,dashed] (0,0) rectangle (2,2);
\end{tikzpicture}
}
\end{array} : mb^* \to n
\end{align}
\end{subequations}
\end{itemize}
For computational reasons, we require each vertex in $\Gamma$ to have valence 3. This is general since higher valence vertices can always be decomposed into trivalent vertices.
\end{definition}

\begin{definition}[Skein local relations] \label{def:local_relations_1}
If $D \subseteq \Sigma$ is a domain wall structure, we define
\begin{align}
Z(\Sigma) = \mathbb{C} \left\{ \text{skein vectors on $D \subseteq \Sigma$}\right\} / \text{local relations}.
\end{align}
We have the following local relations:
\begin{itemize}
\item[] {\bf isotopy:} We can isotope $\Gamma$ inside $\Sigma$.
\item[] {\bf orientation:} We can reverse the orientation on an edge in $\Sigma \backslash D$ and take the dual of the corresponding label.
\item[] {\bf dot:} Moving the dot one sector following the orientation of $\Sigma$ corresponds to transforming the vertex vector using the rigid structure:
\item[]
\begin{subequations}
\begin{align}
&&
\begin{array}{c}
\includeTikz{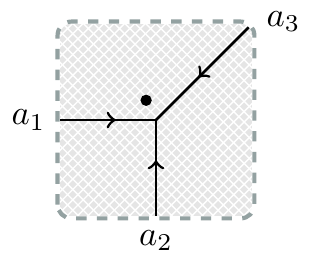}{
\begin{tikzpicture}
\filldraw[fill=bulkcolor,postaction={pattern=crosshatch,pattern color=white}] (-1,-1) rectangle (1,1);
\begin{scope}[thick,decoration={markings,mark=at position 0.5 with {\arrow{<}}}]
\draw[postaction={decorate}] (0,0) -- (-1,0);
\draw[postaction={decorate}] (0,0) -- (0,-1);
\draw[postaction={decorate}] (0,0) -- (1,1);
\end{scope}
\draw[fill=black] (-0.1,0.2) circle (0.05);
\node[left] at (-1,0) {$a_1$};
\node[below] at (0,-1) {$a_2$};
\node[above,right] at (1,1) {$a_3$};
\draw[white,very thick] (-1,-1) rectangle (1,1);
\draw[rounded corners,white,very thick] (-1,-1) rectangle (1,1);
\draw[rounded corners,dashed,boundarycolor,very thick] (-1,-1) rectangle (1,1);
\end{tikzpicture}
}
\end{array} \quad &\to \quad
\begin{array}{c}
\includeTikz{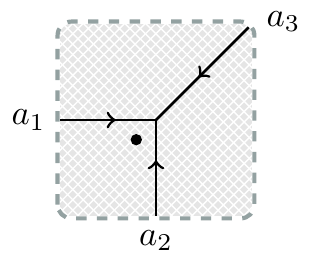}{
\begin{tikzpicture}
\filldraw[fill=bulkcolor,postaction={pattern=crosshatch,pattern color=white}] (-1,-1) rectangle (1,1);
\begin{scope}[thick,decoration={markings,mark=at position 0.5 with {\arrow{<}}}]
\draw[postaction={decorate}] (0,0) -- (-1,0);
\draw[postaction={decorate}] (0,0) -- (0,-1);
\draw[postaction={decorate}] (0,0) -- (1,1);
\end{scope}
\draw[fill=black] (-0.2,-0.2) circle (0.05);
\node[left] at (-1,0) {$a_1$};
\node[below] at (0,-1) {$a_2$};
\node[above,right] at (1,1) {$a_3$};
\draw[white,very thick] (-1,-1) rectangle (1,1);
\draw[rounded corners,white,very thick] (-1,-1) rectangle (1,1);
\draw[rounded corners,dashed,boundarycolor,very thick] (-1,-1) rectangle (1,1);
\end{tikzpicture}
}
\end{array} \\
&&
\begin{array}{c}
\includeTikz{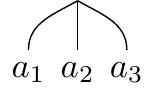}{
\begin{tikzpicture}[scale=0.5]
\draw (0,0) to[out=180+30,in=90] (-1,-1);
\draw (0,0) -- (0,-1);
\draw (0,0) to[out=-30,in=90] (1,-1);
\node[below] at (-1,-1) {$a_1$};
\node[below] at (0,-1) {$a_2$};
\node[below] at (1,-1) {$a_3$};
\end{tikzpicture}
}
\end{array} \quad &\to \quad
\begin{array}{c}
\includeTikz{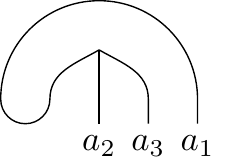}{
\begin{tikzpicture}[scale=0.5]
\draw (0,0) to[out=180+30,in=90] (-1,-1);
\draw (0,0) -- (0,-1.5);
\draw (0,0) to[out=-30,in=90] (1,-1) -- (1,-1.5);
\draw (-1,-1) to[out=-90,in=0] (-1.5,-1.5) to[out=180,in=-90] (-2,-1) to[out=90,in=180] (0,1) to[out=0,in=90] (2,-1) -- (2,-1.5);
\node[below] at (2,-1.5) {$a_1$};
\node[below] at (0,-1.5) {$a_2$};
\node[below] at (1,-1.5) {$a_3$};
\end{tikzpicture}
}
\end{array}\\
{\bf F:}&&
 \begin{array}{c}
\includeTikz{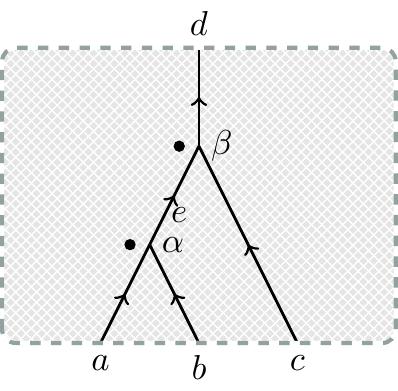}{
\begin{tikzpicture}
\filldraw[fill=bulkcolor,postaction={pattern=crosshatch,pattern color=white}] (-2,-2) rectangle (2,1);
\begin{scope}[thick,decoration={markings,mark=at position 0.5 with {\arrow{>}}}]
\draw[postaction={decorate}] (0,-2) -- (-0.5,-1);
\draw[postaction={decorate}] (-1,-2) -- (-0.5,-1);
\draw[postaction={decorate}] (-0.5,-1) -- (0,0);
\draw[postaction={decorate}] (1,-2) -- (0,0);
\draw[postaction={decorate}] (0,0) -- (0,1);
\end{scope}
\node[below] at (-1,-2) {$a$};
\node[below] at (0,-2) {$b$};
\node[below] at (1,-2) {$c$};
\node[above] at (0,1) {$d$};
\node[right] at (-0.5,-1) {$\alpha$};
\node[right] at (0,0) {$\beta$};
\node at (-0.2,-0.7) {$e$};
 \draw[fill=black] (-0.5-0.2,-1) circle (0.05);
\draw[fill=black] (0-0.2,0) circle (0.05);
\draw[white,very thick] (-2,-2) rectangle (2,1);
\draw[rounded corners,white,very thick] (-2,-2) rectangle (2,1);
\draw[rounded corners,dashed,boundarycolor,very thick] (-2,-2) rectangle (2,1);
\end{tikzpicture}
}
\end{array}
    &=\sum_{\mu\nu}
    F_{\alpha\beta}^{\mu\nu}
\begin{array}{c}
\includeTikz{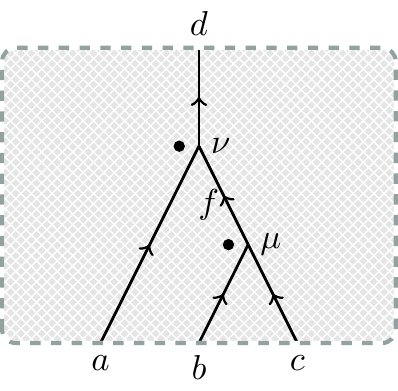}{
\begin{tikzpicture}
\filldraw[fill=bulkcolor,postaction={pattern=crosshatch,pattern color=white}] (-2,-2) rectangle (2,1);
\begin{scope}[thick,decoration={markings,mark=at position 0.5 with {\arrow{>}}}]
\draw[postaction={decorate}] (0,-2) -- (0.5,-1);
\draw[postaction={decorate}] (1,-2) -- (0.5,-1);
\draw[postaction={decorate}] (0.5,-1) -- (0,0);
\draw[postaction={decorate}] (-1,-2) -- (0,0);
\draw[postaction={decorate}] (0,0) -- (0,1);
\end{scope}
\node[below] at (-1,-2) {$a$};
\node[below] at (0,-2) {$b$};
\node[below] at (1,-2) {$c$};
\node[above] at (0,1) {$d$};
\node[right] at (0.5,-1) {$\mu$};
\node[right] at (0,0) {$\nu$};
\node at (0.1,-0.6) {$f$};
\draw[fill=black] (0.5-0.2,-1) circle (0.05);
\draw[fill=black] (0-0.2,0) circle (0.05);
\draw[white,very thick] (-2,-2) rectangle (2,1);
\draw[rounded corners,white,very thick] (-2,-2) rectangle (2,1);
\draw[rounded corners,dashed,boundarycolor,very thick] (-2,-2) rectangle (2,1);
\end{tikzpicture}
}
\end{array} \\
  {\bf G:}&&
  \begin{array}{c}
  \includeTikz{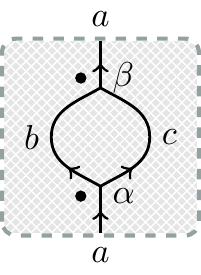}{
  \begin{tikzpicture}
\filldraw[fill=bulkcolor,postaction={pattern=crosshatch,pattern color=white}] (-1,-1) rectangle (1,1);
\begin{scope}[thick,decoration={markings,mark=at position 0.5 with {\arrow{>}}}]
\draw[postaction={decorate}] (0,-1) -- (0,-0.5);
\draw[postaction={decorate}] (0,-0.5) to[out=90+60,in=270] (-0.5,0);
\draw[postaction={decorate}] (0,-0.5) to[out=30,in=270] (0.5,0);
\draw (-0.5,0) to[out=90,in=180+30] (0,0.5);
\draw (0.5,0) to[out=90,in=-30] (0,0.5);
\draw[postaction={decorate}] (0,0.5) -- (0,1);
\end{scope}
\node[below] at (0,-1) {$a$};
\node[above] at (0,1) {$a$};
\node[left] at (-0.5,0) {$b$};
\node[right] at (0.5,0) {$c$};
\draw[fill=black] (-0.2,-0.6) circle (0.05);
\draw[fill=black] (-0.2,0.6) circle (0.05);
\node[right] at (0,-0.6) {$\alpha$};
\node[right] at (0,0.6) {$\beta$};
\draw[very thick,white] (-1,-1) rectangle (1,1);
\draw[rounded corners,very thick,white] (-1,-1) rectangle (1,1);
\draw[dashed,rounded corners,very thick,boundarycolor] (-1,-1) rectangle (1,1);
  \end{tikzpicture}
  }
  \end{array}
  &= G_{\alpha\beta}
  \begin{array}{c}
  \includeTikz{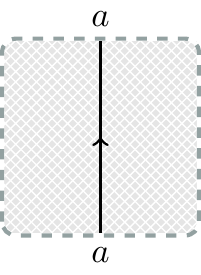}{
  \begin{tikzpicture}
\filldraw[fill=bulkcolor,postaction={pattern=crosshatch,pattern color=white}] (-1,-1) rectangle (1,1);
\begin{scope}[thick,decoration={markings,mark=at position 0.5 with {\arrow{>}}}]
\draw[postaction={decorate}] (0,-1) -- (0,1);
\end{scope}
\node[below] at (0,-1) {$a$};
\node[above] at (0,1) {$a$};
\draw[very thick,white] (-1,-1) rectangle (1,1);
\draw[rounded corners,very thick,white] (-1,-1) rectangle (1,1);
\draw[dashed,rounded corners,very thick,boundarycolor] (-1,-1) rectangle (1,1);
  \end{tikzpicture}
  }
  \end{array}\\
  {\bf L}:&&
\begin{array}{c}
\includeTikz{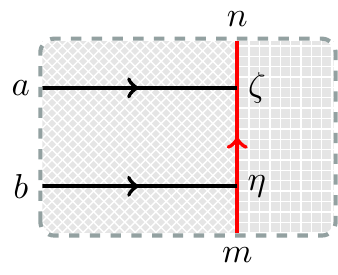}{
\begin{tikzpicture}
\filldraw[fill=bulkcolor,postaction={pattern=crosshatch,pattern color=white}] (-2,-1) rectangle (0,1);
\filldraw[fill=bulkcolor,postaction={pattern=grid,pattern color=white}] (0,-1) rectangle (1,1);
\begin{scope}[very thick,decoration={markings,mark=at position 0.5 with {\arrow{>}}}]
\draw[red,postaction={decorate}] (0,-1) -- (0,1);
\draw[postaction={decorate}] (-2,0.5) -- (0,0.5);
\draw[postaction={decorate}] (-2,-0.5) -- (0,-0.5);
\end{scope}
\draw[very thick,white] (-2,-1) rectangle (1,1);
\draw[very thick,rounded corners,white] (-2,-1) rectangle (1,1);
\draw[very thick,rounded corners,boundarycolor,dashed] (-2,-1) rectangle (1,1);
\node at (-2.2,0.5) {$a$};
\node at (-2.2,-0.5) {$b$};
\node at (0,1.2) {$n$};
\node at (0,-1.2) {$m$};
\node at (0.2,0.5) {$\zeta$};
\node at (0.2,-0.5) {$\eta$};
\end{tikzpicture}
}
\end{array} &= \sum_{\alpha\mu} L_{\eta\zeta}^{\alpha\mu}
\begin{array}{c}
\includeTikz{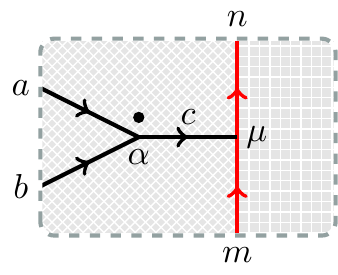}{
\begin{tikzpicture}
\filldraw[fill=bulkcolor,postaction={pattern=crosshatch,pattern color=white}] (-2,-1) rectangle (0,1);
\filldraw[fill=bulkcolor,postaction={pattern=grid,pattern color=white}] (0,-1) rectangle (1,1);
\begin{scope}[very thick,decoration={markings,mark=at position 0.5 with {\arrow{>}}}]
\draw[red,postaction={decorate}] (0,-1) -- (0,0);
\draw[red,postaction={decorate}] (0,0) -- (0,1);
\draw[postaction={decorate}] (-1,0) -- (0,0);
\draw[postaction={decorate}] (-2,-0.5) -- (-1,0);
\draw[postaction={decorate}] (-2,0.5) -- (-1,0);
\end{scope}
\node at (-1,-0.2) {$\alpha$};
\node at (0.2,0) {$\mu$};
\node at (-2.2,0.5) {$a$};
\node at (-2.2,-0.5) {$b$};
\node at (0,1.2) {$n$};
\node at (0,-1.2) {$m$};
\node at (-0.5,0.2) {$c$};
\draw[fill=black] (-1,0.2) circle (0.05);
\draw[very thick,white] (-2,-1) rectangle (1,1);
\draw[very thick,rounded corners,white] (-2,-1) rectangle (1,1);
\draw[very thick,rounded corners,boundarycolor,dashed] (-2,-1) rectangle (1,1);
\end{tikzpicture}
}
\end{array}\\
{\bf R:}&&
\begin{array}{c}
\includeTikz{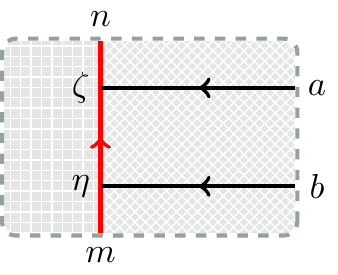}{
\begin{tikzpicture}[xscale=-1]
\filldraw[fill=bulkcolor,postaction={pattern=crosshatch,pattern color=white}] (-2,-1) rectangle (0,1);
\filldraw[fill=bulkcolor,postaction={pattern=grid,pattern color=white}] (0,-1) rectangle (1,1);
\begin{scope}[very thick,decoration={markings,mark=at position 0.5 with {\arrow{>}}}]
\draw[red,postaction={decorate}] (0,-1) -- (0,1);
\draw[postaction={decorate}] (-2,0.5) -- (0,0.5);
\draw[postaction={decorate}] (-2,-0.5) -- (0,-0.5);
\end{scope}
\draw[very thick,white] (-2,-1) rectangle (1,1);
\draw[very thick,rounded corners,white] (-2,-1) rectangle (1,1);
\draw[very thick,rounded corners,boundarycolor,dashed] (-2,-1) rectangle (1,1);
\node at (-2.2,0.5) {$a$};
\node at (-2.2,-0.5) {$b$};
\node at (0,1.2) {$n$};
\node at (0,-1.2) {$m$};
\node at (0.2,0.5) {$\zeta$};
\node at (0.2,-0.5) {$\eta$};
\end{tikzpicture}
}
\end{array} &= \sum_{\alpha\mu} R_{\eta\zeta}^{\alpha\mu}
\begin{array}{c}
\includeTikz{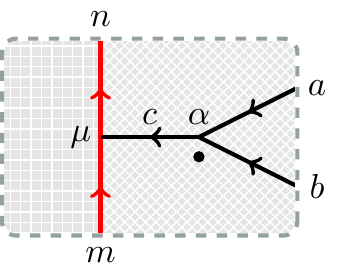}{
\begin{tikzpicture}[xscale=-1]
\filldraw[fill=bulkcolor,postaction={pattern=crosshatch,pattern color=white}] (-2,-1) rectangle (0,1);
\filldraw[fill=bulkcolor,postaction={pattern=grid,pattern color=white}] (0,-1) rectangle (1,1);
\begin{scope}[very thick,decoration={markings,mark=at position 0.5 with {\arrow{>}}}]
\draw[red,postaction={decorate}] (0,-1) -- (0,0);
\draw[red,postaction={decorate}] (0,0) -- (0,1);
\draw[postaction={decorate}] (-1,0) -- (0,0);
\draw[postaction={decorate}] (-2,-0.5) -- (-1,0);
\draw[postaction={decorate}] (-2,0.5) -- (-1,0);
\end{scope}
\node at (-1,0.2) {$\alpha$};
\node at (0.2,0) {$\mu$};
\node at (-2.2,0.5) {$a$};
\node at (-2.2,-0.5) {$b$};
\node at (0,1.2) {$n$};
\node at (0,-1.2) {$m$};
\node at (-0.5,0.2) {$c$};
\draw[fill=black] (-1,-0.2) circle (0.05);
\draw[very thick,white] (-2,-1) rectangle (1,1);
\draw[very thick,rounded corners,white] (-2,-1) rectangle (1,1);
\draw[very thick,rounded corners,boundarycolor,dashed] (-2,-1) rectangle (1,1);
\end{tikzpicture}
}
\end{array}\\
{\bf C:}&&
\begin{array}{c}
\includeTikz{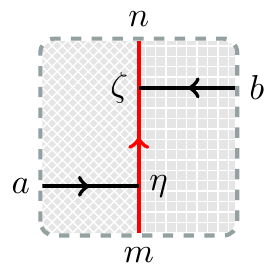}{
\begin{tikzpicture}
\filldraw[fill=bulkcolor,postaction={pattern=crosshatch,pattern color=white}] (-1,-1) rectangle (0,1);
\filldraw[fill=bulkcolor,postaction={pattern=grid,pattern color=white}] (0,-1) rectangle (1,1);
\begin{scope}[very thick,decoration={markings,mark=at position 0.5 with {\arrow{>}}}]
\draw[red,postaction={decorate}] (0,-1) -- (0,1);
\draw[postaction={decorate}] (-1,-0.5) -- (0,-0.5);
\draw[postaction={decorate}] (1,0.5) -- (0,0.5);
\end{scope}
\node at (-1-0.2,-0.5) {$a$};
\node at (1+0.2,0.5) {$b$};
\node at (0,1+0.2) {$n$};
\node at (0,-1-0.2) {$m$};
\node at (-0.2,0.5) {$\zeta$};
\node at (0.2,-0.5) {$\eta$};
\draw[very thick,white] (-1,-1) rectangle (1,1);
\draw[very thick,rounded corners,white] (-1,-1) rectangle (1,1);
\draw[very thick,rounded corners,dashed,boundarycolor] (-1,-1) rectangle (1,1);
\end{tikzpicture}
}
\end{array} &= \sum_{\mu\nu} C_{\eta\zeta}^{\mu\nu}
\begin{array}{c}
\includeTikz{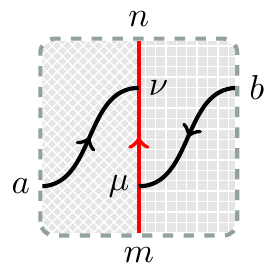}{
\begin{tikzpicture}
\filldraw[fill=bulkcolor,postaction={pattern=crosshatch,pattern color=white}] (-1,-1) rectangle (0,1);
\filldraw[fill=bulkcolor,postaction={pattern=grid,pattern color=white}] (0,-1) rectangle (1,1);
\begin{scope}[very thick,decoration={markings,mark=at position 0.5 with {\arrow{>}}}]
\draw[red,postaction={decorate}] (0,-1) -- (0,1);
\draw[postaction={decorate}] (-1,-0.5) to[out=0,in=180] (0,0.5);
\draw[postaction={decorate}] (1,0.5) to[out=180,in=0] (0,-0.5);
\end{scope}
\node at (-1-0.2,-0.5) {$a$};
\node at (1+0.2,0.5) {$b$};
\node at (0,1+0.2) {$n$};
\node at (0,-1-0.2) {$m$};
\node at (0.2,0.5) {$\nu$};
\node at (-0.2,-0.5) {$\mu$};
\draw[very thick,white] (-1,-1) rectangle (1,1);
\draw[very thick,rounded corners,white] (-1,-1) rectangle (1,1);
\draw[very thick,rounded corners,dashed,boundarycolor] (-1,-1) rectangle (1,1);
\end{tikzpicture}
}
\end{array}\\
{\bf LG:}&&
\begin{array}{c}
\includeTikz{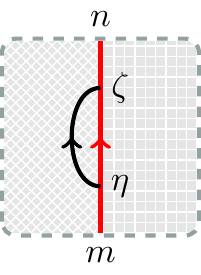}{
\begin{tikzpicture}
\filldraw[fill=bulkcolor,postaction={pattern=crosshatch,pattern color=white}] (-1,-1) rectangle (0,1);
\filldraw[fill=bulkcolor,postaction={pattern=grid,pattern color=white}] (0,-1) rectangle (1,1);
\begin{scope}[very thick,decoration={markings,mark=at position 0.5 with {\arrow{>}}}]
\draw[red,postaction={decorate}] (0,-1) -- (0,1);
\draw[postaction={decorate}] (0,-0.5) to[out=180,in=180] (0,0.5);
\end{scope}
\node at (0.2,-0.5) {$\eta$};
\node at (0.2,0.5) {$\zeta$};
\node at (0,1+0.2) {$n$};
\node at (0,-1-0.2) {$m$};
\draw[very thick,white] (-1,-1) rectangle (1,1);
\draw[very thick,rounded corners,white] (-1,-1) rectangle (1,1);
\draw[very thick,rounded corners,dashed,boundarycolor] (-1,-1) rectangle (1,1);
\end{tikzpicture}
}
\end{array} &= LG_{\eta\zeta} \begin{array}{c}
\includeTikz{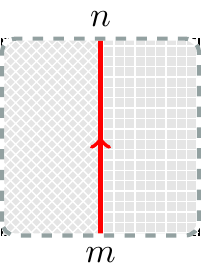}{
\begin{tikzpicture}
\filldraw[fill=bulkcolor,postaction={pattern=crosshatch,pattern color=white}] (-1,-1) rectangle (0,1);
\filldraw[fill=bulkcolor,postaction={pattern=grid,pattern color=white}] (0,-1) rectangle (1,1);
\begin{scope}[very thick,decoration={markings,mark=at position 0.5 with {\arrow{>}}}]
\draw[red,postaction={decorate}] (0,-1) -- (0,1);
\end{scope}
\node at (0,1+0.2) {$n$};
\node at (0,-1-0.2) {$m$};
\draw[very thick,rounded corners,white] (-1,-1) rectangle (1,1);
\draw[very thick,rounded corners,dashed,boundarycolor] (-1,-1) rectangle (1,1);
\end{tikzpicture}
}
\end{array}\\
{\bf RG:}&&
\begin{array}{c}
\includeTikz{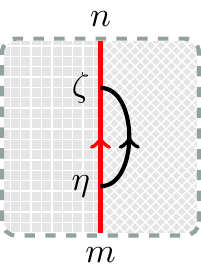}{
\begin{tikzpicture}[xscale=-1]
\filldraw[fill=bulkcolor,postaction={pattern=crosshatch,pattern color=white}] (-1,-1) rectangle (0,1);
\filldraw[fill=bulkcolor,postaction={pattern=grid,pattern color=white}] (0,-1) rectangle (1,1);
\begin{scope}[very thick,decoration={markings,mark=at position 0.5 with {\arrow{>}}}]
\draw[red,postaction={decorate}] (0,-1) -- (0,1);
\draw[postaction={decorate}] (0,-0.5) to[out=180,in=180] (0,0.5);
\end{scope}
\node at (0.2,-0.5) {$\eta$};
\node at (0.2,0.5) {$\zeta$};
\node at (0,1+0.2) {$n$};
\node at (0,-1-0.2) {$m$};
\draw[very thick,white] (-1,-1) rectangle (1,1);
\draw[very thick,rounded corners,white] (-1,-1) rectangle (1,1);
\draw[very thick,rounded corners,dashed,boundarycolor] (-1,-1) rectangle (1,1);
\end{tikzpicture}
}
\end{array} &= RG_{\eta\zeta} \begin{array}{c}
\includeTikz{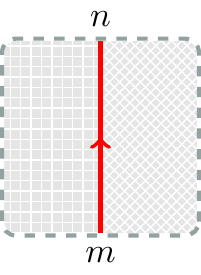}{
\begin{tikzpicture}[xscale=-1]
\filldraw[fill=bulkcolor,postaction={pattern=crosshatch,pattern color=white}] (-1,-1) rectangle (0,1);
\filldraw[fill=bulkcolor,postaction={pattern=grid,pattern color=white}] (0,-1) rectangle (1,1);
\begin{scope}[very thick,decoration={markings,mark=at position 0.5 with {\arrow{>}}}]
\draw[red,postaction={decorate}] (0,-1) -- (0,1);
\end{scope}
\node at (0,1+0.2) {$n$};
\node at (0,-1-0.2) {$m$};
\draw[very thick,white] (-1,-1) rectangle (1,1);
\draw[very thick,rounded corners,white] (-1,-1) rectangle (1,1);
\draw[very thick,rounded corners,dashed,boundarycolor] (-1,-1) rectangle (1,1);
\end{tikzpicture}
}
\end{array}\\
\intertext{From the local relations {\bf G},{\bf LG} and {\bf RG} respectively, we get:}
\nonumber\\
{\text{\bf Inverse G}:}&&
\begin{array}{c}
\includeTikz{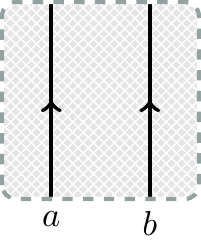}{
\begin{tikzpicture}
\filldraw[fill=bulkcolor,postaction={pattern=crosshatch,pattern color=white}] (-1,-1) rectangle (1,1);
\begin{scope}[very thick,decoration={markings,mark=at position 0.5 with {\arrow{>}}}]
\draw[postaction={decorate}] (-0.5,-1) -- (-0.5,1);
\draw[postaction={decorate}] (0.5,-1) -- (0.5,1);
\end{scope}
\node[below] at (-0.5,-1) {$a$};
\node[below] at (0.5,-1) {$b$};
\draw[very thick,white] (-1,-1) rectangle (1,1);
\draw[rounded corners,very thick,white] (-1,-1) rectangle (1,1);
\draw[dashed,rounded corners,very thick,boundarycolor] (-1,-1) rectangle (1,1);
\end{tikzpicture}
}
\end{array} &= \sum_{\alpha\beta} G^{-1}_{\alpha\beta}
\begin{array}{c}
\includeTikz{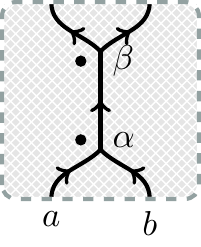}{
\begin{tikzpicture}
\filldraw[fill=bulkcolor,postaction={pattern=crosshatch,pattern color=white}] (-1,-1) rectangle (1,1);
\begin{scope}[very thick,decoration={markings,mark=at position 0.5 with {\arrow{>}}}]
\draw[postaction={decorate}] (-0.5,-1) to[out=90,in=180+45] (0,-0.5);
\draw[postaction={decorate}] (0.5,-1) to[out=90,in=-45] (0,-0.5);
\draw[postaction={decorate}] (0,-0.5) -- (0,0.5);
\draw[postaction={decorate}] (0,0.5) to [out=90+45,in=270] (-0.5,1);
\draw[postaction={decorate}] (0,0.5) to [out=45,in=270] (0.5,1);
\end{scope}
\node[below] at (-0.5,-1) {$a$};
\node[below] at (0.5,-1) {$b$};
\node[right] at (0,-0.4) {$\alpha$};
\node[right] at (0,0.4) {$\beta$};
\draw[fill=black] (-0.2,-0.4) circle (0.05);
\draw[fill=black] (-0.2,0.4) circle (0.05);
\draw[very thick,white] (-1,-1) rectangle (1,1);
\draw[rounded corners,very thick,white] (-1,-1) rectangle (1,1);
\draw[dashed,rounded corners,very thick,boundarycolor] (-1,-1) rectangle (1,1);
\end{tikzpicture}
}
\end{array}\\
{\text{\bf bubble}:}&&
\begin{array}{c}
\includeTikz{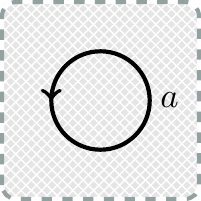}{
\begin{tikzpicture}
\filldraw[fill=bulkcolor,postaction={pattern=crosshatch,pattern color=white}] (-1,-1) rectangle (1,1);
\begin{scope}[very thick,decoration={markings,mark=at position 0.5 with {\arrow{>}}}]
\draw[postaction={decorate}] (0,0) circle (0.5);
\end{scope}
\node at (0.7,0) {$a$};
\draw[very thick,white] (-1,-1) rectangle (1,1);
\draw[rounded corners,very thick,white] (-1,-1) rectangle (1,1);
\draw[dashed,rounded corners,very thick,boundarycolor] (-1,-1) rectangle (1,1);
\end{tikzpicture}
}
\end{array} &= {\rm dim}(a)
\begin{array}{c}
\includeTikz{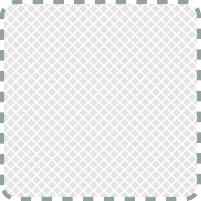}{
\begin{tikzpicture}
\filldraw[fill=bulkcolor,postaction={pattern=crosshatch,pattern color=white}] (-1,-1) rectangle (1,1);
\draw[very thick,white] (-1,-1) rectangle (1,1);
\draw[rounded corners,very thick,white] (-1,-1) rectangle (1,1);
\draw[dashed,rounded corners,very thick,boundarycolor] (-1,-1) rectangle (1,1);
\end{tikzpicture}
}
\end{array}\\
{\bf Inverse LG:}&&
\begin{array}{c}
\includeTikz{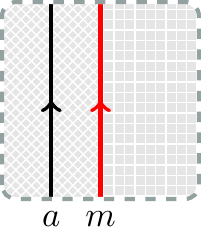}{
\begin{tikzpicture}
\filldraw[fill=bulkcolor,postaction={pattern=crosshatch,pattern color=white}] (-1,-1) rectangle (0,1);
\filldraw[fill=bulkcolor,postaction={pattern=grid,pattern color=white}] (0,-1) rectangle (1,1);
\begin{scope}[very thick,decoration={markings,mark=at position 0.5 with {\arrow{>}}}]
\draw[postaction={decorate}] (-0.5,-1) -- (-0.5,1);
\draw[postaction={decorate},red] (0,-1) -- (0,1);
\end{scope}
\node[below] at (-0.5,-1) {$a$};
\node[below] at (0,-1) {$m$};
\draw[very thick,white] (-1,-1) rectangle (1,1);
\draw[rounded corners,very thick,white] (-1,-1) rectangle (1,1);
\draw[dashed,rounded corners,very thick,boundarycolor] (-1,-1) rectangle (1,1);
\end{tikzpicture}
}
\end{array} &= \sum_{\zeta\eta} LG^{-1}_{\zeta\eta}
\begin{array}{c}
\includeTikz{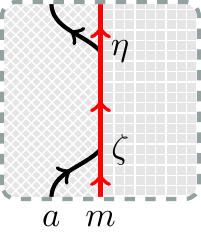}{
\begin{tikzpicture}
\filldraw[fill=bulkcolor,postaction={pattern=crosshatch,pattern color=white}] (-1,-1) rectangle (0,1);
\filldraw[fill=bulkcolor,postaction={pattern=grid,pattern color=white}] (0,-1) rectangle (1,1);
\begin{scope}[very thick,decoration={markings,mark=at position 0.5 with {\arrow{>}}}]
\draw[postaction={decorate}] (-0.5,-1) to[out=90,in=180+45] (0,-0.5);
\draw[postaction={decorate}] (0,0.5) to[out=90+45,in=270] (-0.5,1);
\draw[postaction={decorate},red] (0,-1) -- (0,-0.5);
\draw[postaction={decorate},red] (0,-0.5) -- (0,0.5);
\draw[postaction={decorate},red] (0,0.5) -- (0,1);
\end{scope}
\node[below] at (-0.5,-1) {$a$};
\node[below] at (0,-1) {$m$};
\node at (0.2,-0.5) {$\zeta$};
\node at (0.2,0.5) {$\eta$};
\draw[very thick,white] (-1,-1) rectangle (1,1);
\draw[rounded corners,very thick,white] (-1,-1) rectangle (1,1);
\draw[dashed,rounded corners,very thick,boundarycolor] (-1,-1) rectangle (1,1);
\end{tikzpicture}
}
\end{array}\\
{\bf Inverse RG:}&&
\begin{array}{c}
\includeTikz{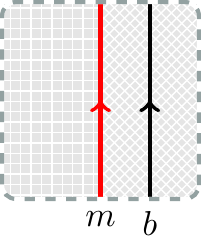}{
\begin{tikzpicture}[xscale=-1]
\filldraw[fill=bulkcolor,postaction={pattern=crosshatch,pattern color=white}] (-1,-1) rectangle (0,1);
\filldraw[fill=bulkcolor,postaction={pattern=grid,pattern color=white}] (0,-1) rectangle (1,1);
\begin{scope}[very thick,decoration={markings,mark=at position 0.5 with {\arrow{>}}}]
\draw[postaction={decorate}] (-0.5,-1) -- (-0.5,1);
\draw[postaction={decorate},red] (0,-1) -- (0,1);
\end{scope}
\node[below] at (-0.5,-1) {$b$};
\node[below] at (0,-1) {$m$};
\draw[very thick,white] (-1,-1) rectangle (1,1);
\draw[rounded corners,very thick,white] (-1,-1) rectangle (1,1);
\draw[dashed,rounded corners,very thick,boundarycolor] (-1,-1) rectangle (1,1);
\end{tikzpicture}
}
\end{array} &= \sum_{\zeta\eta} RG^{-1}_{\zeta\eta}
\begin{array}{c}
\includeTikz{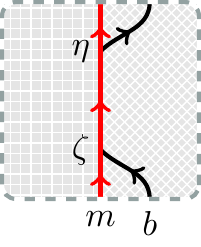}{
\begin{tikzpicture}[xscale=-1]
\filldraw[fill=bulkcolor,postaction={pattern=crosshatch,pattern color=white}] (-1,-1) rectangle (0,1);
\filldraw[fill=bulkcolor,postaction={pattern=grid,pattern color=white}] (0,-1) rectangle (1,1);
\begin{scope}[very thick,decoration={markings,mark=at position 0.5 with {\arrow{>}}}]
\draw[postaction={decorate}] (-0.5,-1) to[out=90,in=180+45] (0,-0.5);
\draw[postaction={decorate}] (0,0.5) to[out=90+45,in=270] (-0.5,1);
\draw[postaction={decorate},red] (0,-1) -- (0,-0.5);
\draw[postaction={decorate},red] (0,-0.5) -- (0,0.5);
\draw[postaction={decorate},red] (0,0.5) -- (0,1);
\end{scope}
\node[below] at (-0.5,-1) {$b$};
\node[below] at (0,-1) {$m$};
\node at (0.2,-0.5) {$\zeta$};
\node at (0.2,0.5) {$\eta$};
\draw[very thick,white] (-1,-1) rectangle (1,1);
\draw[rounded corners,very thick,white] (-1,-1) rectangle (1,1);
\draw[dashed,rounded corners,very thick,boundarycolor] (-1,-1) rectangle (1,1);
\end{tikzpicture}
}
\end{array}
\end{align}
\end{subequations}
\end{itemize}
\end{definition}

\begin{definition}[representation theory of categories]
Let $\cat{C}$ be a category. A {\em representation} of $\cat{C}$ is a functor $V:\cat{C} \to {\bf Vec}$ from $\cat{C}$ into the category of vector spaces. The functor $V$ consists of a vector space $V_a$ for each object $a \in \cat{C}$ and a linear map $V_f : V_a \to V_b$ for each morphism $f : a \to b$ in $\cat{C}$ such that $V_f \circ V_g = V_{f \circ g}$ and $V_{\rm id} = {\rm id}$.
\end{definition}
We are particularly interested in the following computational problem: Let $\cat{C}$ be a category. If $\cat{C}$ is semi-simple, compute a complete list of simple representations of $\cat{C}$. The first step in computing the representations of $\cat{C}$ is computing the Karoubi envelope:
\begin{definition}[Karoubi envelope] \label{def:karoubi_envelope}
Let $\cat{C}$ be a category. The Karoubi envelope of $\cat{C}$ is a category defined as follows: The objects are idempotent endomorphisms in $\cat{C}$. A morphism between two such idempotents $i : a \to a$ and $j : b \to b$ is a morphism $f : a \to b$ such that $jfi = f$. If $\cat{C}$ is semi-simple, we can compute the Karoubi envelope of $\cat{C}$ as follows:
\begin{enumerate}
\item For each object $a \in \cat{C}$, compute an {\em Artin-Wedderburn isomorphism} ${\rm End}(a) \cong \prod_{i=1}^d M_i$ where each $M_i$ is a matrix algebra. The idempotents $e_{11}$ from each factor in this product are non isomorphic simple objects in the Karoubi envelope of $\cat{C}$. Computing Artin-Wedderburn decompositions is computationally hard, but possible for algebras of small ($<50$) dimension. This is demonstrated in the Mathematica notebook \texttt{AWdecomposition.nb} in the ancillary material\cite{anc}.
\item Given two idempotents $i,j$, solving the linear equation $jfi=f$ gives the morphisms between $i$ and $j$.
\end{enumerate}
\end{definition}
The Karoubi envelope of $\cat{C}$ is equivalent to the category of representations of $\cat{C}$. Given a simple object $i : a \to a$ in the Karoubi envelope, the corresponding representation is $\cat{C}(a,-)i$. Computing bases and action matrices for this representation is a linear algebra problem.
\begin{definition}[Annular categories and point defects] \label{def:annular_category_appendix}
Let $D \subseteq \Sigma$ be a domain wall structure. If $\Sigma$ has a boundary, then $Z(\Sigma)$ has more structure than just a vector space. In a neighborhood of a boundary component, the domain wall structure looks like:
\begin{align}
\begin{array}{c}
\includeTikz{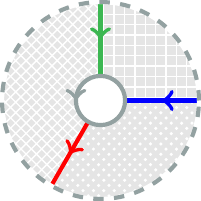}{
\begin{tikzpicture}[scale=0.5]
\path[fill=bulkcolor,postaction={pattern=crosshatch,pattern color=white}] (0,0) -- (90:2) arc(90:180+60:2) -- cycle;
\path[fill=bulkcolor,postaction={pattern=grid,pattern color=white}] (0,0) -- (90:2) arc(90:0:2) -- cycle;
\path[fill=bulkcolor,postaction={pattern=crosshatch dots,pattern color=white}] (0,0) -- (0:2) arc(0:-90-30:2) -- cycle;
\begin{scope}[very thick,decoration={markings,mark=at position 0.5 with {\arrow{>}}}]
\draw[nicegreen,postaction={decorate}] (0,2) -- (0,0.5);
\draw[red,postaction={decorate}] (-0.5*0.5, -0.866025*0.5) -- (-0.5*2, -0.866025*2);
\draw[blue,postaction={decorate}] (2,0) -- (0.5,0);
\draw[boundarycolor,postaction={decorate},fill=white] (0,0) circle (0.5);
\end{scope}
\draw[white,very thick] (0,0) circle (2);
\draw[boundarycolor,very thick,dashed] (0,0) circle (2);
\end{tikzpicture}
}
\end{array}.
\end{align}
In general, there can be an arbitrary number of components of $D$ terminating on the boundary, but for the sake of clarity, we shall work with this explicit example. In a neighborhood of the boundary component, a skein vector looks like:
\begin{align} \label{eq:boundary_component_skein_neighbourhood}
\begin{array}{c}
\includeTikz{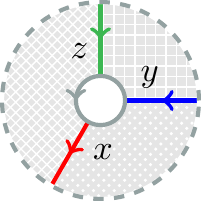}{
\begin{tikzpicture}[scale=0.5]
\path[fill=bulkcolor,postaction={pattern=crosshatch,pattern color=white}] (0,0) -- (90:2) arc(90:180+60:2) -- cycle;
\path[fill=bulkcolor,postaction={pattern=grid,pattern color=white}] (0,0) -- (90:2) arc(90:0:2) -- cycle;
\path[fill=bulkcolor,postaction={pattern=crosshatch dots,pattern color=white}] (0,0) -- (0:2) arc(0:-90-30:2) -- cycle;
\begin{scope}[very thick,decoration={markings,mark=at position 0.5 with {\arrow{>}}}]
\draw[nicegreen,postaction={decorate}] (0,2) -- (0,0.5);
\draw[red,postaction={decorate}] (-0.5*0.5, -0.866025*0.5) -- (-0.5*2, -0.866025*2);
\draw[blue,postaction={decorate}] (2,0) -- (0.5,0);
\draw[boundarycolor,postaction={decorate},fill=white] (0,0) circle (0.5);
\end{scope}
\node[left] at (0,1) {$z$};
\node[above] at (1,0) {$y$};
\node[right] at (-0.5*1.2+0.2, -0.866025*1.2) {$x$};
\draw[white,very thick] (0,0) circle (2);
\draw[boundarycolor,very thick,dashed] (0,0) circle (2);
\end{tikzpicture}
}
\end{array}.
\end{align}
We define a domain wall structure
\begin{align}
A =\begin{array}{c}
\includeTikz{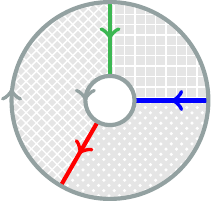}{
\begin{tikzpicture}[scale=0.5]
\path[fill=bulkcolor,postaction={pattern=crosshatch,pattern color=white}] (0,0) -- (90:2) arc(90:180+60:2) -- cycle;
\path[fill=bulkcolor,postaction={pattern=grid,pattern color=white}] (0,0) -- (90:2) arc(90:0:2) -- cycle;
\path[fill=bulkcolor,postaction={pattern=crosshatch dots,pattern color=white}] (0,0) -- (0:2) arc(0:-90-30:2) -- cycle;
\begin{scope}[very thick,decoration={markings,mark=at position 0.5 with {\arrow{>}}}]
\draw[nicegreen,postaction={decorate}] (0,2) -- (0,0.5);
\draw[red,postaction={decorate}] (-0.5*0.5, -0.866025*0.5) -- (-0.5*2, -0.866025*2);
\draw[blue,postaction={decorate}] (2,0) -- (0.5,0);
\draw[boundarycolor,postaction={decorate},fill=white] (0,0) circle (0.5);
\end{scope}
\begin{scope}[very thick,decoration={markings,mark=at position 0.5 with {\arrow{<}}}]
\draw[boundarycolor,postaction={decorate}] (0,0) circle (2);
\end{scope}
\node at (-2,0) {};
\end{tikzpicture}
}
\end{array}.
\end{align}
Every skein vector on $A$ is equivalent a sum of skein vectors of the following form:
\begin{align} \label{eq:annular_morphism}
\begin{array}{c}
\includeTikz{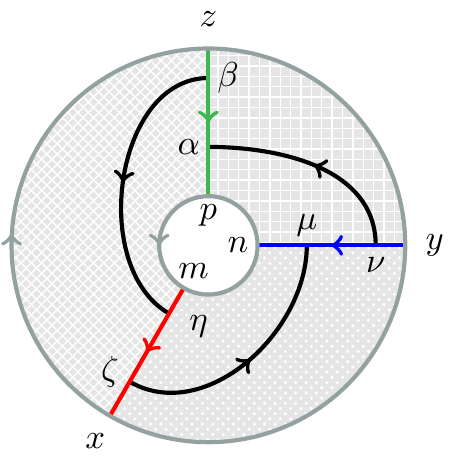}{
\begin{tikzpicture}
\path[fill=bulkcolor,postaction={pattern=crosshatch,pattern color=white}] (0,0) -- (90:2) arc(90:180+60:2) -- cycle;
\path[fill=bulkcolor,postaction={pattern=grid,pattern color=white}] (0,0) -- (90:2) arc(90:0:2) -- cycle;
\path[fill=bulkcolor,postaction={pattern=crosshatch dots,pattern color=white}] (0,0) -- (0:2) arc(0:-90-30:2) -- cycle;
\begin{scope}[very thick,decoration={markings,mark=at position 0.5 with {\arrow{>}}}]
\draw[postaction={decorate}] (-0.5*1.6, -0.866025*1.6) to[in=270,out=-30] (1,0);
\draw[postaction={decorate}] (1.7,0) to[out=90,in=0] (0,1);
\draw[postaction={decorate}] (0,1.7) to[out=180,in=180-30] (-0.5*0.8, -0.866025*0.8);
\draw[nicegreen,postaction={decorate}] (0,2) -- (0,0.5);
\draw[red,postaction={decorate}] (-0.5*0.5, -0.866025*0.5) -- (-0.5*2, -0.866025*2);
\draw[blue,postaction={decorate}] (2,0) -- (0.5,0);
\draw[boundarycolor,postaction={decorate},fill=white] (0,0) circle (0.5);
\end{scope}
\begin{scope}[very thick,decoration={markings,mark=at position 0.5 with {\arrow{<}}}]
\draw[boundarycolor,postaction={decorate}] (0,0) circle (2);
\end{scope}
\node at (-2,0) {};
\node at (0,2.3) {$z$};
\node at (2.3,0) {$y$};
\node at (-0.5*2.3, -0.866025*2.3) {$x$};
\node at (0,0.3) {$p$};
\node at (0.3,0) {$n$};
\node at (-0.5*0.3, -0.866025*0.3) {$m$};
\node at (-0.5*1.6-0.2, -0.866025*1.6+0.1) {$\zeta$};
\node at (-0.5*0.6+0.2, -0.866025*0.6-0.3) {$\eta$};
\node at (1,0.2) {$\mu$};
\node at (1.7,-0.2) {$\nu$};
\node at (-0.2,1) {$\alpha$};
\node at (0.2,1.7) {$\beta$};
\end{tikzpicture}
}
\end{array} \in Z(A).
\end{align}
The space $Z(A)$ forms a category. The objects are tuples of module category objects $(x,y,z)$ and the morphisms are skein vectors \eqref{eq:annular_morphism}. Composition is given by gluing one skein vector inside the other, subject to matching labels. We call $Z(A)$ an {\em annular category}. We can glue \eqref{eq:annular_morphism} into \eqref{eq:boundary_component_skein_neighbourhood}. Therefore, $Z(\Sigma)$ has an action of the annular category associated to each boundary component. Suppose that $V$ is a representation of the annular category $Z(A)$. We also call $V$ a {\em point defect}. Vectors in $V$ are depicted as follows:
\begin{align} \label{eq:annular_cat_rep_vector}
\begin{array}{c}
\includeTikz{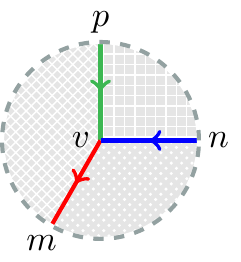}{
\begin{tikzpicture}
\path[fill=bulkcolor,postaction={pattern=crosshatch,pattern color=white}] (0,0) -- (90:1) arc(90:180+60:1) -- cycle;
\path[fill=bulkcolor,postaction={pattern=grid,pattern color=white}] (0,0) -- (90:1) arc(90:0:1) -- cycle;
\path[fill=bulkcolor,postaction={pattern=crosshatch dots,pattern color=white}] (0,0) -- (0:1) arc(0:-90-30:1) -- cycle;
\begin{scope}[very thick,decoration={markings,mark=at position 0.5 with {\arrow{>}}}]
\draw[nicegreen,postaction={decorate}] (0,1) -- (0,0);
\draw[red,postaction={decorate}] (0,0) -- (-0.5, -0.866025);
\draw[blue,postaction={decorate}] (1,0) -- (0,0);
\end{scope}
\node at (-0.2,0) {$v$};
\node at (1.2,0) {$n$};
\node at (0,1.2) {$p$};
\node at (-0.5*1.2, -0.866025*1.2) {$m$};
\draw[white,very thick] (0,0) circle (1);
\draw[boundarycolor,very thick,dashed] (0,0) circle (1);
\end{tikzpicture}
}
\end{array} \in V_{m,n,p}.
\end{align}
Diagrammaticallys, the annular category action of \eqref{eq:annular_morphism} on \eqref{eq:annular_cat_rep_vector} looks like:
\begin{align} \label{eq:annular_category_local_relation}
\begin{array}{c}
\includeTikz{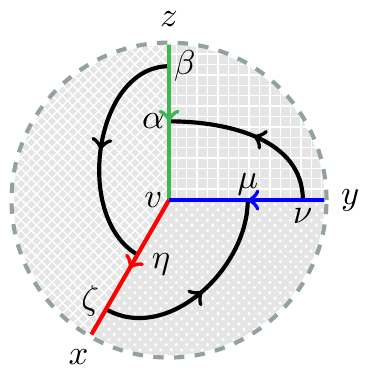}{
\begin{tikzpicture}[scale=0.8]
\path[fill=bulkcolor,postaction={pattern=crosshatch,pattern color=white}] (0,0) -- (90:2) arc(90:180+60:2) -- cycle;
\path[fill=bulkcolor,postaction={pattern=grid,pattern color=white}] (0,0) -- (90:2) arc(90:0:2) -- cycle;
\path[fill=bulkcolor,postaction={pattern=crosshatch dots,pattern color=white}] (0,0) -- (0:2) arc(0:-90-30:2) -- cycle;
\begin{scope}[very thick,decoration={markings,mark=at position 0.5 with {\arrow{>}}}]
\draw[postaction={decorate}] (-0.5*1.6, -0.866025*1.6) to[in=270,out=-30] (1,0);
\draw[postaction={decorate}] (1.7,0) to[out=90,in=0] (0,1);
\draw[postaction={decorate}] (0,1.7) to[out=180,in=180-30] (-0.5*0.8, -0.866025*0.8);
\draw[nicegreen,postaction={decorate}] (0,2) -- (0,0);
\draw[red,postaction={decorate}] (0,0) -- (-0.5*2, -0.866025*2);
\draw[blue,postaction={decorate}] (2,0) -- (0,0);
\end{scope}
\draw[very thick,white] (0,0) circle (2);
\draw[very thick,boundarycolor,dashed] (0,0) circle (2);
\node at (-2,0) {};
\node at (-0.2,0) {$v$};
\node at (0,2.3) {$z$};
\node at (2.3,0) {$y$};
\node at (-0.5*2.3, -0.866025*2.3) {$x$};
\node at (-0.5*1.6-0.2, -0.866025*1.6+0.1) {$\zeta$};
\node at (-0.5*0.6+0.2, -0.866025*0.6-0.3) {$\eta$};
\node at (1,0.2) {$\mu$};
\node at (1.7,-0.2) {$\nu$};
\node at (-0.2,1) {$\alpha$};
\node at (0.2,1.7) {$\beta$};
\end{tikzpicture}
}
\end{array} = \sum_w A_{\eta\zeta\mu\nu\alpha\beta v}^w
\begin{array}{c}
\includeTikz{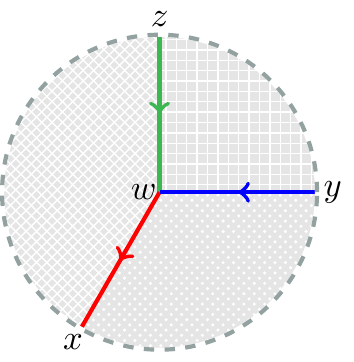}{
\begin{tikzpicture}[scale=0.8]
\path[fill=bulkcolor,postaction={pattern=crosshatch,pattern color=white}] (0,0) -- (90:2) arc(90:180+60:2) -- cycle;
\path[fill=bulkcolor,postaction={pattern=grid,pattern color=white}] (0,0) -- (90:2) arc(90:0:2) -- cycle;
\path[fill=bulkcolor,postaction={pattern=crosshatch dots,pattern color=white}] (0,0) -- (0:2) arc(0:-90-30:2) -- cycle;
\begin{scope}[very thick,decoration={markings,mark=at position 0.5 with {\arrow{>}}}]
\draw[nicegreen,postaction={decorate}] (0,2) -- (0,0);
\draw[red,postaction={decorate}] (0,0) -- (-0.5*2, -0.866025*2);
\draw[blue,postaction={decorate}] (2,0) -- (0,0);
\end{scope}
\node at (-0.2,0) {$w$};
\node at (2.2,0) {$y$};
\node at (0,2.2) {$z$};
\node at (-0.5*2.2, -0.866025*2.2) {$x$};
\draw[white,very thick] (0,0) circle (2);
\draw[boundarycolor,dashed,very thick] (0,0) circle (2);
\end{tikzpicture}
}
\end{array}
\end{align}

For the special case in Example~\ref{ex:anyons}, the annular category coincides with the tube algebra\cite{ocneanu,Bultinck2015,SETPaper}.
\end{definition}
\begin{example}[Anyons] \label{ex:anyons}
Consider the following annular category for a fusion category $\cat{C}$:
\begin{align}
\begin{array}{c}
\includeTikz{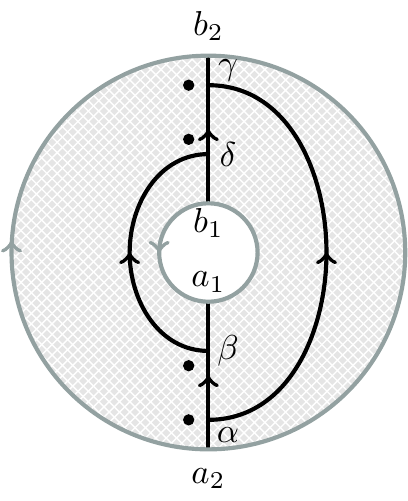}{
\begin{tikzpicture}
\filldraw[fill=bulkcolor,postaction={pattern=crosshatch,pattern color=white}] (0,0) circle (2);
\begin{scope}[very thick,decoration={markings,mark=at position 0.5 with {\arrow{>}}}]
\draw[postaction={decorate}] (0,-2) -- (0,-0.5);
\draw[postaction={decorate}] (0,0.5) -- (0,2);
\draw[postaction={decorate}] (0,-1) to[out=180,in=270] (-0.8,0) to[out=90,in=180] (0,1);
\draw[postaction={decorate}] (0,-1.7) to[out=0,in=270] (1.2,0) to[out=90,in=0] (0,1.7);
\draw[boundarycolor,postaction={decorate},fill=white] (0,0) circle (0.5);
\end{scope}
\begin{scope}[very thick,decoration={markings,mark=at position 0.5 with {\arrow{<}}}]
\draw[boundarycolor,postaction={decorate}] (0,0) circle (2);
\end{scope}
\node at (-2,0) {};
\node at (0,-2.3) {$a_2$};
\node at (0,-0.3) {$a_1$};
\node at (0,2.3) {$b_2$};
\node at (0,0.3) {$b_1$};
\node at (0.2,-1.85) {$\alpha$};
\node at (0.2,-1) {$\beta$};
\node at (0.2,1.85) {$\gamma$};
\node at (0.2,1) {$\delta$};
\draw[fill=black] (-0.2,-1.7) circle (0.05);
\draw[fill=black] (-0.2,-1.15) circle (0.05);
\draw[fill=black] (-0.2,1.7) circle (0.05);
\draw[fill=black] (-0.2,1.15) circle (0.05);
\end{tikzpicture}
}
\end{array} : (a_1,b_1) \to (a_2,b_2)
\end{align}
Representations $V$ of this annular category parameterize anyons in the Levin-Wen phase associated to $\cat{C}$. Vectors
\begin{align} \label{eq:anyon_internal_state}
\begin{array}{c}
\includeTikz{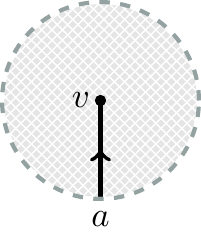}{
\begin{tikzpicture}
\filldraw[fill=bulkcolor,postaction={pattern=crosshatch,pattern color=white}] (0,0) circle (1);
\begin{scope}[very thick,decoration={markings,mark=at position 0.5 with {\arrow{>}}}]
\draw[postaction={decorate}] (0,-1) -- (0,0);
\end{scope}
\draw[fill=black] (0,0) circle (0.05);
\draw[white,very thick] (0,0) circle (1);
\draw[very thick,dashed,boundarycolor] (0,0) circle (1);
\node at (-0.2,0) {$v$};
\node at (0.2,0) {};
\node at (0,-1.2) {$a$};
\end{tikzpicture}
}
\end{array} \in V_{(a,1)}
\end{align}
are important because they can terminate a fusion category string in a compound defect skein vector (defined below).
\end{example}
\begin{definition}[Compound defect] A {\em compound defect} consists of a domain wall structure $D \subseteq \Sigma$ and an annular category representation for each boundary component of $\Sigma$.
\end{definition}
\begin{definition}[Compound defect skein vector] Let $D \subseteq \Sigma$ be a compound defect. A compound defect skein vector consists of a skein vector for the domain wall structure and a vector in each boundary component representation. We require the object labels on the vector and the boundary component to match. 
\end{definition}
\begin{definition}[Local relations]
Let $\Sigma$ be a compound defect. Again, we define
\begin{align}
Z(\Sigma) = \mathbb{C} \left\{ \text{skein vectors}\right\} / \text{local relations}.
\end{align}
We have all the local relations from Definition~\ref{def:local_relations_1}. There is one additional local relation:
\begin{itemize}
\item[] {\bf annular category action:} A vector which was used to fill a boundary component of $\Sigma$ can absorb a skein vector on the surrounding annulus and transform according to its annular category representation. This is depicted in \eqref{eq:annular_category_local_relation}.
\end{itemize}
An important consequence of the annular category action local relation is the bubble action local relation:
\begin{align} {\rm dim}(a)
\begin{array}{c}
\includeTikz{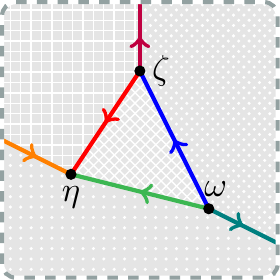}{
\begin{tikzpicture}[scale=0.7]
\path[fill=bulkcolor,postaction={pattern=crosshatch,pattern color=white}] (0,1) -- (-1,-0.5) -- (1,-1) -- cycle;
\path[fill=bulkcolor,postaction={pattern=grid,pattern color=white}] (-2,2) -- (0,2) -- (0,1) -- (-1,-0.5) -- (-2,0) -- cycle;
\path[fill=bulkcolor,postaction={pattern=crosshatch dots,pattern color=white}] (0,2) -- (0,1) -- (1,-1) -- (2,-1.5) -- (2,2) -- cycle;
\path[fill=bulkcolor,postaction={pattern=dots,pattern color=white}] (-2,0) -- (-1,-0.5) -- (1,-1) -- (2,-1.5) -- (2,-2) -- (-2,-2) -- cycle;
\begin{scope}[very thick,decoration={markings,mark=at position 0.5 with {\arrow{>}}}]
\draw[red,postaction={decorate}] (0,1) -- (-1,-0.5);
\draw[nicegreen,postaction={decorate}] (1,-1) -- (-1,-0.5);
\draw[blue,postaction={decorate}] (1,-1) -- (0,1);
\draw[purple,postaction={decorate}] (0,1) -- (0,2);
\draw[orange,postaction={decorate}] (-2,0) -- (-1,-0.5);
\draw[teal,postaction={decorate}] (1,-1) -- (2,-1.5);
\end{scope}
\node[right] at (0,1) {$\zeta$};
\node[below] at (-1,-0.5) {$\eta$};
\node[above] at (1.1,-1) {$\omega$};
\draw[very thick,white] (-2,-2) rectangle (2,2);
\draw[very thick,rounded corners,white] (-2,-2) rectangle (2,2);
\draw[very thick,rounded corners,boundarycolor,dashed] (-2,-2) rectangle (2,2);
\filldraw (0,1) circle (0.07);
\filldraw (-1,-0.5) circle (0.07);
\filldraw (1,-1) circle (0.07);
\end{tikzpicture}
}
\end{array} &= 
\begin{array}{c}
\includeTikz{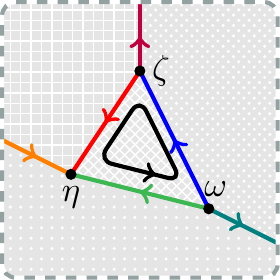}{
\begin{tikzpicture}[scale=0.7]
\path[fill=bulkcolor,postaction={pattern=crosshatch,pattern color=white}] (0,1) -- (-1,-0.5) -- (1,-1) -- cycle;
\path[fill=bulkcolor,postaction={pattern=grid,pattern color=white}] (-2,2) -- (0,2) -- (0,1) -- (-1,-0.5) -- (-2,0) -- cycle;
\path[fill=bulkcolor,postaction={pattern=crosshatch dots,pattern color=white}] (0,2) -- (0,1) -- (1,-1) -- (2,-1.5) -- (2,2) -- cycle;
\path[fill=bulkcolor,postaction={pattern=dots,pattern color=white}] (-2,0) -- (-1,-0.5) -- (1,-1) -- (2,-1.5) -- (2,-2) -- (-2,-2) -- cycle;
\begin{scope}[very thick,decoration={markings,mark=at position 0.5 with {\arrow{>}}}]
\draw[red,postaction={decorate}] (0,1) -- (-1,-0.5);
\draw[nicegreen,postaction={decorate}] (1,-1) -- (-1,-0.5);
\draw[blue,postaction={decorate}] (1,-1) -- (0,1);
\draw[purple,postaction={decorate}] (0,1) -- (0,2);
\draw[orange,postaction={decorate}] (-2,0) -- (-1,-0.5);
\draw[teal,postaction={decorate}] (1,-1) -- (2,-1.5);
\draw[postaction={decorate},rounded corners] (0,0.6*1) -- (-1*0.6,-0.5*0.6) -- (1*0.6,-1*0.6) -- cycle;
\end{scope}
\node[right] at (0,1) {$\zeta$};
\node[below] at (-1,-0.5) {$\eta$};
\node[above] at (1.1,-1) {$\omega$};
\draw[very thick,white] (-2,-2) rectangle (2,2);
\draw[very thick,rounded corners,white] (-2,-2) rectangle (2,2);
\draw[very thick,rounded corners,boundarycolor,dashed] (-2,-2) rectangle (2,2);
\filldraw (0,1) circle (0.07);
\filldraw (-1,-0.5) circle (0.07);
\filldraw (1,-1) circle (0.07);
\end{tikzpicture}
}
\end{array}
=  \sum_{\alpha\beta\gamma\delta\theta\phi} RG^{-1}_{\alpha\beta}RL^{-1}_{\gamma\delta} RL^{-1}_{\theta\phi}
\begin{array}{c}
\includeTikz{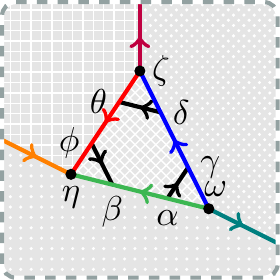}{
\begin{tikzpicture}[scale=0.7]
\path[fill=bulkcolor,postaction={pattern=crosshatch,pattern color=white}] (0,1) -- (-1,-0.5) -- (1,-1) -- cycle;
\path[fill=bulkcolor,postaction={pattern=grid,pattern color=white}] (-2,2) -- (0,2) -- (0,1) -- (-1,-0.5) -- (-2,0) -- cycle;
\path[fill=bulkcolor,postaction={pattern=crosshatch dots,pattern color=white}] (0,2) -- (0,1) -- (1,-1) -- (2,-1.5) -- (2,2) -- cycle;
\path[fill=bulkcolor,postaction={pattern=dots,pattern color=white}] (-2,0) -- (-1,-0.5) -- (1,-1) -- (2,-1.5) -- (2,-2) -- (-2,-2) -- cycle;
\begin{scope}[very thick,decoration={markings,mark=at position 0.5 with {\arrow{>}}}]
\draw[postaction={decorate}] (-0.7, -0.05) -- (-0.4, -0.65);
\draw[postaction={decorate}] (0.4,-0.85) -- (0.7,-0.4);
\draw[postaction={decorate}] (0.3,0.4) -- (-0.3,0.55);
\draw[red,postaction={decorate}] (0,1) -- (-1,-0.5);
\draw[nicegreen,postaction={decorate}] (1,-1) -- (-1,-0.5);
\draw[blue,postaction={decorate}] (1,-1) -- (0,1);
\draw[purple,postaction={decorate}] (0,1) -- (0,2);
\draw[orange,postaction={decorate}] (-2,0) -- (-1,-0.5);
\draw[teal,postaction={decorate}] (1,-1) -- (2,-1.5);
\end{scope}
\node[left] at (-0.3,0.55) {$\theta$};
\node[left] at (-0.7, -0.05) {$\phi$};
\node[below] at (-0.4, -0.65) {$\beta$};
\node[below] at (0.4,-0.85) {$\alpha$};
\node[right] at (0.7,-0.4) {$\gamma$};
\node[right] at (0.3,0.4) {$\delta$};
\node[right] at (0,1) {$\zeta$};
\node[below] at (-1,-0.5) {$\eta$};
\node[above] at (1.1,-1) {$\omega$};
\draw[very thick,white] (-2,-2) rectangle (2,2);
\draw[very thick,rounded corners,white] (-2,-2) rectangle (2,2);
\draw[very thick,rounded corners,boundarycolor,dashed] (-2,-2) rectangle (2,2);
\filldraw (0,1) circle (0.07);
\filldraw (-1,-0.5) circle (0.07);
\filldraw (1,-1) circle (0.07);
\end{tikzpicture}
} 
\end{array} \\
&=  \sum_{\alpha\beta\gamma\delta\theta\phi\eta'\zeta'\omega'} RG^{-1}_{\alpha\beta}RL^{-1}_{\gamma\delta} RL^{-1}_{\theta\phi} A_{\eta\beta\phi}^{\eta'} A_{\omega\alpha\gamma}^{\omega'}A_{\zeta\theta\delta}^{\zeta'}
\begin{array}{c}
\includeTikz{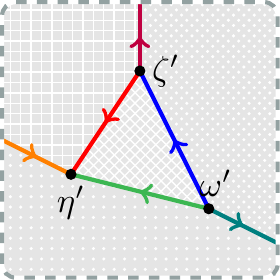}{
\begin{tikzpicture}[scale=0.7]
\path[fill=bulkcolor,postaction={pattern=crosshatch,pattern color=white}] (0,1) -- (-1,-0.5) -- (1,-1) -- cycle;
\path[fill=bulkcolor,postaction={pattern=grid,pattern color=white}] (-2,2) -- (0,2) -- (0,1) -- (-1,-0.5) -- (-2,0) -- cycle;
\path[fill=bulkcolor,postaction={pattern=crosshatch dots,pattern color=white}] (0,2) -- (0,1) -- (1,-1) -- (2,-1.5) -- (2,2) -- cycle;
\path[fill=bulkcolor,postaction={pattern=dots,pattern color=white}] (-2,0) -- (-1,-0.5) -- (1,-1) -- (2,-1.5) -- (2,-2) -- (-2,-2) -- cycle;
\begin{scope}[very thick,decoration={markings,mark=at position 0.5 with {\arrow{>}}}]
\draw[red,postaction={decorate}] (0,1) -- (-1,-0.5);
\draw[nicegreen,postaction={decorate}] (1,-1) -- (-1,-0.5);
\draw[blue,postaction={decorate}] (1,-1) -- (0,1);
\draw[purple,postaction={decorate}] (0,1) -- (0,2);
\draw[orange,postaction={decorate}] (-2,0) -- (-1,-0.5);
\draw[teal,postaction={decorate}] (1,-1) -- (2,-1.5);
\end{scope}
\node[right] at (0,1) {$\zeta'$};
\node[below] at (-1,-0.5) {$\eta'$};
\node[above] at (1.1,-1) {$\omega'$};
\draw[very thick,white] (-2,-2) rectangle (2,2);
\draw[very thick,rounded corners,white] (-2,-2) rectangle (2,2);
\draw[very thick,rounded corners,boundarycolor,dashed] (-2,-2) rectangle (2,2);
\filldraw (0,1) circle (0.07);
\filldraw (-1,-0.5) circle (0.07);
\filldraw (1,-1) circle (0.07);
\end{tikzpicture}
}
\end{array}
\end{align}
This local relation is closely related to the bubble term in the Levin-Wen Hamiltonian from \onlinecite{Levin2005}. We have
\begin{align} \label{eq:levin_wen_operator}
\begin{array}{c}
\includeTikz{bubble_action_1}{
\begin{tikzpicture}[scale=0.7]
\path[fill=bulkcolor,postaction={pattern=crosshatch,pattern color=white}] (0,1) -- (-1,-0.5) -- (1,-1) -- cycle;
\path[fill=bulkcolor,postaction={pattern=grid,pattern color=white}] (-2,2) -- (0,2) -- (0,1) -- (-1,-0.5) -- (-2,0) -- cycle;
\path[fill=bulkcolor,postaction={pattern=crosshatch dots,pattern color=white}] (0,2) -- (0,1) -- (1,-1) -- (2,-1.5) -- (2,2) -- cycle;
\path[fill=bulkcolor,postaction={pattern=dots,pattern color=white}] (-2,0) -- (-1,-0.5) -- (1,-1) -- (2,-1.5) -- (2,-2) -- (-2,-2) -- cycle;
\begin{scope}[very thick,decoration={markings,mark=at position 0.5 with {\arrow{>}}}]
\draw[red,postaction={decorate}] (0,1) -- (-1,-0.5);
\draw[nicegreen,postaction={decorate}] (1,-1) -- (-1,-0.5);
\draw[blue,postaction={decorate}] (1,-1) -- (0,1);
\draw[purple,postaction={decorate}] (0,1) -- (0,2);
\draw[orange,postaction={decorate}] (-2,0) -- (-1,-0.5);
\draw[teal,postaction={decorate}] (1,-1) -- (2,-1.5);
\end{scope}
\node[right] at (0,1) {$\zeta$};
\node[below] at (-1,-0.5) {$\eta$};
\node[above] at (1.1,-1) {$\omega$};
\draw[very thick,white] (-2,-2) rectangle (2,2);
\draw[very thick,rounded corners,white] (-2,-2) rectangle (2,2);
\draw[very thick,rounded corners,boundarycolor,dashed] (-2,-2) rectangle (2,2);
\filldraw (0,1) circle (0.07);
\filldraw (-1,-0.5) circle (0.07);
\filldraw (1,-1) circle (0.07);
\end{tikzpicture}
}
\end{array} &= \frac{1}{{\rm dim}(\mathcal{C})} \sum_a {\rm dim}(a)^2
\begin{array}{c}
\includeTikz{bubble_action_1}{
\begin{tikzpicture}[scale=0.7]
\path[fill=bulkcolor,postaction={pattern=crosshatch,pattern color=white}] (0,1) -- (-1,-0.5) -- (1,-1) -- cycle;
\path[fill=bulkcolor,postaction={pattern=grid,pattern color=white}] (-2,2) -- (0,2) -- (0,1) -- (-1,-0.5) -- (-2,0) -- cycle;
\path[fill=bulkcolor,postaction={pattern=crosshatch dots,pattern color=white}] (0,2) -- (0,1) -- (1,-1) -- (2,-1.5) -- (2,2) -- cycle;
\path[fill=bulkcolor,postaction={pattern=dots,pattern color=white}] (-2,0) -- (-1,-0.5) -- (1,-1) -- (2,-1.5) -- (2,-2) -- (-2,-2) -- cycle;
\begin{scope}[very thick,decoration={markings,mark=at position 0.5 with {\arrow{>}}}]
\draw[red,postaction={decorate}] (0,1) -- (-1,-0.5);
\draw[nicegreen,postaction={decorate}] (1,-1) -- (-1,-0.5);
\draw[blue,postaction={decorate}] (1,-1) -- (0,1);
\draw[purple,postaction={decorate}] (0,1) -- (0,2);
\draw[orange,postaction={decorate}] (-2,0) -- (-1,-0.5);
\draw[teal,postaction={decorate}] (1,-1) -- (2,-1.5);
\end{scope}
\node[right] at (0,1) {$\zeta$};
\node[below] at (-1,-0.5) {$\eta$};
\node[above] at (1.1,-1) {$\omega$};
\draw[very thick,white] (-2,-2) rectangle (2,2);
\draw[very thick,rounded corners,white] (-2,-2) rectangle (2,2);
\draw[very thick,rounded corners,boundarycolor,dashed] (-2,-2) rectangle (2,2);
\filldraw (0,1) circle (0.07);
\filldraw (-1,-0.5) circle (0.07);
\filldraw (1,-1) circle (0.07);
\end{tikzpicture}
}
\end{array} \\
&= \frac{1}{{\rm dim}(\mathcal{C})} \sum_a {\rm dim}(a) \; \;
\begin{array}{c}
\includeTikz{bubble_action_2}{
\begin{tikzpicture}[scale=0.7]
\path[fill=bulkcolor,postaction={pattern=crosshatch,pattern color=white}] (0,1) -- (-1,-0.5) -- (1,-1) -- cycle;
\path[fill=bulkcolor,postaction={pattern=grid,pattern color=white}] (-2,2) -- (0,2) -- (0,1) -- (-1,-0.5) -- (-2,0) -- cycle;
\path[fill=bulkcolor,postaction={pattern=crosshatch dots,pattern color=white}] (0,2) -- (0,1) -- (1,-1) -- (2,-1.5) -- (2,2) -- cycle;
\path[fill=bulkcolor,postaction={pattern=dots,pattern color=white}] (-2,0) -- (-1,-0.5) -- (1,-1) -- (2,-1.5) -- (2,-2) -- (-2,-2) -- cycle;
\begin{scope}[very thick,decoration={markings,mark=at position 0.5 with {\arrow{>}}}]
\draw[red,postaction={decorate}] (0,1) -- (-1,-0.5);
\draw[nicegreen,postaction={decorate}] (1,-1) -- (-1,-0.5);
\draw[blue,postaction={decorate}] (1,-1) -- (0,1);
\draw[purple,postaction={decorate}] (0,1) -- (0,2);
\draw[orange,postaction={decorate}] (-2,0) -- (-1,-0.5);
\draw[teal,postaction={decorate}] (1,-1) -- (2,-1.5);
\draw[postaction={decorate},rounded corners] (0,0.6*1) -- (-1*0.6,-0.5*0.6) -- (1*0.6,-1*0.6) -- cycle;
\end{scope}
\node[right] at (0,1) {$\zeta$};
\node[below] at (-1,-0.5) {$\eta$};
\node[above] at (1.1,-1) {$\omega$};
\draw[very thick,white] (-2,-2) rectangle (2,2);
\draw[very thick,rounded corners,white] (-2,-2) rectangle (2,2);
\draw[very thick,rounded corners,boundarycolor,dashed] (-2,-2) rectangle (2,2);
\filldraw (0,1) circle (0.07);
\filldraw (-1,-0.5) circle (0.07);
\filldraw (1,-1) circle (0.07);
\end{tikzpicture}
}
\end{array}
\end{align}
\end{definition}
\begin{definition}[renormalization] All the local relations described above apply to a fixed domain wall structure and are sufficient to construct $Z(\Sigma)$. Now we describe a procedure which can modify the compound defect $\Sigma$, which we call {\em renormalization}. Choose an open ball in $\Sigma$. For example it might look like:
\begin{align} \label{eq:compound_defect_ref}
	 C =\begin{array}{c}
	\includeTikz{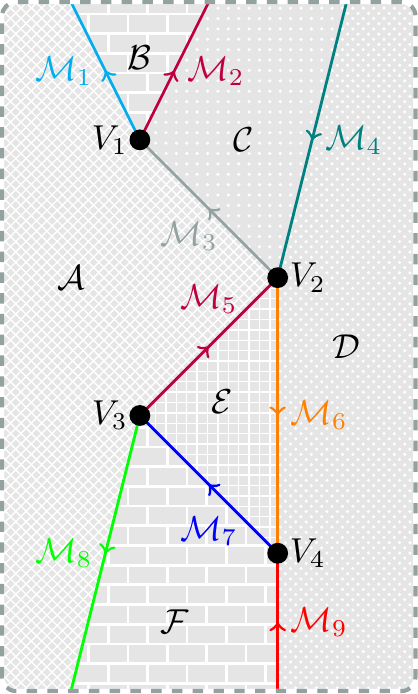}{
		\begin{tikzpicture}[scale=0.7,every node/.style={scale=1}]
    \path[fill=bulkcolor,postaction={pattern=crosshatch,pattern color=white}] (0,0) -- (1,0) -- (2,4) -- (4,6) -- (2,8) -- (1,10) -- (0,10) -- cycle;
    \path[fill=bulkcolor,postaction={pattern=grid,pattern color=white}] (4,2) -- (4,6) -- (2,4) -- cycle;
    \path[fill=bulkcolor,postaction={pattern=bricks,pattern color=white}] (1,10) -- (3,10) -- (2,8) -- cycle;
\path[fill=bulkcolor,postaction={pattern=crosshatch dots,pattern color=white}] (6,0) -- (6,10) -- (5,10) -- (4,6) -- (4,2) -- (4,0) -- cycle;
\path[fill=bulkcolor,postaction={pattern=bricks,pattern color=white}] (1,0) -- (4,0) -- (4,2) -- (2,4) -- cycle;
\path[fill=bulkcolor,postaction={pattern=dots,pattern color=white}] (3,10) -- (5,10) -- (4,6) -- (2,8) -- cycle;
\begin{scope}[thick,decoration={markings,mark=at position 0.5 with {\arrow{>}}}]
    \draw[red,postaction={decorate}] (4,0) -- (4,2) node[midway,right] {$\cat{M}_9$};
		\draw[blue,postaction={decorate}] (4,2) -- (2,4) node[midway,below,yshift=-2mm] {$\cat{M}_7$};
		\draw[green,postaction={decorate}]  (2,4) -- (1,0) node[midway,left] {$\cat{M}_8$};
		\draw[orange,postaction={decorate}] (4,6) -- (4,2) node[midway, right] {$\cat{M}_6$};
		\draw[purple,postaction={decorate}] (2,4) -- (4,6) node[midway, above,yshift=2mm] {$\cat{M}_5$};
		\draw[teal,postaction={decorate}] (5,10) -- (4,6) node[midway, right] {$\cat{M}_4$};
		\draw[boundarycolor,postaction={decorate}] (4,6) -- (2,8) node[midway, below,xshift=-2mm] {$\cat{M}_3$};
		\draw[cyan,postaction={decorate}] (2,8) -- (1,10) node[midway,left] {$\cat{M}_1$};
		\draw[purple,postaction={decorate}] (2,8) -- (3,10) node[midway, right] {$\cat{M}_2$};
    \end{scope}
		\filldraw (4,2) circle (4pt) node[right]{$V_4$};
		\filldraw (2,4) circle (4pt) node[left]{$V_3$};
		\filldraw (4,6) circle (4pt) node[right]{$V_2$};
		\filldraw (2,8) circle (4pt) node[left]{$V_1$};
		\node at (1,6) {$\cat{A}$};
		\node at (2.5,1) {$\cat{F}$};
		\node at (3.2,4.2) {$\cat{E}$};
		\node at (3.5,8) {$\cat{C}$};
		\node at (2,9.2) {$\cat{B}$};
		\node at (5,5) {$\cat{D}$};
    \draw[white,very thick] (0,0) rectangle (6,10);
    \draw[white,very thick,rounded corners] (0,0) rectangle (6,10);
		\draw[boundarycolor,very thick,dashed,rounded corners] (0,0) rectangle (6,10);
		\end{tikzpicture}
	}\end{array}
\end{align}
The vector space $Z(C)$ forms a representation $W$ of the boundary annular category. Define
\begin{align} C' =  
\begin{array}{c}
\includeTikz{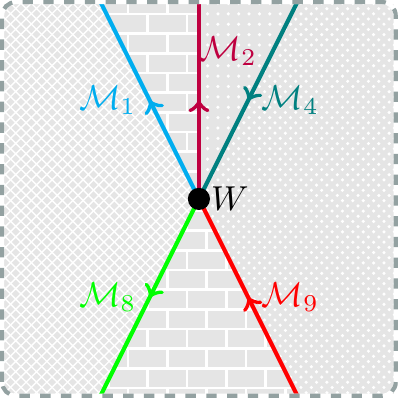}{
\begin{tikzpicture}
\path[fill=bulkcolor,postaction={pattern=crosshatch,pattern color=white}] (-2,-2) -- (-1,-2) -- (0,0) -- (-1,2) -- (-2,2) --  cycle;
\path[fill=bulkcolor,postaction={pattern=bricks,pattern color=white}] (0,0) -- (-1,2) -- (0,2) -- cycle;
\path[fill=bulkcolor,postaction={pattern=dots,pattern color=white}] (0,0) -- (0,2) -- (1,2) -- cycle;
\path[fill=bulkcolor,postaction={pattern=crosshatch dots,pattern color=white}] (0,0) -- (1,2) -- (2,2) -- (2,-2) -- (1,-2) -- cycle;
\path[fill=bulkcolor,postaction={pattern=bricks,pattern color=white}] (-1,-2) -- (0,0) -- (1,-2) -- cycle;
\begin{scope}[very thick,decoration={markings,mark=at position 0.5 with {\arrow{>}}}]
\draw[cyan,postaction={decorate}] (0,0) -- (-1,2) node[midway,left] {$\cat{M}_1$};
\draw[purple,postaction={decorate}] (0,0) -- (0,2);
\draw[teal,postaction={decorate}] (1,2) -- (0,0) node[midway, right] {$\cat{M}_4$};
\draw[green,postaction={decorate}]  (0,0) -- (-1,-2) node[midway,left] {$\cat{M}_8$};
\draw[red,postaction={decorate}] (1,-2) -- (0,0) node[midway,right] {$\cat{M}_9$};
\end{scope}
\node[purple] at (0.3,1.5) {$\cat{M}_2$};
\filldraw (0,0) circle (3pt) node[right]{$W$};
\draw[very thick,white] (-2,-2) rectangle (2,2);
\draw[very thick,rounded corners,white] (-2,-2) rectangle (2,2);
\draw[very thick,rounded corners,boundarycolor,dashed] (-2,-2) rectangle (2,2);
\end{tikzpicture}
}
\end{array}
\end{align}
Renormalization modifies $\Sigma$ by replacing $C$ with $C'$ and applies the annular category intertwining isomorphism $Z(C) \to Z(C')$ to the skein vectors on $C$. This is the justification for calling annular categories {\em ``local observable algebras''}. If there is an isomorphism of annular category representations which sends one skein vector to another, then they are equivalent up to renormalization. 
\end{definition}
\begin{example}[Anyon braidings]
Let $V$ be a representation of the annular category from Example~\ref{ex:anyons}. The central idempotents in an annular category measure anyon charge. We have the following isomorphism of annular category representations because the left and right sides both have charge $V$:
\begin{align}
\begin{array}{c}
\includeTikz{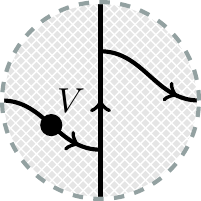}{
\begin{tikzpicture}
\draw[fill=bulkcolor,postaction={pattern=crosshatch,pattern color=white}] (0,0) circle (1);
\begin{scope}[very thick,decoration={markings,mark=at position 0.5 with {\arrow{>}}}]
\draw[postaction={decorate}] (0,-1) -- (0,1);
\end{scope}
\begin{scope}[very thick,decoration={markings,mark=at position 0.8 with {\arrow{>}}}]
\draw[postaction={decorate}] (-1,0) to[out=0,in=180] (0,-0.5);
\draw[postaction={decorate}] (0,0.5) to[out=0,in=180] (1,0);
\end{scope}
\filldraw (-0.5,-0.25) circle (3pt);
\node at (-0.3,0) {$V$};
\draw[very thick,white] (0,0) circle (1);
\draw[very thick,boundarycolor,dashed] (0,0) circle (1);
\end{tikzpicture}
}
\end{array} \cong
\begin{array}{c}
\includeTikz{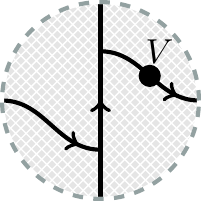}{
\begin{tikzpicture}
\draw[fill=bulkcolor,postaction={pattern=crosshatch,pattern color=white}] (0,0) circle (1);
\begin{scope}[very thick,decoration={markings,mark=at position 0.5 with {\arrow{>}}}]
\draw[postaction={decorate}] (0,-1) -- (0,1);
\end{scope}
\begin{scope}[very thick,decoration={markings,mark=at position 0.8 with {\arrow{>}}}]
\draw[postaction={decorate}] (-1,0) to[out=0,in=180] (0,-0.5);
\draw[postaction={decorate}] (0,0.5) to[out=0,in=180] (1,0);
\end{scope}
\filldraw (0.5,0.25) circle (3pt);
\node at (0.6,0.5) {$V$};
\draw[very thick,white] (0,0) circle (1);
\draw[very thick,boundarycolor,dashed] (0,0) circle (1);
\end{tikzpicture}
}
\end{array}.
\end{align}
Applied to the vector in \eqref{eq:anyon_internal_state}, the isomorphism has the following form:
\begin{align}
\begin{array}{c}
\includeTikz{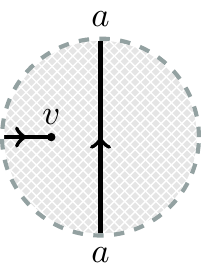}{
\begin{tikzpicture}
\draw[fill=bulkcolor,postaction={pattern=crosshatch,pattern color=white}] (0,0) circle (1);
\begin{scope}[very thick,decoration={markings,mark=at position 0.5 with {\arrow{>}}}]
\draw[postaction={decorate}] (0,-1) -- (0,1);
\draw[postaction={decorate}] (-1,0) -- (-0.5,0);
\end{scope}
\filldraw (-0.5,0) circle (1pt) node[above]{$v$};
\node at (0,-1.2) {$a$};
\node at (0,1.2) {$a$};
\draw[very thick,white] (0,0) circle (1);
\draw[very thick,boundarycolor,dashed] (0,0) circle (1);
\end{tikzpicture}
}
\end{array} \mapsto \sum_{w\alpha\beta} H_{w\alpha\beta}^{v}
\begin{array}{c}
\includeTikz{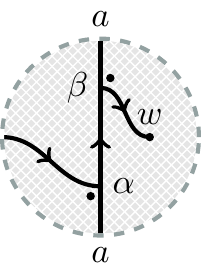}{
\begin{tikzpicture}
\draw[fill=bulkcolor,postaction={pattern=crosshatch,pattern color=white}] (0,0) circle (1);
\begin{scope}[very thick,decoration={markings,mark=at position 0.5 with {\arrow{>}}}]
\draw[postaction={decorate}] (0,-1) -- (0,1);
\draw[postaction={decorate}] (-1,0) to[out=0,in=180] (0,-0.5);
\draw[postaction={decorate}] (0,0.5) to[out=0,in=180] (0.5,0);
\end{scope}
\filldraw (0.5,0) circle (1pt) node[above]{$w$};
\node[right] at (0,-0.5) {$\alpha$};
\node[left] at (0,0.5) {$\beta$};
\node at (0,-1.2) {$a$};
\node at (0,1.2) {$a$};
\filldraw (0.1,0.6) circle (1pt);
\filldraw (-0.1,-0.6) circle (1pt);
\draw[very thick,white] (0,0) circle (1);
\draw[very thick,boundarycolor,dashed] (0,0) circle (1);
\end{tikzpicture}
}
\end{array}.
\end{align}
Notice that the internal state of the anyon can change as it passes through the $a$ string. 
\end{example}

\begin{definition}[Ladder category]\label{def:laddercat}
Consider the following domain wall structure:
\begin{align}
L=\begin{array}{c}
\includeTikz{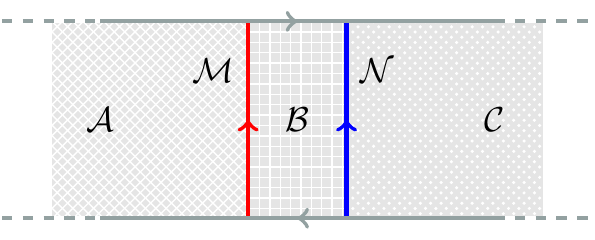}{
\begin{tikzpicture}
\draw[white,fill=bulkcolor,postaction={pattern=crosshatch,pattern color=white}] (-2.5,-1) rectangle (-0.5,1);
\draw[white,fill=bulkcolor,postaction={pattern=grid,pattern color=white}] (-0.5,-1) rectangle (0.5,1);
\draw[white,fill=bulkcolor,postaction={pattern=crosshatch dots,pattern color=white}] (0.5,-1) rectangle (2.5,1);
\begin{scope}[very thick,decoration={markings,mark=at position 0.5 with {\arrow{>}}}]
\draw[red,postaction={decorate}] (-0.5,-1) -- (-0.5,1);
\draw[blue,postaction={decorate}] (0.5,-1) -- (0.5,1);
\node[left] at (-0.5,0.5) {$\cat{M}$};
\node[right] at (0.5,0.5) {$\cat{N}$};
\node at (-2,0) {$\cat{A}$};
\node at (0,0) {$\cat{B}$};
\node at (2,0) {$\cat{C}$};
\draw[white] (2,1) -- (3,1);
\draw[white] (-3,1) -- (-2,1);
\draw[white] (2,-1) -- (3,-1);
\draw[white] (-3,-1) -- (-2,-1);
\draw[boundarycolor,postaction={decorate}] (-2,1) -- (2,1);
\draw[boundarycolor,dashed] (2,1) -- (3,1);
\draw[boundarycolor,dashed] (-3,1) -- (-2,1);
\draw[boundarycolor,dashed] (2,-1) -- (3,-1);
\draw[boundarycolor,dashed] (-3,-1) -- (-2,-1);
\draw[boundarycolor,postaction={decorate}] (2,-1) -- (-2,-1);
\end{scope}
\node at (0,1.1) {};
\node at (0,-1.1) {};
\end{tikzpicture}
}
\end{array}
\end{align}
The skein vectors in $Z(L)$ assemble into a category. The objects are pairs of objects $(m \in \cat{M},n \in \cat{N})$. The morphisms are skein vectors of the following form:
\begin{align} \label{eq:ladder_category_morphism}
\begin{array}{c}
\includeTikz{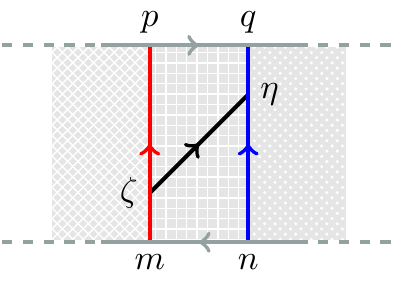}{
\begin{tikzpicture}
\draw[white,fill=bulkcolor,postaction={pattern=crosshatch,pattern color=white}] (-1.5,-1) rectangle (-0.5,1);
\draw[white,fill=bulkcolor,postaction={pattern=grid,pattern color=white}] (-0.5,-1) rectangle (0.5,1);
\draw[white,fill=bulkcolor,postaction={pattern=crosshatch dots,pattern color=white}] (0.5,-1) rectangle (1.5,1);
\begin{scope}[very thick,decoration={markings,mark=at position 0.5 with {\arrow{>}}}]
\draw[postaction={decorate}] (-0.5,-0.5) -- (0.5,0.5);
\draw[red,postaction={decorate}] (-0.5,-1) -- (-0.5,1);
\draw[blue,postaction={decorate}] (0.5,-1) -- (0.5,1);
\draw[boundarycolor,postaction={decorate}] (-1,1) -- (1,1);
\draw[white] (1,1) -- (2,1);
\draw[white] (-2,1) -- (-1,1);
\draw[white] (1,-1) -- (2,-1);
\draw[white] (-2,-1) -- (-1,-1);
\draw[boundarycolor,dashed] (1,1) -- (2,1);
\draw[boundarycolor,dashed] (-2,1) -- (-1,1);
\draw[boundarycolor,dashed] (1,-1) -- (2,-1);
\draw[boundarycolor,dashed] (-2,-1) -- (-1,-1);
\draw[boundarycolor,postaction={decorate}] (1,-1) -- (-1,-1);
\end{scope}
\node[below] at (-0.5,-1) {$m$};
\node[below] at (0.5,-1) {$n$};
\node[above] at (-0.5,1) {$p$};
\node[above] at (0.5,1) {$q$};
\node[left] at (-0.5,-0.5) {$\zeta$};
\node[right] at (0.5,0.5) {$\eta$};
\end{tikzpicture}
}
\end{array}
\end{align}
Composition is given by vertical gluing of the skein vectors and then reducing back to the form in \eqref{eq:ladder_category_morphism} using local relations. The Karoubi envelope of $Z(L)$ is equivalent to the bimodule category $\cat{M} \otimes_{\cat{B}} \cat{N}$. As explained in Definition~\ref{def:karoubi_envelope}, objects in the Karoubi envelope are idempotents
\begin{align}
i=\sum_{\zeta\eta}i_{\zeta\eta}
\begin{array}{c}
\includeTikz{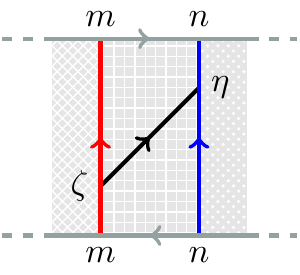}{
\begin{tikzpicture}
\draw[white,fill=bulkcolor,postaction={pattern=crosshatch,pattern color=white}] (-1,-1) rectangle (-0.5,1);
\draw[white,fill=bulkcolor,postaction={pattern=grid,pattern color=white}] (-0.5,-1) rectangle (0.5,1);
\draw[white,fill=bulkcolor,postaction={pattern=crosshatch dots,pattern color=white}] (0.5,-1) rectangle (1,1);
\begin{scope}[very thick,decoration={markings,mark=at position 0.5 with {\arrow{>}}}]
\draw[postaction={decorate}] (-0.5,-0.5) -- (0.5,0.5);
\draw[red,postaction={decorate}] (-0.5,-1) -- (-0.5,1);
\draw[blue,postaction={decorate}] (0.5,-1) -- (0.5,1);
\draw[boundarycolor,postaction={decorate}] (-1,1) -- (1,1);
\draw[white] (1,1) -- (1.5,1);
\draw[white] (-1.5,1) -- (-1,1);
\draw[white] (1,-1) -- (1.5,-1);
\draw[white] (-1.5,-1) -- (-1,-1);
\draw[boundarycolor,dashed] (1,1) -- (1.5,1);
\draw[boundarycolor,dashed] (-1.5,1) -- (-1,1);
\draw[boundarycolor,dashed] (1,-1) -- (1.5,-1);
\draw[boundarycolor,dashed] (-1.5,-1) -- (-1,-1);
\draw[boundarycolor,postaction={decorate}] (1,-1) -- (-1,-1);
\end{scope}
\node[below] at (-0.5,-1) {$m$};
\node[below] at (0.5,-1) {$n$};
\node[above] at (-0.5,1) {$m$};
\node[above] at (0.5,1) {$n$};
\node[left] at (-0.5,-0.5) {$\zeta$};
\node[right] at (0.5,0.5) {$\eta$};
\end{tikzpicture}
}
\end{array} \quad
j=\sum_{\theta\phi}j_{\theta\phi}
\begin{array}{c}
\includeTikz{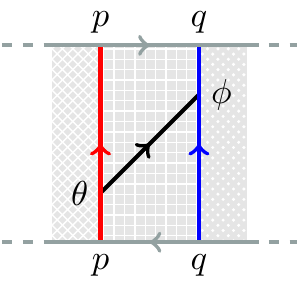}{
\begin{tikzpicture}
\draw[white,fill=bulkcolor,postaction={pattern=crosshatch,pattern color=white}] (-1,-1) rectangle (-0.5,1);
\draw[white,fill=bulkcolor,postaction={pattern=grid,pattern color=white}] (-0.5,-1) rectangle (0.5,1);
\draw[white,fill=bulkcolor,postaction={pattern=crosshatch dots,pattern color=white}] (0.5,-1) rectangle (1,1);
\begin{scope}[very thick,decoration={markings,mark=at position 0.5 with {\arrow{>}}}]
\draw[postaction={decorate}] (-0.5,-0.5) -- (0.5,0.5);
\draw[red,postaction={decorate}] (-0.5,-1) -- (-0.5,1);
\draw[blue,postaction={decorate}] (0.5,-1) -- (0.5,1);
\draw[boundarycolor,postaction={decorate}] (-1,1) -- (1,1);
\draw[white] (1,1) -- (1.5,1);
\draw[white] (-1.5,1) -- (-1,1);
\draw[white] (1,-1) -- (1.5,-1);
\draw[white] (-1.5,-1) -- (-1,-1);
\draw[boundarycolor,dashed] (1,1) -- (1.5,1);
\draw[boundarycolor,dashed] (-1.5,1) -- (-1,1);
\draw[boundarycolor,dashed] (1,-1) -- (1.5,-1);
\draw[boundarycolor,dashed] (-1.5,-1) -- (-1,-1);
\draw[boundarycolor,postaction={decorate}] (1,-1) -- (-1,-1);
\end{scope}
\node[below] at (-0.5,-1) {$p$};
\node[below] at (0.5,-1) {$q$};
\node[above] at (-0.5,1) {$p$};
\node[above] at (0.5,1) {$q$};
\node[left] at (-0.5,-0.5) {$\theta$};
\node[right] at (0.5,0.5) {$\phi$};
\end{tikzpicture}
}
\end{array}.
\end{align}
Left trivalent vertices in the Karoubi envelope are skein vectors
\begin{align} \label{eq:inflated_vertex}
\begin{array}{c}
\includeTikz{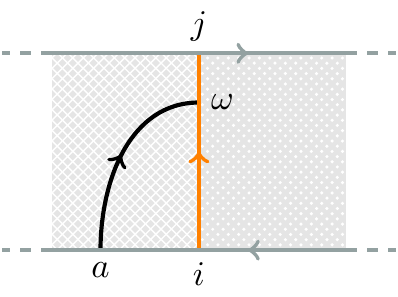}{
\begin{tikzpicture}
\draw[white,fill=bulkcolor,postaction={pattern=crosshatch,pattern color=white}] (-2,-1) rectangle (-0.5,1);
\draw[white,fill=bulkcolor,postaction={pattern=crosshatch dots,pattern color=white}] (-0.5,-1) rectangle (1,1);
\begin{scope}[very thick,decoration={markings,mark=at position 0.5 with {\arrow{>}}}]
\draw[postaction={decorate}] (-1.5,-1) to[out=90,in=180] (-0.5,0.5);
\draw[orange,postaction={decorate}] (-0.5,-1) -- (-0.5,1);
\draw[boundarycolor,postaction={decorate}] (-1,1) -- (1,1);
\draw[boundarycolor] (-2,1) -- (-1,1);
\draw[boundarycolor,dashed] (1,1) -- (1.5,1);
\draw[boundarycolor,dashed] (-2,1) -- (-2.5,1);
\draw[boundarycolor,dashed] (1,-1) -- (1.5,-1);
\draw[boundarycolor,dashed] (-2,-1) -- (-2.5,-1);
\draw[boundarycolor,postaction={decorate}] (1,-1) -- (-1,-1);
\draw[boundarycolor] (-1,-1) -- (-2,-1);
\end{scope}
\node[below] at (-0.5,-1) {$i$};
\node[above] at (-0.5,1) {$j$};
\node[right] at (-0.5,0.5) {$\omega$};
\node[below] at (-1.5,-1) {$a$};
\end{tikzpicture}
}
\end{array}
:=\sum_{\alpha\beta\gamma} l_{\alpha\beta\gamma}
\begin{array}{c}
\includeTikz{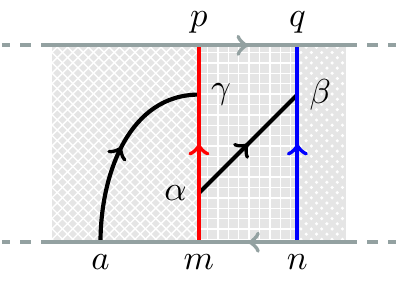}{
\begin{tikzpicture}
\draw[white,fill=bulkcolor,postaction={pattern=crosshatch,pattern color=white}] (-2,-1) rectangle (-0.5,1);
\draw[white,fill=bulkcolor,postaction={pattern=grid,pattern color=white}] (-0.5,-1) rectangle (0.5,1);
\draw[white,fill=bulkcolor,postaction={pattern=crosshatch dots,pattern color=white}] (0.5,-1) rectangle (1,1);
\begin{scope}[very thick,decoration={markings,mark=at position 0.5 with {\arrow{>}}}]
\draw[postaction={decorate}] (-0.5,-0.5) -- (0.5,0.5);
\draw[postaction={decorate}] (-1.5,-1) to[out=90,in=180] (-0.5,0.5);
\draw[red,postaction={decorate}] (-0.5,-1) -- (-0.5,1);
\draw[blue,postaction={decorate}] (0.5,-1) -- (0.5,1);
\draw[boundarycolor,postaction={decorate}] (-1,1) -- (1,1);
\draw[boundarycolor] (-2,1) -- (-1,1);
\draw[boundarycolor,dashed] (1,1) -- (1.5,1);
\draw[boundarycolor,dashed] (-2,1) -- (-2.5,1);
\draw[boundarycolor,dashed] (1,-1) -- (1.5,-1);
\draw[boundarycolor,dashed] (-2,-1) -- (-2.5,-1);
\draw[boundarycolor,postaction={decorate}] (1,-1) -- (-1,-1);
\draw[boundarycolor] (-1,-1) -- (-2,-1);
\end{scope}
\node[below] at (-0.5,-1) {$m$};
\node[below] at (0.5,-1) {$n$};
\node[above] at (-0.5,1) {$p$};
\node[above] at (0.5,1) {$q$};
\node[left] at (-0.5,-0.5) {$\alpha$};
\node[right] at (0.5,0.5) {$\beta$};
\node[right] at (-0.5,0.5) {$\gamma$};
\node[below] at (-1.5,-1) {$a$};
\end{tikzpicture}
}
\end{array}
\end{align}
which absorb the idempotent $j$ on the top and the idempotent
\begin{align}
\sum_{\zeta\eta} i_{\zeta\eta}
\begin{array}{c}
\includeTikz{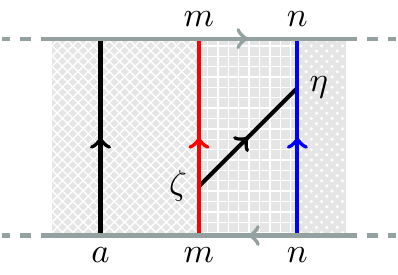}{
\begin{tikzpicture}
\draw[white,fill=bulkcolor,postaction={pattern=crosshatch,pattern color=white}] (-2,-1) rectangle (-0.5,1);
\draw[white,fill=bulkcolor,postaction={pattern=grid,pattern color=white}] (-0.5,-1) rectangle (0.5,1);
\draw[white,fill=bulkcolor,postaction={pattern=crosshatch dots,pattern color=white}] (0.5,-1) rectangle (1,1);
\begin{scope}[very thick,decoration={markings,mark=at position 0.5 with {\arrow{>}}}]
\draw[postaction={decorate}] (-0.5,-0.5) -- (0.5,0.5);
\draw[postaction={decorate}] (-1.5,-1) -- (-1.5,1);
\draw[red,postaction={decorate}] (-0.5,-1) -- (-0.5,1);
\draw[blue,postaction={decorate}] (0.5,-1) -- (0.5,1);
\draw[boundarycolor,postaction={decorate}] (-1,1) -- (1,1);
\draw[boundarycolor] (-2,1) -- (-1,1);
\draw[boundarycolor,dashed] (1,1) -- (1.5,1);
\draw[boundarycolor,dashed] (-2,1) -- (-2.5,1);
\draw[boundarycolor,dashed] (1,-1) -- (1.5,-1);
\draw[boundarycolor,dashed] (-2,-1) -- (-2.5,-1);
\draw[boundarycolor,postaction={decorate}] (1,-1) -- (-1,-1);
\draw[boundarycolor] (-1,-1) -- (-2,-1);
\end{scope}
\node[below] at (-0.5,-1) {$m$};
\node[below] at (0.5,-1) {$n$};
\node[above] at (-0.5,1) {$m$};
\node[above] at (0.5,1) {$n$};
\node[left] at (-0.5,-0.5) {$\zeta$};
\node[right] at (0.5,0.5) {$\eta$};
\node[below] at (-1.5,-1) {$a$};
\end{tikzpicture}
}
\end{array}
\end{align}
on the bottom. Right vertices are defined analogously. Computing these Karoubi envelope trivalent vertices is a linear algebra problem, and the solutions form a vector space. If we also require that the vertices \eqref{eq:inflated_vertex} have left, center and right associators matching bimodules from our original list, then we cut the solution space down to a finite set, which may have more than one element. When we talk about vertices in the Karoubi envelope of a ladder category, we will always assume that they have been chosen to have associators matching our original models. Computing a complete set of trivalent vertices in the Karoubi envelope of the ladder category gives us the decompositions $\cat{M} \otimes_{\cat{B}} \cat{N} \cong \oplus_i \cat{P}_i$ which are tabulated in the Brauer-Picard tables.
\end{definition}

\begin{definition}[Inflation trick] \label{def:inflation_trick}
We are interested in computing explicit representations of annular categories with three bimodule strings. This is a computationally hard problem even for annular categories with two bimodule strings. Moreover, we have isomorphisms
\begin{align}
\begin{array}{c}
\includeTikz{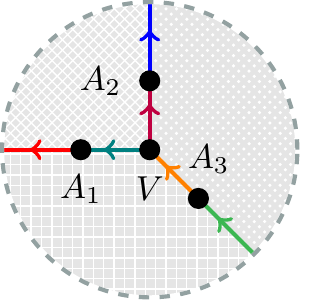}{
\begin{tikzpicture}
\path[fill=bulkcolor,postaction={pattern=crosshatch,pattern color=white}] (0,0) -- (0,1.5) arc(90:180:1.5) (-1.5,0) -- cycle;
\path[fill=bulkcolor,postaction={pattern=grid,pattern color=white}] (0,0) -- (-1.5,0) arc(180:270+45:1.5) (270+45:1.5) -- cycle;
\path[fill=bulkcolor,postaction={pattern=crosshatch dots,pattern color=white}] (0,0) -- (270+45:1.5) arc(-45:90:1.5) (0,1.5) -- cycle;
\begin{scope}[very thick,decoration={markings,mark=at position 0.5 with {\arrow{<}}}]
\draw[teal,postaction={decorate}] (-0.7,0) -- (0,0);
\draw[red,postaction={decorate}] (-1.5,0) -- (-0.7,0);
\draw[blue,postaction={decorate}] (0,1.5) -- (0,0.7);
\draw[purple,postaction={decorate}] (0,0.7) -- (0,0);
\draw[orange,postaction={decorate}] (0,0) -- (0.7*0.707107, -0.7*0.707107);
\draw[nicegreen,postaction={decorate}] (0.7*0.707107, -0.7*0.707107) -- (1.5*0.707107, -1.5*0.707107);
\draw[white] (0,0) circle (1.5);
\draw[boundarycolor,dashed] (0,0) circle (1.5);
\end{scope}
\filldraw (0,0) circle (0.1);
\filldraw (-0.7,0) circle (0.1);
\filldraw (0,0.7) circle (0.1);
\filldraw (0.7*0.707107, -0.7*0.707107) circle (0.1);
\node at (0,-0.4) {$V$};
\node at (-0.7,-0.4) {$A_1$};
\node at (0.6,-0.1) {$A_3$};
\node at (-0.5,0.7) {$A_2$};
\end{tikzpicture}
}
\end{array} \cong
\begin{array}{c}
\includeTikz{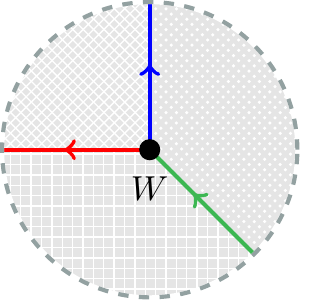}{
\begin{tikzpicture}
\path[fill=bulkcolor,postaction={pattern=crosshatch,pattern color=white}] (0,0) -- (0,1.5) arc(90:180:1.5) (-1.5,0) -- cycle;
\path[fill=bulkcolor,postaction={pattern=grid,pattern color=white}] (0,0) -- (-1.5,0) arc(180:270+45:1.5) (270+45:1.5) -- cycle;
\path[fill=bulkcolor,postaction={pattern=crosshatch dots,pattern color=white}] (0,0) -- (270+45:1.5) arc(-45:90:1.5) (0,1.5) -- cycle;
\begin{scope}[very thick,decoration={markings,mark=at position 0.5 with {\arrow{<}}}]
\draw[red,postaction={decorate}] (-1.5,0) -- (0,0);
\draw[blue,postaction={decorate}] (0,1.5) -- (0,0);
\draw[nicegreen,postaction={decorate}] (0,0) -- (1.5*0.707107, -1.5*0.707107);
\draw[white] (0,0) circle (1.5);
\draw[boundarycolor,dashed] (0,0) circle (1.5);
\end{scope}
\filldraw (0,0) circle (0.1);
\node at (0,-0.4) {$W$};
\end{tikzpicture}
}
\end{array}
\end{align}
which gives us an equivalence relation on three bimodule string annular category representations. The {\em inflation trick} produces one representation from each equivalence class without needing to compute the representation theory of algebras with large dimension. It works as follows: If $\cat{A} \curvearrowright \cat{M} \curvearrowleft \cat{B}$ is a bimodule, then defects interfacing between $\cat{M}$ and itself form a (multi) fusion category ${\rm End}(\cat{M})$:
\begin{align}
\begin{array}{c}
\includeTikz{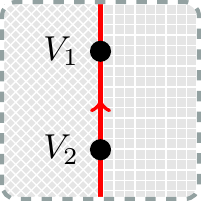}{
\begin{tikzpicture}
\draw[white,fill=bulkcolor,postaction={pattern=crosshatch,pattern color=white}] (-1,-1) rectangle (0,1);
\draw[white,fill=bulkcolor,postaction={pattern=grid,pattern color=white}] (0,-1) rectangle (1,1);
\begin{scope}[very thick,decoration={markings,mark=at position 0.5 with {\arrow{>}}}]
\draw[red,postaction={decorate}] (0,-1) -- (0,1);
\end{scope}
\filldraw (0,0.5) circle (0.1);
\filldraw (0,-0.5) circle (0.1);
\node at (-0.4,0.5) {$V_1$};
\node at (-0.4,-0.5) {$V_2$};
\draw[white,very thick] (-1,-1) rectangle (1,1);
\draw[rounded corners,white,very thick] (-1,-1) rectangle (1,1);
\draw[rounded corners,boundarycolor,very thick,dashed] (-1,-1) rectangle (1,1);
\end{tikzpicture}
}
\end{array} \cong \bigoplus_i
\begin{array}{c}
\includeTikz{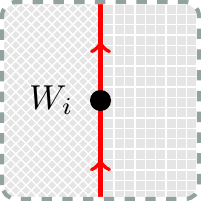}{
\begin{tikzpicture}
\draw[white,fill=bulkcolor,postaction={pattern=crosshatch,pattern color=white}] (-1,-1) rectangle (0,1);
\draw[white,fill=bulkcolor,postaction={pattern=grid,pattern color=white}] (0,-1) rectangle (1,1);
\begin{scope}[very thick,decoration={markings,mark=at position 0.2 with {\arrow{>}},mark=at position 0.8 with {\arrow{>}}}]
\draw[red,postaction={decorate}] (0,-1) -- (0,1);
\end{scope}
\node at (-0.5,0) {$W_i$};
\filldraw (0,0) circle (0.1);
\draw[white,very thick] (-1,-1) rectangle (1,1);
\draw[rounded corners,white,very thick] (-1,-1) rectangle (1,1);
\draw[rounded corners,boundarycolor,very thick,dashed] (-1,-1) rectangle (1,1);
\end{tikzpicture}
}
\end{array}
\end{align}
If $\cat{M}$ is indecomposable, then ${\rm End}(\cat{M})$ is a fusion category Morita equivalent to $\cat{A} \otimes \cat{B}^{\rm rev}$. In particular, ${\rm End}(\cat{M})$ has a unit object, represented as an idempotent:
\begin{align} \sum_{\zeta\eta\theta\phi} E_{\zeta\eta\theta\phi}
\begin{array}{c}
\includeTikz{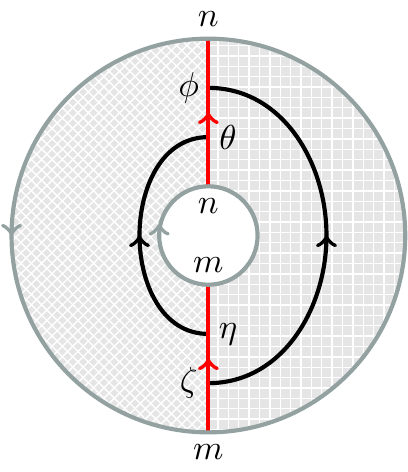}{
\begin{tikzpicture}
\path[fill=bulkcolor,postaction={pattern=crosshatch,pattern color=white}] (0,2) arc(90:270:2) (0,-2) -- cycle;
\path[fill=bulkcolor,postaction={pattern=grid,pattern color=white}] (0,-2) arc(-90:90:2) (0,2) -- cycle;
\begin{scope}[very thick,decoration={markings,mark=at position 0.5 with {\arrow{>}}}]
\draw[postaction={decorate}] (0,-1) to[out=180,in=270] (-0.7,0) to[out=90,in=180] (0,1);
\draw[postaction={decorate}] (0,-1.5) to[out=0,in=270] (1.2,0) to[out=90,in=0] (0,1.5);
\draw[red,postaction={decorate}]  (0,-2) -- (0,-0.5);
\draw[red,postaction={decorate}]  (0,0.5) -- (0,2);
\draw[boundarycolor,postaction={decorate}] circle (2);
\end{scope}
\node at (-2,0) {};
\begin{scope}[very thick,decoration={markings,mark=at position 0.5 with {\arrow{<}}}]
\draw[boundarycolor,postaction={decorate},fill=white] circle (0.5);
\end{scope}
\node at (0,-2.2) {$m$};
\node at (0,2.2) {$n$};
\node at (0,-0.3) {$m$};
\node at (0,0.3) {$n$};
\node at (-0.2,-1.5) {$\zeta$};
\node at (0.2,-1) {$\eta$};
\node at (-0.2,1.5) {$\phi$};
\node at (0.2,1) {$\theta$};
\end{tikzpicture}
}
\end{array}
\end{align}
If we replace the trivalent vertices on the top (bottom) half of the idempotent with trivalent vertices in the Karoubi envelope of a ladder category corresponding to an embedding $\cat{M} \hookrightarrow \cat{P} \otimes_{\cat{B}} \cat{Q}$, then we get an idempotent in a 3 bimodule string annular category which encodes the functor $\cat{M} \hookrightarrow \cat{P} \otimes_{\cat{B}} \cat{Q}$.
\end{definition}

\begin{definition}[Associators for vertical defect fusion] \label{def:vertical_defect_fusion_associator}
We shall denote annular category intertwining maps
\begin{align} \Theta :
\begin{array}{c}
\includeTikz{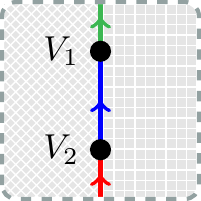}{
\begin{tikzpicture}
\draw[white,fill=bulkcolor,postaction={pattern=crosshatch,pattern color=white}] (-1,-1) rectangle (0,1);
\draw[white,fill=bulkcolor,postaction={pattern=grid,pattern color=white}] (0,-1) rectangle (1,1);
\begin{scope}[very thick,decoration={markings,mark=at position 0.5 with {\arrow{>}}}]
\draw[red,postaction={decorate}] (0,-1) -- (0,-0.5);
\draw[blue,postaction={decorate}] (0,-0.5) -- (0,0.5);
\end{scope}
\begin{scope}[very thick,decoration={markings,mark=at position 0.7 with {\arrow{>}}}]
\draw[nicegreen,postaction={decorate}] (0,0.5) -- (0,1);
\end{scope}
\filldraw (0,0.5) circle (0.1);
\filldraw (0,-0.5) circle (0.1);
\node at (-0.4,0.5) {$V_1$};
\node at (-0.4,-0.5) {$V_2$};
\draw[white,very thick] (-1,-1) rectangle (1,1);
\draw[rounded corners,white,very thick] (-1,-1) rectangle (1,1);
\draw[rounded corners,boundarycolor,very thick,dashed] (-1,-1) rectangle (1,1);
\end{tikzpicture}
}
\end{array} \to
\begin{array}{c}
\includeTikz{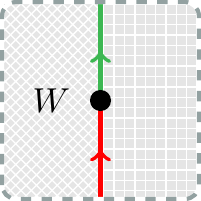}{
\begin{tikzpicture}
\draw[white,fill=bulkcolor,postaction={pattern=crosshatch,pattern color=white}] (-1,-1) rectangle (0,1);
\draw[white,fill=bulkcolor,postaction={pattern=grid,pattern color=white}] (0,-1) rectangle (1,1);
\begin{scope}[very thick,decoration={markings,mark=at position 0.5 with {\arrow{>}}}]
\draw[red,postaction={decorate}] (0,-1) -- (0,0);
\draw[nicegreen,postaction={decorate}] (0,0) -- (0,1);
\end{scope}
\node at (-0.5,0) {$W$};
\filldraw (0,0) circle (0.1);
\draw[white,very thick] (-1,-1) rectangle (1,1);
\draw[rounded corners,white,very thick] (-1,-1) rectangle (1,1);
\draw[rounded corners,boundarycolor,very thick,dashed] (-1,-1) rectangle (1,1);
\end{tikzpicture}
}
\end{array}
\end{align}
using squiggly trivalent vertices:
\begin{align} 
\begin{array}{c}
\includeTikz{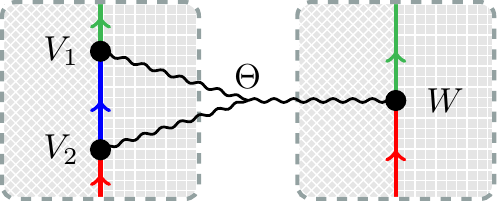}{
\begin{tikzpicture}
\draw[white,fill=bulkcolor,postaction={pattern=crosshatch,pattern color=white}] (-1,-1) rectangle (0,1);
\draw[white,fill=bulkcolor,postaction={pattern=grid,pattern color=white}] (0,-1) rectangle (1,1);
\begin{scope}[very thick,decoration={markings,mark=at position 0.5 with {\arrow{>}}}]
\draw[red,postaction={decorate}] (0,-1) -- (0,-0.5);
\draw[blue,postaction={decorate}] (0,-0.5) -- (0,0.5);
\end{scope}
\begin{scope}[very thick,decoration={markings,mark=at position 0.7 with {\arrow{>}}}]
\draw[nicegreen,postaction={decorate}] (0,0.5) -- (0,1);
\end{scope}
\filldraw (0,0.5) circle (0.1);
\filldraw (0,-0.5) circle (0.1);
\node at (-0.4,0.5) {$V_1$};
\node at (-0.4,-0.5) {$V_2$};
\draw[white,very thick] (-1,-1) rectangle (1,1);
\draw[rounded corners,white,very thick] (-1,-1) rectangle (1,1);
\draw[rounded corners,boundarycolor,very thick,dashed] (-1,-1) rectangle (1,1);
\draw[white,fill=bulkcolor,postaction={pattern=crosshatch,pattern color=white}] (-1+3,-1) rectangle (0+3,1);
\draw[white,fill=bulkcolor,postaction={pattern=grid,pattern color=white}] (0+3,-1) rectangle (1+3,1);
\begin{scope}[very thick,decoration={markings,mark=at position 0.5 with {\arrow{>}}}]
\draw[red,postaction={decorate}] (0+3,-1) -- (0+3,0);
\draw[nicegreen,postaction={decorate}] (0+3,0) -- (0+3,1);
\end{scope}
\node at (0.5+3,0) {$W$};
\filldraw (0+3,0) circle (0.1);
\draw[white,very thick] (-1+3,-1) rectangle (1+3,1);
\draw[rounded corners,white,very thick] (-1+3,-1) rectangle (1+3,1);
\draw[rounded corners,boundarycolor,very thick,dashed] (-1+3,-1) rectangle (1+3,1);
\begin{scope}[thick,decoration={snake, segment length=2mm,amplitude=0.2mm}]
\draw[decorate]  (0,-0.5) -- (1.5,0);
\draw[decorate] (0,0.5) -- (1.5,0);
\draw[decorate] (1.5,0) -- (3,0);
\end{scope}
\node[above] at (1.5,0) {$\Theta$};
\end{tikzpicture}
}
\end{array}
\end{align}
The squiggly lines represent a renormalization process. The change of basis matrix $F_{\Theta\Phi}^{\Xi\Psi}$ between the following intertwining maps
\begin{subequations}
\begin{align}
%
\label{eqn:verticalB}
\end{align}
\end{subequations}
is the associator for vertical defect fusion. This allows us to compute the associator for ${\rm End}(\cat{M})$.
\end{definition}

\clearpage

\clearpage

\section{Domain wall definitions} \label{appendix:domain_wall_definitions}

\renewcommand{\arraystretch}{1.3}

\begin{table}[h]\centering
	\begin{tabular}{!{\vrule width 1pt}>{\columncolor[gray]{.9}[\tabcolsep]}c!{\vrule width 1pt}c !{\color[gray]{.8}\vrule}  c !{\color[gray]{.8}\vrule}c!{\vrule width 1pt}}
		\toprule[1pt]
		\rowcolor[gray]{.9}[\tabcolsep]Name &Subgroup $\leqslant S_3$&\# inequiv. cocycles&Cocycle formula\\
		\toprule[1pt]
		$\mathcal{A}_1$&$\{0\}$&1&$+1$\\
		\toprule[1pt]
		$\mathcal{A}_2$&$\ZZ{3}\cong\langle(0,1)\rangle$&1&$+1$\\
		\toprule[1pt]
		$\mathcal{A}_3$&$\ZZ{2}\cong\langle(1,0)\rangle$&1&$+1$\\
		\toprule[1pt]
		$\mathcal{A}_4$&$S_3\cong\langle(1,0),(0,1)\rangle$&1&$+1$\\
		\toprule[1pt]
	\end{tabular}
  \caption{Defining data for $\vvec{}|\vvec{S_3}$ bimodules. Names $\mathcal{A}_i$ consistent with \onlinecite{PhysRevB.96.195129}}
\end{table}

\begin{table}[h]\centering
	%
  \caption{Defining data for $\vvec{\ZZ{2}}|\vvec{S_3}$ bimodules. Names $K_i$ consistent with \onlinecite{1811.06738}, with the left and right versions of each domain wall given the same name.}
\end{table}
\begin{table}[h!]\centering
	%
  \caption{Defining data for $\vvec{\vvec{\ZZ{3}}}|\vvec{S_3}$ bimodules. Names $Q_i$ chosen such that the left and right versions of each bimodule given the same name.}
\end{table}

\begin{table}[h]\centering
	\resizebox{\textwidth}{!}{
	%
}
	\caption{Defining data for $\vvec{S_3}|\vvec{S_3}$ bimodules}\label{tab:S3S3dws}
\end{table}

\clearpage

\section{Brauer-Picard tables (domain wall fusion)} \label{ap:domain_wall_fusions}

In this appendix, we provide complete tables of the fusion rules for bimodules between $\vvec{S_3}$ and $\vvec{}$, $\vvec{\ZZ{2}}$, $\vvec{\ZZ{3}}$, and $\vvec{S_3}$. 

\begin{table}[h]\centering
%
\caption{Domain wall fusion for $\vvec{S_3}|\vvec{}$ domain walls with $\vvec{}|\vvec{S_3}$ domain walls.}\label{tab:vacstrip}
\end{table}
Any $\vvec{S_3}$ bimodule $\cat{B}$ that occurs in Table~\ref{tab:vacstrip} is actually the composite of two module categories.
\begin{align}
%
.
\end{align}
Similar statements can be made for the remaining tables in this appendix. This allows us to identify bimodules that `factor through' a subcategory.

\begin{table}[h]\centering
%
}
\caption{Domain wall fusion tables for $\vvec{S_3}|\vvec{\ZZ{2}}$ domain walls with $\vvec{\ZZ{2}}|\vvec{S_3}$ domain walls. Red color indicates domain walls that do not factor through vacuum.}
\end{table}

\begin{table}\centering
%
}
\caption{Domain wall fusion tables for $\vvec{S_3}|\vvec{\ZZ{3}}$ domain walls with $\vvec{\ZZ{3}}|\vvec{S_3}$ domain walls. Blue color indicates domain walls that do factor through vacuum.}
\end{table}

\begin{sidewaystable}
\renewcommand{\arraystretch}{1.5}
\resizebox{\textwidth}{!}
{
	%
}
\caption{Domain wall fusions for $\vvec{S_3}\lvert\vvec{S_3}$ domain walls. Entries are colored according to their origin. Domain walls colored black factor as $\vvec{S_3}|\vvec{}|\vvec{S_3}$, red domain walls factor as $\vvec{S_3}|\vvec{\ZZ{2}}|\vvec{S_3}$, blue domain walls factor as $\vvec{S_3}|\vvec{\ZZ{3}}|\vvec{S_3}$, and green domain walls do not factor through a smaller phase.}\label{tab:S3BPR}
\end{sidewaystable}

\clearpage

\section{Vertical fusion of binary interface defects}

In this appendix, we study the fusion rings of ${\rm End}(\cat{M})$ for bimodules $\cat{C}\curvearrowright\cat{M}\curvearrowleft\vvec{S_3}$ that do not factor through a smaller fusion category.

\subsection{$\vvec{}|\vvec{S_3}$ (boundaries)}
The fusion categories ${\rm End}(\cat{A}_1)$ and ${\rm End}(\cat{A}_2)$ are equivalent to $\vvec{S_3}$. Using this equivalence, we denote the binary interface defects of $\mathcal{A}_1$ and $\mathcal{A}_2$ by elements of $S_3=\{(),(12),(13),(23),(123),(132)\}$. The fusion categories ${\rm End}(\cat{A}_3)$ and ${\rm End}(\cat{A}_4)$ are equivalent to $\rrep{S_3}$. Using these equivalences, we denote the binary interface defects of $\mathcal{A}_3$ and $\mathcal{A}_4$ by irreducible representations of $S_3$, labeled $1,\sigma,\psi$ for the trivial, sign and 2D representation.

\begin{table}[h!]\centering
	\resizebox{\textwidth}{!}
	{

	\caption{Vertical fusion of $\cat{A}_i$ binary interface defects}
\end{table}

\subsection{$\vvec{\ZZ{2}}|\vvec{S_3}$}

The endomorphism category of all 4 $\vvec{\ZZ{2}}|\vvec{S_3}$ domain walls is $\rrep{\ZZ{2}\times S_3}$. We identify the binary interface defects with irreducible representations of $Z_2\times S_3$ labeled $1,1^\prime,\sigma,\sigma^\prime,\psi,\psi^\prime$.
\begin{table}[h!]\centering
	%
	\caption{Vertical defect fusion for all four $\vvec{\ZZ{2}}|\vvec{S_3}$ domain walls.}
\end{table}
\clearpage
\subsection{$\vvec{\ZZ{3}}|\vvec{S_3}$}

\begin{table}[h!]\centering
%
	\caption{Vertical defect fusion for both $\vvec{\ZZ{3}}|\vvec{S_3}$ domain walls. This is the fusion ring of a $\ZZ{3}\times\ZZ{3}$ Tambara-Yamagami fusion category. Together with Appendix~\ref{appendix:endM}, this establishes Proposition~\ref{prop:TY}.}
\end{table}

\subsection{$\vvec{S_3}|\vvec{S_3}$}
\begin{table}[h!]\centering
	%
	\caption{Vertical fusion of $G_1$ binary interface defects. Notice that ${\rm End}(G_1)\cong Z(\vvec{S_3})$.}
\end{table}
\clearpage

\section{Fusing anyons into binary interface defects}\label{ap:BID}

In this appendix, we study the action of fusing anyons from the bulk into binary interface defects. 

\subsection{$\vvec{}|\vvec{S_3}$ (boundaries)}

\begin{table}[h!]\centering
	%
	\caption{Action of $S_3$ anyons on defects occurring at interfaces between boundary types (corners).}
\end{table}

The (right) action of the $\vvec{S_3}$ symmetry on the boundaries is
\begin{align}
%
.
\end{align}

\clearpage
\subsection{$\vvec{\ZZ{2}}|\vvec{S_3}$}

\begin{table}[h!]
	\centering
	%
	\caption{Action of anyons on defects occurring at the interface of $\vvec{\ZZ{2}}|\vvec{S_3}$ domain walls. The remaining $K_\cdot$ actions can be inferred using the symmetry actions of Eqn.~\ref{eqn:symmetryKs}}
\end{table}

Symmetry action on the domain walls
\begin{align}
%
.\label{eqn:symmetryKs}
\end{align}

\clearpage
\subsection{$\vvec{\ZZ{3}}|\vvec{S_3}$}
\vspace*{-5mm}
\begin{table}[h!]\centering
	%
	}
	\caption{Action of anyons on defects occurring at the interface of $\vvec{\ZZ{3}}|\vvec{S_3}$ domain walls $Q_8$ and $Q_{9,1}$. The remaining actions can be inferred using the symmetry actions of Eqn.~\ref{eqn:symmetryQs}}
\end{table}

Symmetry action on the domain walls
\begin{align}
%
.\label{eqn:symmetryQs}
\end{align}

\clearpage
\subsection{$\vvec{S_3}|\vvec{S_3}$}

\begin{table}[h!]\centering
%
	\caption{Action of anyons on defects occurring at the interface of the $\vvec{S_3}|\vvec{S_3}$ domain walls $G_1$ and $I$  (twists).}\label{tab:anyonontwist}
\end{table}

\clearpage

\clearpage

\section{Associator for ${\rm End}(Q_8)$}
\label{appendix:endM}

We now provide an example computation of the associator for objects in ${\rm End}(Q_8)$. Recall that the left category is $\vvec{\ZZ{3}}$ and the right category is $\vvec{S_3}$. A basis for the annular category representations is given by
\begin{align}
	%
,
\end{align}
where node labels indicate irreps, $a,y\in \{0,1\}$, $b,c,g,h,x,z\in \{0,1,2\}$. We remark that at all times, the left entry of the string label should be taken modulo 2 and the right entry modulo 3.

From these representations, we construct the compound defects associated to vertical fusion. By studying the annular action on these compound defects, we obtain the following decompositions into simple defects
\begin{subequations}
\begin{align}
%
.
\end{align}
\end{subequations}

Given these isomorphisms, the computation of the associator is straightforward. We illustrate for $(\sigma\circ\sigma)\circ\sigma=\sum_{(g_0,h_0),(g_1,h_1)}\left[F_{\sigma,\sigma,\sigma}^{\sigma}\right]_{(g_0,h_0),(g_1,h_1)}\sigma\circ(\sigma\circ\sigma)$.
\begin{align}
%
.
\end{align}
From this, we obtain
\begin{align}
\left[F_{\sigma,\sigma,\sigma}^{\sigma}\right]_{(g_0,h_0),(g_1,h_1)}\id_\sigma=
%
=\frac{\omega^{2 (g_1 h_0 + g_0 (h_0 + h_1))}}{3}\id_\sigma.
\end{align}
The remaining associator elements can be found in a similar way. After regauging the trivalent vertices, we obtain

\begin{subequations}
\begin{align}
\left[F_{(g_0,h_0),\sigma,(g_1,h_1)}^{\sigma}\right]_{\sigma,\sigma}&=\chi((g_0,h_0),(g_1,h_1))\\
\left[F_{\sigma,(g_0,h_0),\sigma}^{(g_1,h_1)}\right]_{\sigma,\sigma}&=\chi((g_0,h_0),(g_1,h_1))\\
\left[F_{\sigma,\sigma,\sigma}^{\sigma}\right]_{(g_0,h_0),(g_1,h_1)}&=\frac{\nu}{3 \chi((g_0,h_0),(g_1,h_1))},
\end{align}
\end{subequations}
with $\nu=1$ and $\chi()$ the symmetric bicharacter defined by
\begin{align}
\chi((g_0,h_0),(g_1,h_1))&=\omega^{g_0h_1+h_0g_1}.
\end{align}

\clearpage

\section{$\vvec{\ZZ{p}}$ bimodule data}

In this appendix, we reproduce the defining data for $\vvec{\ZZ{p}}$ bimodule categories. Additionally, we reproduce the Brauer-Picard tables for these bimodules and physical interpretations.

\vspace{5mm}


\begin{table}[h]\centering
	\begin{tabular}{!{\vrule width 1pt}>{\columncolor[gray]{.9}[\tabcolsep]}c!{\vrule width 1pt}c !{\color[gray]{.8}\vrule} c !{\color[gray]{.8}\vrule} c!{\vrule width 1pt}}
		\toprule[1pt]
		\rowcolor[gray]{.9}[\tabcolsep]Bimodule label & Left action & Right action & Associator \\
		\toprule[1pt]
		$T$ &
		$
		\begin{array}{c}
		\includeTikz{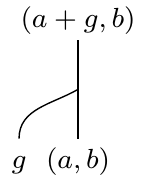}
		{
			\begin{tikzpicture}
			\Laction[$g$][$(a,b)$][$(a+g,b)$];
			\end{tikzpicture}
		}
		\end{array}
		$
		&
		$
		\begin{array}{c}
		\includeTikz{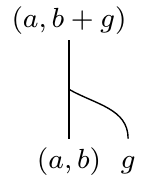}
		{
			\begin{tikzpicture}
			\Raction[$g$][$(a,b)$][$(a,b+g)$];
			\end{tikzpicture}
		}
		\end{array}
		$&
		$
		\begin{array}{c}
		\includeTikz{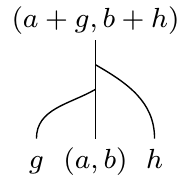}
		{
			\begin{tikzpicture}
			\Lassociator[$g$][$h$][$(a,b)$][$(a+g,b+h)$];
			\end{tikzpicture}
		}
		\end{array}
		=
		\begin{array}{c}
		\includeTikz{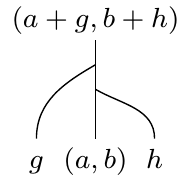}
		{
			\begin{tikzpicture}
			\Rassociator[$g$][$h$][$(a,b)$][$(a+g,b+h)$];
			\end{tikzpicture}
		}
		\end{array}
		$
		\\
		\greycline{2-3}
		$L$ &
		$
		\begin{array}{c}
		\includeTikz{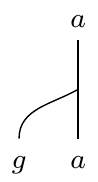}
		{
			\begin{tikzpicture}
			\Laction[$g$][$a$][$a$];
			\end{tikzpicture}
		}
		\end{array}
		$
		&
		$
		\begin{array}{c}
		\includeTikz{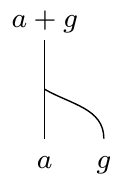}
		{
			\begin{tikzpicture}
			\Raction[$g$][$a$][$a+g$];
			\end{tikzpicture}
		}
		\end{array}
		$&
		$
		\begin{array}{c}
		\includeTikz{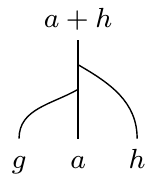}
		{
			\begin{tikzpicture}
			\Lassociator[$g$][$h$][$a$][$a+h$];
			\end{tikzpicture}
		}
		\end{array}
		=
		\begin{array}{c}
		\includeTikz{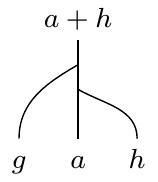}
		{
			\begin{tikzpicture}
			\Rassociator[$g$][$h$][$a$][$a+h$];
			\end{tikzpicture}
		}
		\end{array}
		$
		\\
		\greycline{2-3} 
		$R$ &
		$
		\begin{array}{c}
		\includeTikz{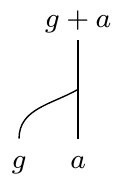}
		{
			\begin{tikzpicture}
			\Laction[$g$][$a$][$g+a$];
			\end{tikzpicture}
		}
		\end{array}
		$
		&
		$
		\begin{array}{c}
		\includeTikz{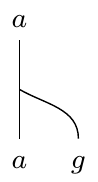}
		{
			\begin{tikzpicture}
			\Raction[$g$][$a$][$a$];
			\end{tikzpicture}
		}
		\end{array}
		$&
		$
		\begin{array}{c}
		\includeTikz{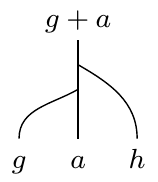}
		{
			\begin{tikzpicture}
			\Lassociator[$g$][$h$][$a$][$g+a$];
			\end{tikzpicture}
		}
		\end{array}
		=
		\begin{array}{c}
		\includeTikz{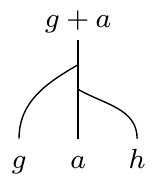}
		{
			\begin{tikzpicture}
			\Rassociator[$g$][$h$][$a$][$g+a$];
			\end{tikzpicture}
		}
		\end{array}
		$
		\\
		\greycline{2-3} 
		$F_0$ &
		$
		\begin{array}{c}
		\includeTikz{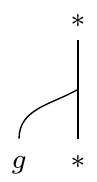}
		{
			\begin{tikzpicture}
			\Laction[$g$][$*$][$*$];
			\end{tikzpicture}
		}
		\end{array}
		$
		&
		$
		\begin{array}{c}
		\includeTikz{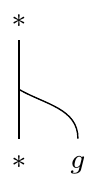}
		{
			\begin{tikzpicture}
			\Raction[$g$][$*$][$*$];
			\end{tikzpicture}
		}
		\end{array}
		$&
		$
		\begin{array}{c}
		\includeTikz{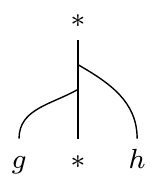}
		{
			\begin{tikzpicture}
			\Lassociator[$g$][$h$][$*$][$*$];
			\end{tikzpicture}
		}
		\end{array}
		=
		\begin{array}{c}
		\includeTikz{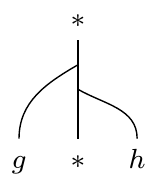}
		{
			\begin{tikzpicture}
			\Rassociator[$g$][$h$][$*$][$*$];
			\end{tikzpicture}
		}
		\end{array}
		$
		\\		
		\toprule[1pt]
		$X_k$ &
		$
		\begin{array}{c}
		\includeTikz{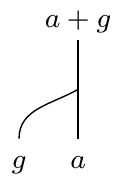}
		{
			\begin{tikzpicture}
			\Laction[$g$][$a$][$a+g$];
			\end{tikzpicture}
		}
		\end{array}
		$
		&
		$
		\begin{array}{c}
		\includeTikz{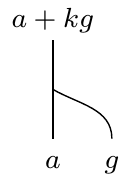}
		{
			\begin{tikzpicture}
			\Raction[$g$][$a$][$a+kg$];
			\end{tikzpicture}
		}
		\end{array}
		$&
		$
		\begin{array}{c}
		\includeTikz{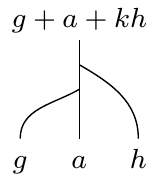}
		{
			\begin{tikzpicture}
			\Lassociator[$g$][$h$][$a$][$g+a+kh$];
			\end{tikzpicture}
		}
		\end{array}
		=
		\begin{array}{c}
		\includeTikz{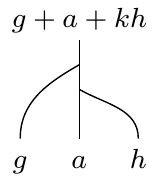}
		{
			\begin{tikzpicture}
			\Rassociator[$g$][$h$][$a$][$g+a+kh$];
			\end{tikzpicture}
		}
		\end{array}
		$
		\\
		\greycline{2-3}
		$F_q$&
		$
		\begin{array}{c}
		\includeTikz{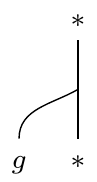}
		{
			\begin{tikzpicture}
			\Laction[$g$][$*$][$*$];
			\end{tikzpicture}
		}
		\end{array}
		$
		&
		$
		\begin{array}{c}
		\includeTikz{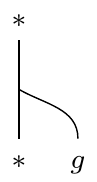}
		{
			\begin{tikzpicture}
			\Raction[$g$][$*$][$*$];
			\end{tikzpicture}
		}
		\end{array}
		$&
		$
		\begin{array}{c}
		\includeTikz{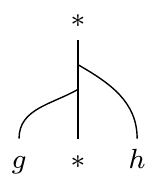}
		{
			\begin{tikzpicture}
			\Lassociator[$g$][$h$][$*$][$*$];
			\end{tikzpicture}
		}
		\end{array}
		=
		e^{\frac{2\pi i}{p} q g h}
		\begin{array}{c}
		\includeTikz{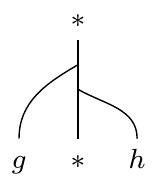}
		{
			\begin{tikzpicture}
			\Rassociator[$g$][$h$][$*$][$*$];
			\end{tikzpicture}
		}
		\end{array}
		$
		\\
		\hline
	\end{tabular}
	\caption{Data for all $\vvec{\ZZ{p}}-\vvec{\ZZ{p}}$ bimodules. $q \in H^2(\ZZ{p},U(1)) \cong \ZZ{p}$. Bimodules below the thick line are invertible. Note that in the main text, these bimodule labels are supplemented with a superscript indicating for which $\vvec{\ZZ{p}}$ they are defined. Reproduced from \onlinecite{1806.01279}.}\label{tab:zpdata}
\end{table}

\begin{table}\centering
	\begin{tabular}{!{\vrule width 1pt}>{\columncolor[gray]{.9}[\tabcolsep]}c!{\vrule width 1pt}c c c c!{\vrule width 1pt}c c!{\vrule width 1pt}}
		\toprule[1pt]
		\rowcolor[gray]{.9}[\tabcolsep]$\otimes_{\vvec{\ZZ{p}}}$&$T$&$L$&$R$&$F_0$&$X_l$&$F_r$\\
		\toprule[1pt]
		$T$&$p\cdot T$&$T$&$p\cdot R$&$R$&$T$&$R$\\
		$L$&$p\cdot L$&$L$&$p\cdot F_0$&$F_0$&$L$&$F_0$\\
		$R$&$T$&$p\cdot T$&$R$&$p\cdot R$&$R$&$T$\\
		$F_0$&$L$&$p\cdot L$&$F_0$&$p\cdot F_0$&$F_0$&$L$\\
		\toprule[1pt]
		$X_k$&$T$&$L$&$R$&$F_0$&$X_{kl}$&$F_{k^{-1}r}$\\
		$F_q$&$L$&$T$&$F_0$&$R$&$F_{ql}$&$X_{q^{-1}r}$\\
		\toprule[1pt]
	\end{tabular}
	\caption{Multiplication table for $\protect\bpr{\protect\vvec{\ZZ{p}}}$, reproduced from \onlinecite{1806.01279}.}\label{tab:zpBPRtable}
\end{table}
\begin{table}\centering
	\begin{tabular}{!{\vrule width 1pt}c!{\vrule width 1pt}c|c!{\vrule width 1pt}}
		\toprule[1pt]
		\rowcolor[gray]{.9}[\tabcolsep]Bimodule label & Domain wall & Action on particles\\
		\toprule[1pt]
		$T$&$\begin{array}{c}\includeTikz{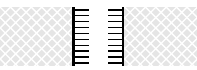}{
			\begin{tikzpicture}[yscale=.3]
			\draw[white](0,-1.1)--(0,1.21);
			\fill[fill=bulkcolor,postaction={pattern=crosshatch,pattern color=white}](-1,-1) rectangle (-.25,1);
			\fill[fill=bulkcolor,postaction={pattern=crosshatch,pattern color=white}](1,-1) rectangle (.25,1);
			\draw[thick](-.25,-1)--(-.25,1);
			\draw[thick](.25,-1)--(.25,1);
			\foreach \x in {0,...,9}{\draw (.1,-.9+.2*\x)--(.25,-.9+.2*\x);};
			\foreach \x in {0,...,9}{\draw (-.1,-.9+.2*\x)--(-.25,-.9+.2*\x);};
			\end{tikzpicture}}
		\end{array}$&Condenses $e$ on both sides\\
		$L$&$\begin{array}{c}\includeTikz{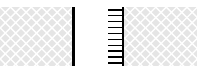}{\begin{tikzpicture}[yscale=.3]
			\draw[white](0,-1.1)--(0,1.21);
			\fill[fill=bulkcolor,postaction={pattern=crosshatch,pattern color=white}](-1,-1) rectangle (-.25,1);
			\fill[fill=bulkcolor,postaction={pattern=crosshatch,pattern color=white}](1,-1) rectangle (.25,1);
			\draw[thick](-.25,-1)--(-.25,1);
			\draw[thick](.25,-1)--(.25,1);
			\foreach \x in {0,...,9}{\draw (.1,-.9+.2*\x)--(.25,-.9+.2*\x);};
			\end{tikzpicture}}\end{array}$&Condenses $m$ on left and $e$ on right\\
		$R$&$\begin{array}{c}\includeTikz{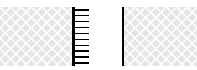}{\begin{tikzpicture}[yscale=.3]
			\draw[white](0,-1.1)--(0,1.21);
			\fill[fill=bulkcolor,postaction={pattern=crosshatch,pattern color=white}](-1,-1) rectangle (-.25,1);
			\fill[fill=bulkcolor,postaction={pattern=crosshatch,pattern color=white}](1,-1) rectangle (.25,1);
			\draw[thick](-.25,-1)--(-.25,1);
			\draw[thick](.25,-1)--(.25,1);
			\foreach \x in {0,...,9}{\draw (-.1,-.9+.2*\x)--(-.25,-.9+.2*\x);};
			\end{tikzpicture}}\end{array}$&Condenses $e$ on left and $m$ on right\\
		$F_0$&$\begin{array}{c}\includeTikz{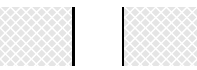}{\begin{tikzpicture}[yscale=.3]
			\draw[white](0,-1.1)--(0,1.21);
			\fill[fill=bulkcolor,postaction={pattern=crosshatch,pattern color=white}](-1,-1) rectangle (-.25,1);
			\fill[fill=bulkcolor,postaction={pattern=crosshatch,pattern color=white}](1,-1) rectangle (.25,1);
			\draw[thick](-.25,-1)--(-.25,1);
			\draw[thick](.25,-1)--(.25,1);
			\end{tikzpicture}}\end{array}$&Condenses $m$ on both sides\\
		\toprule[1pt]
		$X_k$&$\begin{array}{c}\includeTikz{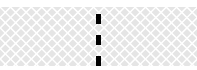}{\begin{tikzpicture}[yscale=.3]
			\draw[white](0,-1.1)--(0,1.21);
			\fill[fill=bulkcolor,postaction={pattern=crosshatch,pattern color=white}](-1,-1) rectangle (1,1);
			\draw[ultra thick,dashed] (0,-1)--(0,1);
			\end{tikzpicture}}\end{array}$&$X_k:e^am^b\mapsto e^{ka}m^{k^{-1} b}$\\
		$F_{q}=F_1 X_q$&$\begin{array}{c}\includeTikz{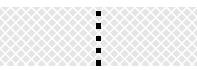}{\begin{tikzpicture}[yscale=.3]
			\draw[white](0,-1.1)--(0,1.21);
			\fill[fill=bulkcolor,postaction={pattern=crosshatch,pattern color=white}](-1,-1) rectangle (1,1);
			\draw[ultra thick,dotted] (0,-1)--(0,1);\end{tikzpicture}}\end{array}$&
		$F_1:e^am^b\mapsto e^{b}m^{a}$\\
		\toprule[1pt]
	\end{tabular}
	\caption{Domain walls on the lattice corresponding to bimodules. Reproduced from \onlinecite{1806.01279}.}\label{tab:Zpbimodinterp}
\end{table}

\end{document}